\documentclass{aa}  

\usepackage{graphicx}
\usepackage[colorlinks,pdfusetitle,urlcolor=blue,citecolor=blue,linkcolor=blue,bookmarksnumbered,plainpages=false]{hyperref}
\usepackage{adjustbox}

\usepackage{txfonts}
\usepackage{subfigure}

\def\R23{\mbox{$\rm R_{23}$}}

\begin{document}

 \title{The VLT-MUSE and ALMA view of the MACS 1931.8-2635 brightest cluster galaxy}


   \author{Ciocan B. I.
          \inst{1}
          \and
          Ziegler, B. L.
          \inst{1}
          \and
           Verdugo, M.
          \inst{1}
           \and
       Papaderos, P.
       \inst{1,2,3}
          \and
        Fogarty, K.          
            \inst{4,5}
             \and
       Donahue, M.
       \inst{6} 
       \and
       Postman, M.
       \inst{4}}

   \institute{Institute for Astronomy (IfA), University of Vienna,
              T\"urkenschanzstrasse 17, A-1180 Vienna\\
              \email{ciocan.bianca-iulia@univie.ac.at}
        \and
           Instituto de Astrof\'{i}sica e Ci\^{e}ncias do Espa\c{c}o, Universidade de Lisboa, OAL, Tapada da Ajuda, PT1349-018 Lisboa, Portugal 
         \and
         Departamento de F\'{i}sica, Faculdade de Ci\^{e}ncias da Universidade de Lisboa, Edifício C8, Campo Grande, PT1749-016 Lisboa, Portugal
	\and
             Division of Physics, Math, and Astronomy, California Institute of Technology, Pasadena, CA, USA
            \and
            Space Telescope Science Institute, Baltimore, MD, USA
                        \and
            Physics and Astronomy Dept., Michigan State University, East Lansing, MI 48824, USA
            }
   \date{Received 11.2020 ; accepted 01.2021}

\abstract 
{
We reveal the importance of ongoing in-situ star formation in the Brightest Cluster Galaxy (BCG) in the massive cool-core CLASH cluster MACS 1931.8-2635 at a redshift of z=0.35 by analysing archival VLT-MUSE optical integral field spectroscopy. Using a multi-wavelength approach, we assess the stellar and warm ionized medium components, spatially resolved by the VLT-MUSE spectroscopy, and link them to the molecular gas by incorporating sub-mm ALMA observations.\\
We measure the fluxes of strong emission lines such as:   $\rm{[O\textsc{ii}]}\:\lambda 3727$, $\rm{H\beta}$,  $\rm{[O\textsc{iii}]}\:\lambda 5007$,  $\rm{H\alpha}$, $\rm{[N\textsc{ii}]}\: \ \lambda 6584 $ and $\rm{[S\textsc{ii}]}\:\lambda 6718, 6732 $, allowing us to determine the physical conditions of the warm ionized gas, such as electron temperature, electron density, extinction, ionization parameter, (O/H) gas metallicities, star formation rates and gas kinematics, as well as the star formation history of the system. Our analysis reveals the ionizing sources in different regions of the galaxy.\\
The ionized gas flux brightness peak corresponds to the location of the supermassive black hole in the BCG and the system shows a diffuse warm ionized gas tail extending 30 kpc in N-E direction. The ionized and molecular gas are co-spatial and co-moving, with the gaseous component in the tail likely falling inward, providing fuel for star formation and  accretion-powered nuclear activity. The gas is ionized by a mix of star formation and other energetic processes which give rise to LINER-like emission, with active galactic nuclei emission dominant
only in the BCG core. We measure a star formation rate of  $\sim 97\: \rm{M_{\odot}/yr}$, with its peak at the BCG core. However, star formation accounts for only 50-60\% of the energetics needed to ionize the warm gas. In situ star formation generated by thermally unstable intracluster medium cooling and/or  dry mergers dominate the stellar mass growth of the BCG at z<0.5 and these mechanisms account for the build-up of  20\% of the stellar mass of the system. Our measurements reveal that the most central regions of the BCG contain the lowest gas phase oxygen abundance whereas the $\rm{H\alpha}$ arm exhibits slightly more elevated values, suggesting the transport of gas out to large distances from the centre due to active galactic nuclei outbursts. The galaxy is a dispersion dominated system,  typical for massive, elliptical galaxies. The gas and stellar kinematics are decoupled, with the gaseous velocity fields being more closely related to the bulk motions of the intracluster medium.}


\maketitle



\setcounter{section}{0}
\section{Introduction}
The centres of massive clusters are often dominated by massive elliptical galaxies (brightest cluster galaxies, BCGs hereafter), suggesting a strong link between the formation and evolution of the BCG and that of the host cluster.  For example, \cite{lauer14} have  observationally  demonstrated that the structural properties of the BCG depend on its specific location within the cluster, such that BCGs which are closer to the X-ray centre or which have smaller peculiar velocities (relative to the cluster mean), have more extended envelopes. This suggests that the inner  regions of BCGs are formed outside the cluster but interactions, both gravitative and hydrodynamical, in the heart of the cluster lead to the growth of the envelopes of these systems.\\ The unique properties of BCGs were proposed to have arisen through a few special mechanisms. Such a mechanism was proposed by  \cite{f77} to explain the formation and evolution of these systems as the result of cooling flows.  Another explanation for the formation of these galaxies involves galactic cannibalism due to dynamical friction \citep{o77}. More recent theoretical models favour a two–phase hierarchical formation scenario for BCGs: rapid cooling and in-situ star formation at high redshifts followed by a growth through repeated mergers (e.g.  \citealt{d07}, \citealt{bell16}, \citealt{lavoie16}, \citealt{burke13}, \citealt{cerulo19}). 
Several observations confirm the existence of BCGs in a state of an ongoing/recent merger phase, exhibiting, by e.g., close companion galaxies \citep{rines07}. 
However, many BCGs located in cool-core clusters, still exhibit signatures of significant star formation (SF) ( e.g.  \citealt{dh10},  \citealt{tremb12}) and also harbour significant amounts of molecular gas even at lower redshifts  (e.g. \citealt{mcn14}, \citealt{olivares19}, \citealt{fogarty19}), indicating that their stellar mass can also be built up via in-situ SF at later epochs.  Thermally unstable residual intracluster medium (ICM) cooling  was shown to explain the on-going SF  in cluster cores  (e.g.\citealt{voit15b}).  However,  in the absence of a source of external heating, cluster cores would experience rapid cooling with $\tau_{cool} < 1$ Gyr  and extremely high rates of mass deposition - up to $\sim$ 1000 $\rm{M_{\odot}/yr}$ - onto the BCG, promoting starbursts (e.g \citealt{mcn07}). The lack of such observational signatures hints to the fact that a source of heating must be present, and the best candidate is the central  active galactic nucleus (AGN). Thus, in recent years it has become clear that galaxy cluster cores are particularly well suited to study the feedback processes that are thought to inhibit the cooling of gas. Cosmological hydrodynamical simulations have shown that AGN feedback can regulate the cooling in the ICM, such that some ICM condensation occurs, but  the overall energy injected by the AGN into the ICM offsets radiative cooling and prevents the formation of the $1000\:\rm{s\:M\odot/yr}$ cooling flows (\citealt{li15},  \citealt{li17}, \citealt{gaspari18}). The modelling of mechanical-mode AGN feedback  has  demonstrated that, as feedback acts on the system, the outflows it drives can promote condensation of the hot, low-entropy ambient medium by raising some of it to greater altitudes. This adiabatic uplift promotes condensation by lowering the cooling time - free fall time  $\rm{(t_{cool} /t_{ff})}$ ratio of the uplifted gas. The diffuse X-ray emitting gas in the cluster centre is expected to become highly susceptible to local thermal instabilities once $\rm{(t_{cool} /t_{ff})}$ drops below a threshold  of $\sim$10. 
Once this happens, the condensates can rain down toward the bottom of the potential well, giving rise to kpc long cold filaments threading the BCG and its outskirts. This rain of cold gas into the galaxy, at first, provides additional fuel for  star formation and AGN feedback and temporarily boosts the strength of the outflows, but eventually, those strengthening outflows add enough heat to the ambient medium to raise $\rm{(t_{cool} /t_{ff})}$ high enough to stop the condensation. Therefore, this  precipitation-regulated feedback  is a “cold feedback” mechanism that fuels a central black hole through “chaotic cold accretion” (\citealt{gaspari13}, \citealt{voit15b}, \citealt{ps15}, \citealt{voit17}).\\
For example, \cite{dh15} used CLASH-HST  UV photometric data to study the UV morphologies and SFRs of 25 CLASH BCGs with redshifts $z \sim 0.2 - 0.9$. They demonstrated that  the only  cluster cores hosting BCGs with detectable SF are those with low entropy X-ray emitting gas in their centres, in accordance with theoretical models. These galaxies exhibit a wide diversity of morphologies, with strong UV excesses systems showing distinctive knots, multiple elongated clumps, and extended filaments of emission that differ from the smooth profiles of the UV-quiet BCGs. These filamentary structures, which are similar to those seen in SF BCGs at lower z,  suggest  bi-polar streams of clumpy star formation. The unobscured star formation rates  (SFRs) estimated from the UV images are in the order of  $80 \: M_{\odot} yr^{-1}$ in the most extended and highly structured systems. The morphology of the star-forming UV structures is very similar to the cold-gas structures, which are produced in simulations of precipitation-driven AGN feedback \citep{li14}. \\ Likewise, \cite{tremb15}  have analysed the star-forming clouds and filaments in the central regions of 16  lower redshift z < 0.3 cool core BCGs based on a multi-wavelength approach. The systems exhibit SFRs from effectively 0 up to $\sim 150 M_{\odot}/yr$. Most of the filaments they study  are FUV-bright and star forming  and the authors suggest that they have either been uplifted by the radio lobe or buoyant X-ray cavity or have formed in situ by jet-triggered star formation or rapid cooling in the cavities' compressed shell.  For the great majority of the BCGs, the maximal projected radius at which FUV emission is observed corresponds to a  $\rm{(t_{cool} /t_{ff})} \sim10$, in accordance  with theoretical predictions.\\
To reveal the importance of ongoing in-situ star formation to the total mass build-up in the most massive galaxies of our universe, we present our investigation of the warm ionised gas in the BCG of MACS 1931.8-2635 using VLT-MUSE integral field spectroscopic (IFS) observations.  The measurement of strong spectral line fluxes enables the investigation of the properties of the ionised gas  and stellar component. ALMA sub-mm observations of the M1931 BCG from \cite{fogarty19} allow us to link the ionised gas properties to those of the cold molecular gas. \\
The paper is structured as follows: in section \ref{dataaaa}, we describe the main physical properties of the MACS 1931.8-2635  galaxy cluster and its BCG;  in section \ref{data processing}, we present the archival MUSE observations of MACS 931.8-2635 and describe the additional sources of  data used in this work. This section also describes  the main steps of the data reduction and processing.  Section \ref{ana} describes the analysis of the MUSE IFS observations based on different tools and pipelines. Section \ref{results} presents  the spatially resolved properties of the ionised gas (gas maps, kinematic maps, ionisation sources, electron temperature, electron density, extinction, ionisation parameter, SFRs, gas phase metallicities). In this section, we also present the comparisons between ionised and molecular gas properties. The star formation history of the system, as well as the kinematics of the stellar component, are also discussed in this section. In Section \ref{discussion}, we discuss implications for the formation of multiphase gas in the  MACS 1931.8-2635 BCG, and the possible relationships between  features observed in the optical, X-ray, and radio. We summarise our conclusions in section \ref{conclusion}.\\
Throughout this study,  we have used the concordance  $\rm{\Lambda CDM}$ cosmology with $ \rm{H_{0}} = 70\: \rm{km \:s^{-1}\: Mpc^{-1}}$, $\Omega_{0}  = 0.32$, $\Omega_{\Lambda} = 0.68$.  With these cosmological parameters, 1'' subtends  $\sim 5$ kpc at the redshift of z=0.352 of the BCG.

\section{Data}
\label{dataaaa}
\subsection{The MACS 1931.8-2635  galaxy cluster and its BCG}
\begin{figure*}
    \centering
    \includegraphics[width=0.5\textwidth,angle=0,clip=true]{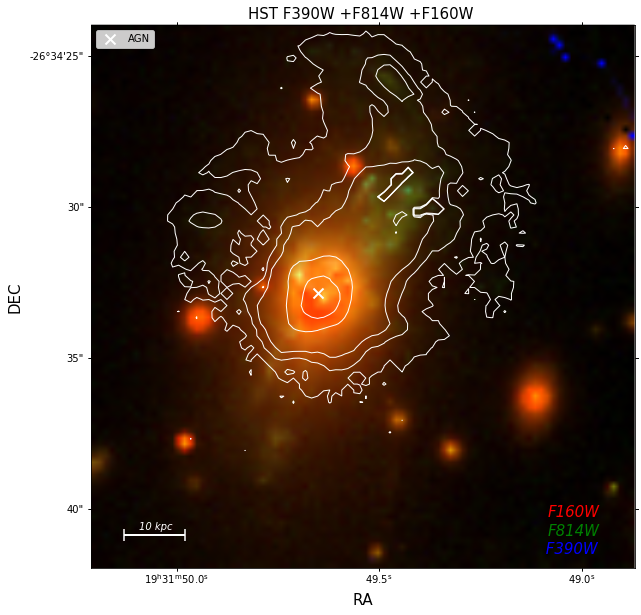}
    \centering   
    \caption{HST composite RGB image of the M1931 BCG:  the F160W image is shown in red, the F814W   in green and the F390W  in blue. The white contours show to the $\rm{H\alpha}$ flux intensity, as measured from MUSE.  The cross shows the location of the AGN, as identified according to different diagnostic diagrams, see sect. \ref{BPTsect}.}  
\label{hst}
\end{figure*}

MACS 1931.8-2635 (hereafter M1931) is a massive, X-ray luminous, cool-core  galaxy cluster at a redshift $z\sim0.35$. 
The system was  observed as part of the Cluster Lensing And Supernova survey with Hubble - CLASH - \citep{postman12} and CLASH-VLT survey \citep{rosati14}.  According to \cite{postman12}, the galaxy cluster has a X-ray temperature $ \rm{k\cdot T_{x}}=6.7$ keV and a bolometric luminosity $\rm{L_{b}=20.9\cdot 10^{44} \:ergs\:s^{-1}}$.  \cite{merten15} reported a virial radius for  the  M1931 cluster of  $\rm{r_{vir}}=1.61$ Mpc/h and a virial mass  $\rm{M_{vir}=0.83\pm0.06\cdot10^{15}\:M\odot/h}$ from their lensing analysis.  On the other hand, \cite{umetsu16} derived a virial mass for the M1931 cluster, based on  comprehensive analysis of strong-lensing, weak-lensing shear and magnification data of   $\rm{M_{vir}=1.802\pm0.9\cdot10^{15}\:M\odot/h}$. \\
Figure \ref{hst} displays a composite Hubble Space Telescope (HST) image of the M1931 BCG at a redshift of z=0.352, showing  the F160W image in red, the F814W   in green and the F390W in blue. The white contours display the $\rm{H\alpha}$ flux intensity, as measured from the MUSE data cube. This figure reveals a filamentary system with intense nebular emission. The extended emission in this galaxy seen in the HST data was  reported and characterised  by  \cite{fogarty19}. The M1931 BCG has been shown to have elevated SFR ranging from  $\sim 90 \:\rm{M_{\odot}/yr}$ from UV data \citep{dh15} to  $\sim150\:\rm{M_{\odot}/yr}$ from IR data \citep{santos16} up to $\sim 250\:\rm{M_{\odot}/yr}$  from UV through far-IR \citep{fogarty17}. Such high levels of star formation are not only atypical for "red and dead" elliptical galaxies but, they  also imply a phase of significant ongoing stellar mass growth.  The M1931 BCG  has stellar mass of $\rm{M_{*} \sim 5.9 \pm 1.1 \cdot 10^{11}\:M_{\odot}}$ \citep{bell16}.  Observations have also demonstrated that this galaxy harbours one of the most X-ray luminous cool-cores yet discovered, with an equivalent mass cooling-rate of $\sim 165 \: \rm{M_{\odot}/yr}$ according to \cite{eh11}, hinting that it might be undergoing a similar phase of  ICM condensation like for e.g. in the Phoenix galaxy cluster \citep{mc12b}. The central ICM entropy is estimated to be  $\rm{K_{0} = k_{B}\cdot T_{X}\cdot n_{e}^{-2/3}=14\pm4\: keV \:cm^3}$ according to  \cite{dh15} and hence, the M1931 BCG has a  low core entropy $(\rm{K_{0}<30\: keV \:cm^3})$, which is a necessary condition for a
 multi-phase, star forming  system \citep{voit15b}. 
\cite{eh11} has demonstrated that on scales of $r \sim 200$ kpc, a spiral of cooler, denser  X-ray gas  is observed to wrap around the core. Such spiral structures arise naturally from mergers and subsequent sloshing.
Both the X-ray and optical data reveal oscillatory motion of the cool core along a roughly north–south direction as well as extended Intra Cluster Light (ICL), suggesting  that the BCG likely experienced a merger within the north–south direction. \\ Sub-mm ALMA observations \citep{fogarty19} of M1931 have revealed that this BCG harbours one of  the largest known reservoirs of cold gas in a cluster core ($\rm{M_{H_2}=1.9 \pm 0.3\cdot 10^{10} M_{\odot}}$) as well as large amounts of dust, with several dust clumps having temperatures less than 10 K.\\ The  M1931 BCG  represents an  example of a cluster with a rapidly cooling core and powerful AGN feedback. The AGN outburst combined with merger-induced motion has most likely led to the cool-core undergoing destruction.   This system is probably transitioning  between two dominant modes of fuelling for star formation and feedback.  It might be  evolving from a quasar-mode cooling and feedback typical for higher-redshift  cool cores to a weaker and less efficient feedback and cooling mode typical of lower-redshift cluster cores.
\\
\section{ Data processing}
\label{data processing}
\subsection{Observations and data reduction}
M1931 was observed in June 2015 with the ESO-VLT MUSE integral field spectrograph \citep{bacon14}  under the GTO program 095.A-0525 (PI: Kneib, J.-P.). The MUSE pointing consists of three exposures of t=2924 s each, all centred on the cluster core. Both the raw and reduced data can be found on the ESO Science Archive Facility. We have used the raw data and reduced it using the standard calibrations provided by the ESO-MUSE pipeline, version 1.2.1 \citep{pipeline}. To reduce the sky residuals, we have used an additional tool, the Zurich Atmosphere Purge  version 1.0 - \texttt{ZAP} - \citep{soto} on the calibrated cubes.  \texttt{ZAP}  is a tool specially developed for  the reduction of  IFU data and its  core functionality is  sky subtraction based on principal component analysis (PCA). The tool employs filtering and data segmentation to enhance the inherent capabilities of PCA for sky subtraction, it constructs a sky residual spectrum for each  spaxel which can then be subtracted from the original data cube and it, therefore, reduces sky emission residuals while preserving the flux and line shapes of the astronomical objects. After accounting for the sky residuals for the three exposures, we combined them using the MUSE Python Data Analysis Framework - \texttt{MPDAF}- \citep{mpdaf} into a single data cube. The final calibrated data cube has a FoV of $1.1 \:\rm{arcmin^{2}} $, a spatial sampling of $0.2"$ in the wavelength range $4750 -9350 \:\AA$ and a spectral resolution of $\sim2.5 \:\AA$. \\
The optical IFU data is supplemented by sub-mm ALMA observations from \cite{fogarty19}, allowing us to link the  warm ionised gas to  the cold molecular gas component. This sub-mm data set contains Band 3 (beam size: 0.82 arcsec x 0.53 arcsec), Band 6 (beam size: 0.87 arcsec x 0.72 arcsec) and Band 7 (beam size: 0.94 arcsec x 0.78 arcsec) ALMA observations. 
For the astrometric mismatch correction, we also made use of the CLASH SUBARU and HST photometric data for the M1931 cluster \citep{postman12}.

\subsection{Astrometric alignment MUSE ALMA}
\label{astrometry}
Our analysis of the MUSE and ALMA data requires that the images are aligned to a common astrometric reference frame.
For this, we have used the white-light image of the MUSE cube  and corrected it for the astrometric mismatch to the  SUBARU Suprime-Cam r-band image available from the CLASH project website\footnote{https://archive.stsci.edu/prepds/clash/}, which has excellent alignment with the HST observations.  According to \cite{fogarty19}, the ALMA data is aligned with the HST data, and any systematic astrometric alignment errors between  the two data sets  is not of significant concern. Thus, accounting for the astrometric mismatch between the MUSE image and  the Subaru image is equivalent to accounting for the astrometric mismatch between MUSE and ALMA. For the correction of the astrometric mismatch, we have developed a \texttt{PYTHON} code using different \texttt{MPDAF} routines. To test the validity of the alignment between the MUSE and SUBARU images, we have used  an additional \texttt{PYTHON} package  -  the \texttt{image registration} package - and we measure an offset between the 2  images, using the DFT upsampling methods, of  [0.072, -0.17] pixels (i.e. [0.014,-0.034] acrcsec). These values are smaller than one third of the point spread function (PSF) of the MUSE IFS,  which is 3.9 pixels, and hence, the two images are almost perfectly aligned. To further test the quality of our astrometric correction, we used the catalogue with the coordinates of Two Micron All-Sky Survey stars which fall in the field of view of MUSE and investigated their position in our images. We observe a very good agreement with a minimal offset. Moreover, the  position we derive for the AGN from the BPT analysis (see Sect. \ref{BPTsect}) from the MUSE data coincides with the location of the sub-mm continuum point source, as observed in the ALMA data. 

\subsection{MUSE ALMA ratio maps}
We have created $\rm{H\alpha}$-CO(1-0) and $\rm{H\alpha}$-CO(3-2)  flux, velocity and velocity dispersion ratio (or difference) maps by dividing (or subtracting) the ALMA moment maps from the corresponding MUSE maps.   
To be able to compare the two data sets, we have smoothed  the MUSE maps to match the resolution of the ALMA data, using the \texttt{PYTHON astropy.convolution} function. Then, we have applied the \texttt{reproject} task from the \texttt{PYTHON astropy} package to resample the ALMA data onto the MUSE data pixel grids. This function resamples the  data to a new projection using interpolation  and it  essentially tells the user which pixels in the new image had a corresponding pixel in the old image. Then, the re-projected ALMA image was divided (or subtracted) from the MUSE map. It is worth mentioning that due to the lower spectral resolution of the ALMA data, molecular gas kinematics were measured only in the most central regions of the system.

\section{Analysis}
\label{ana}
In the following section, we describe the analysis of the MUSE data, which allows us to study both the gaseous and stellar component of the BCG. 
Using both the population spectral synthesis code Fitting Analysis using Differential Evolution Optimisation - \texttt{FADO} - \citep{fado} and  \texttt{PORTO3D} (\citealt{polis13}, \citealt{gomes16}), as well as the  Muse Python Data Analysis Framework - \texttt{MPDAF} - \citep{mpdaf}, we reliably measure the fluxes of strong emission lines in the optical spectrum such as  $\rm{[O\textsc{ii}]}\:\lambda 3727$, $\rm{H\beta}$, $\rm{[O\textsc{iii}]}\:\lambda 5007$,  $\rm{H\alpha}$, $\rm{[N\textsc{ii}]}\:\lambda 6584 $ and $\rm{[S\textsc{ii}]}\:\lambda 6718, 6732 $, allowing us to investigate the ionising sources, as well as to derive the electron temperature, electron density, colour-excess, ionisation parameter, star formation rates, (O/H) gas metallicities and gas kinematics. The star formation history was recovered by employing  \texttt{FADO} and \texttt{PORTO3D}. The stellar kinematics are probed using the Galaxy IFU Spectroscopy Tool -  \texttt{GIST} - \citep{gist}, which implements the Voronoi binning  routine \citep{vor} and  the Python implementation of the Penalized PiXel-Fitting routine - \texttt{pPXF} - \citep{ppxf}.

\subsection{Emission line flux measurements and star formation history with \texttt{FADO} }
\begin{figure*}
    \centering
    \includegraphics[width=0.5\textwidth,angle=0,clip=true]{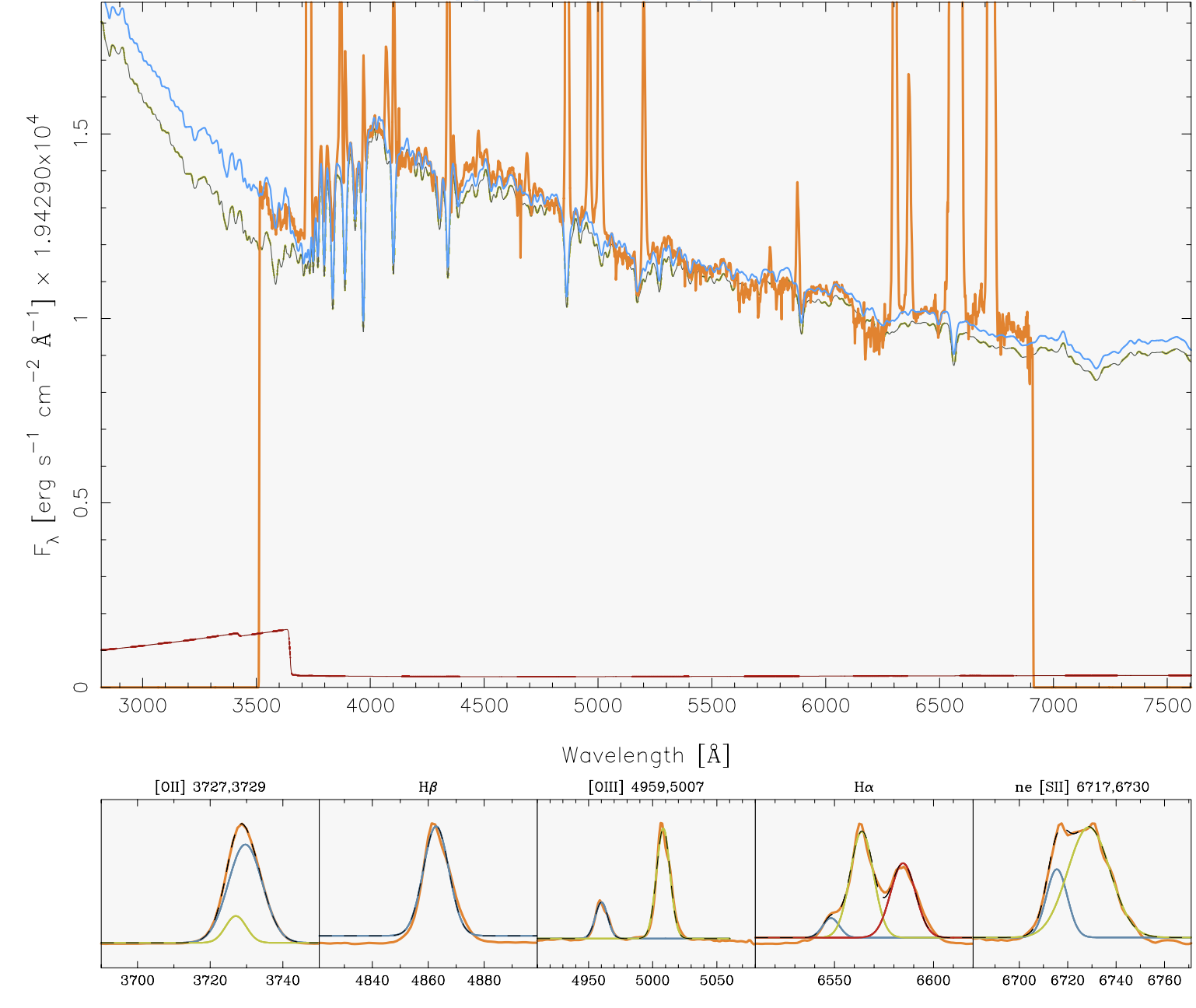}
    \centering
\caption{Graphical output of \texttt{FADO}.  The panel from the \textit{upper side}  shows the integrated spectrum of the  M1931 BCG orange, as recovered from MUSE. The best-fitting synthetic SED is shown in light-blue, composed of the  stellar and nebular continuum emission (dark green and red, respectively).
The panel from the  \textit{lower side}  displays the fits to the strongest emission lines, namely, $\rm{[O\textsc{ii}]}\:\lambda 3727, \lambda 3729$,  $\rm{H\beta}$, $\rm{[O\textsc{iii}]}\:\lambda4959, 5007$, $\rm{H\alpha}$,   $\rm{[N\textsc{ii}]}\:\lambda 6584$ and $\rm{[S\textsc{ii}]}\:\lambda 6718, 6732$. }  
\label{fado}
\end{figure*}

From the MUSE data-cube, whose field of view encloses the central regions of the whole M1931 cluster, we extracted a sub-cube centred on the BCG using different \texttt{MPDAF} routines. This sub-cube consists of  90x90 spectral pixels (spaxels), i.e. in total 8100 pixels. \\ Each spaxel  of this MUSE sub-cube is fitted  with the \texttt{FADO} pipeline \citep{fado}, which is a tool specially designed to perform population spectral synthesis (PSS), with the additional capability of automatically deriving emission line fluxes and equivalent widths (EWs). This tool can  identify  the star formation history (SFH) that reproduces  self-consistently both the observed nebular characteristics of a star forming  galaxy and the stellar SED.  \texttt{FADO} is the first PSS (i.e., 'reverse')  code  to employ genetic optimisation under self-consistency boundary conditions. This tool uses an advanced variant of the genetic Differential Evolution Optimization (DEO) algorithm of \cite{storn97} which has the advantage of permitting reliable convergence at an affordable expense of computational time.  
Further improvements of \texttt{FADO} in comparison to other PSS codes include  i)  the use of artificial intelligence  (AI) concepts for an initial spectroscopic classification and optimisation of the library of  simple stellar populations (SSP) spectra, ii) the consideration of nebular emission in spectral fits  and  iii) the consistency between the best-fitting SFH and the observed nebular emission characteristics of a star-forming  galaxy. \\
The main modules of FADO are the following: 1) pre-processing of spectral data, 2) spectral synthesis through genetic DEO and 3) computation and storage of the model output.  After the  initial guess for the fitting strategy, which depends on the spectral classification and signal-to-noise of the input spectra, the SSP library is optimised through AI concepts. There are three fitting strategies: Full-Consistency (FC) mode, meaning that the spectral modelling aims at consistency between observed and predicted SED continuum  and Balmer emission-line luminosities and EWs - this being the  default  fitting mode of \texttt{FADO} and the one we have used for the fitting -  the nebular-continuum mode and the stellar mode.
In the fitting module,  \texttt{FADO}  incorporates various  quality checks such as  the auto-determination of the emission-line ratios prior to fitting, automatic clipping of spurious spectral features and examination of the supplied error spectrum. At  the first stage, emission-line fluxes and EWs are measured and their quality is investigated. The quality control first involves a sequential check of various quantities and their errors inferred from DEO-based Gaussian line fitting and de-blending, such as the full width at half maximum (FWHM) and the difference between the central wavelength of emission lines. This first step is devised to identify the spurious spectral features (i.e.  residuals from cosmics, from sky correction, noise peaks) as outliers and to reject them. \texttt{FADO}  also checks whether various emission-line ratios  fall within the range of theoretically expected values. This  second module also deals with the decision-tree based choices of fitting strategy and convergence schemes, with the computation of predicted Balmer-line luminosities and nebular continuum and with the estimation of the uncertainties.
The third FADO module deals with the final measurement of emission-line fluxes and  EWs (the widths are not fixed to the same value for all lines, but determined individually)  and with the computation of secondary evolutionary quantities, such as light- and mass-weighted stellar age and metallicity.\\
Before the fitting, all the spaxels were corrected for Galactic foreground extinction with the \texttt{noao onedspec deredden IRAF} routine, using the empirical selective extinction function of \cite{c89}.
For the fitting routine, we have used the library of  SSP spectra from \cite{bc03}. The SSPs have ages between $1\: x \:10^5$ and $1\: x \:10^{10}$ yrs, a resolution of 3 $\AA$ across the  wavelength range 3200 to 9500 $\AA$ and a wide range of metallicities  (1/200, 1/50, 1/5, 1/2.5, 1 and 2.5 $Z_{\odot}$) for Padova 1994 \citep{96} evolutionary tracks. \texttt{FADO} also allows the user to specify which extinction law to be used and we have chosen the Calzetti law extended to the FUV \citep{calz} for the fitting. It is worth mentioning that the fluxes offered by \texttt{FADO} are corrected for underlying stellar absorption.\\
Figure \ref{fado} shows in the upper panel the integrated spectrum of the 90x90 spaxels MUSE cube, centred on the M1931 BCG  in orange, revealing intense nebular emission. The best-fitting synthetic SED is shown in light-blue and it is composed of stellar and nebular continuum emission - the dark green and red spectra, respectively. The 5 panels from the lower side show the Gaussian fits to the strongest emission lines in the spectrum. These panels allow the user to inspect the quality of the kinematical fitting. The code is designed such as to fit the  $\rm{[O\textsc{ii}]}$ as a doublet, even if the lines are completely blended, as in our case. To overcome this problem, we simply added the fluxes offered by \texttt{FADO} for the $\rm{[O\textsc{ii}]}$ emission lines. It is clear from Fig. \ref{fado}, that the tool  manages to properly fit the  $\rm{H\alpha}$ and $\rm{[N\textsc{ii}]}$ emission lines, which are blended in the spectrum of the M1931 BCG. \texttt{FADO} rejects the $\rm{H\alpha}$ + $\rm{[N\textsc{ii}]}$  de-blending solution 
 if the $\rm{[N\textsc{ii}]} \:6548/6584$ lines differ in their FWHM by more than an error-dependent tolerance bound defined by the error of the individual line fluxes, or in the case that the redshift-corrected difference between the central wavelength between the $\rm{H\alpha}$ and $\rm{[N\textsc{ii}]}$6548/6584 lines do not match the nominal value. We recover a median value for the flux errors for $\rm{log(H\alpha/}\rm{[N\textsc{ii}]}\:\lambda 6584)$ of $\pm0.029$. It is worth mentioning that  the uncertainties in the correction for underlying absorption are included in the errors quoted by FADO for the flux measurements of the emission lines. 
 Therefore, the flux measurements for the blended  $\rm{H\alpha}$ + $\rm{[N\textsc{ii}]}$  lines can be considered to be robust.\\
Additionally, we used the IFU data analysis pipeline \texttt{Porto3D} (\citealt{polis13}, \citealt{gomes16})  to double-check emission-line fluxes, EWs  and the SFH, finding an overall good agreement with \texttt{FADO}. Whereas \texttt{Porto3D} uses the PSS code \texttt{Starlight} \citep{starlight} for fitting the stellar continuum, it shares with \texttt{FADO} essential aspects in the analysis of the residual nebular emission, the computation of spectral synthesis byproducts (e.g., mass- and light-weighted stellar age, luminosity fraction of stellar populations younger than various age cutoffs) and the storage and graphical output of the results. Additionally, it integrates a rectification technique that suppresses residuals between observed and synthetic stellar SED, this way improving on the extraction and analysis of the nebular component in weak-line sources, such as early-type galaxies \citep{gomes16} and galaxy bulges \citep{breda}.\\
 As a final consistency check, we have used   \texttt{MPDAF} \citep{mpdaf}, for the measurements of the emission line fluxes.  We develop a \texttt{PYTHON} code performing for each spaxel simultaneous Gaussian line fitting for the emission lines of interest, after subtracting the stellar continuum. The fit to each emission line is automatically weighted by the variance of the spectrum. The free parameters of the code are the peak position of the Gaussians, their standard deviation  and the amplitude.\\
We find a  very good agreement between the flux measurements offered by the three different tools, within the errors. Nevertheless, we chose to use the \texttt{FADO} measurements to study the properties of the ionised gas, because  this tool aims at self-consistency between observed and predicted SED continuum  and Balmer emission-line luminosities and EWs. \\  For our analysis, we  consider only the spaxels which have a SNR>10 in the emission line of interest, as well as a $\rm{SNR_{emission\:line} > SNR_{20\:\AA\:blue \:continuum\:  window}}$ and  $\rm{SNR_{emission\:line} > SNR_{20\:\AA\:red\:continuum\:  window}}$. All the emission lines which have a  $\rm{SNR}>10$ also have EWs$>10\:\AA$, giving us the confidence, that there is a true line detection in the spectrum. \\

\subsection{Determination of the ionising sources}
\label{ioni}
The ionisation sources in the BCG of M1931 were investigated by means of different diagnostic diagrams. Using a set of four strong emission lines, one can reliably distinguish between SF galaxies, Seyfert II galaxies, LINERs and composite galaxies, where gas excitation is powered both by SF regions and an AGN. We have used three different diagnostic diagrams on our data set: the classical BPT diagram \citep{bald81} as well as the  $\rm{[O\textsc{iii}]/H\beta}$ vs $\rm{[S\textsc{ii}]/H\alpha}$  and the  $\rm{[O\textsc{iii}]/H\beta}$ vs $\rm{[O\textsc{i}]/H\alpha}$ diagnostic diagram of \cite{vel87}. \\ 
We have also  tested predictions from different fully radiative shock models calculated with the shock and photoionisation code \texttt{MAPPINGS V} from the  Mexican Million Models database of \cite{3mdb} on the data set, to see whether large scale shocks are responsible for ionising the gas. The database contains models based on previous projects such as an replica of the \cite{allen08} grids including incomplete shock models, an extension of the \cite{allen08} grids computed for low metallicities using the abundances of \cite{gutkin16}, an extension of the \cite{allen08} grids computed for different shock ages and grids for low shock velocities  by \cite{di14}. All the different grids were tested on M1931 using different metallicities, pre-shock densities,  and shock velocities.\\
To more thoroughly  investigate the ionising sources in the BCG, we have also used the spectral decomposition method of \cite{davies17}. The authors have introduced a method to isolate the contributions of star formation,  AGN activity and LINER emission (in their case LINER emission is associated to shock excitation, but LINER-like emission can arise due to a manifold of mechanisms) to the emission line luminosities of individual spatially resolved regions in galaxies. The method works as follows: from the distribution of the spaxels in the diagnostic diagrams, one should select three ‘basis spectra’, one representative of pure star formation, one for AGN emission, and one for LINER emission. Then, one assumes that the observed  luminosity L of any emission line i in the spectrum of any spaxel j from the IFU data cube can be expressed as a linear superposition of the line luminosities of the SF region basis spectrum, the LINER basis spectrum and the AGN basis spectrum, through the following equation:
\begin{equation}
\rm{L_{i}(j)=m(j) \cdot L_{i}(SF)+n(j)\cdot L_{i}(AGN)+k(j)\cdot L_{i}(LINER)} \label{davies}
\end{equation}
For each spaxel of the M1931 data cube centred on the BCG, we calculate the superposition coefficients m(j), n(j) and k(j) by performing least-squares minimization on equation \ref{davies}  applied to the extinction-corrected fluxes of the four strongest emission lines observed in our spectra, namely  $\rm{[O\textsc{iii}]}\:\lambda 5007$,  $\rm{H\alpha}$,  $\rm{[N\textsc{ii}]}\:\lambda 6584 $ and $\rm{[S\textsc{ii}]}\:\lambda 6718, 6732 $. Then, we use the computed superposition coefficients to calculate the  luminosities of the emission lines of interest associated with star formation, LINER-like excitation and AGN activity for each spaxel of the data cube. The observed luminosities seem to be well reproduced by linear superpositions of the line luminosities extracted from the three basis spectra.  \\  However, the selection of the 3 basis spectra  characteristic for SF, AGN and LINER emission is quite subjective and the computed model luminosities are highly dependent on the choice of these spectra. Additionally, these basis spectra, although clearly classifiable as SF, AGN or LINER, may contain contributions by other gas excitation mechanisms. Therefore, spectral decomposition into these three basis elements may be regarded as a first approximation.
For this reason, we have re-calculated the superposition parameters m(j), n(j) and k(j) and the model luminosities by performing a second iteration. After the first iteration, we calculated the luminosity of the $\rm{[O\textsc{iii}]}$,  $\rm{H\alpha}$,  $\rm{[N\textsc{ii}]}$ and $\rm{[S\textsc{ii}]}$ emission lines associated purely with SF for each spaxel of the data-cube. This was done  by subtracting from the observed luminosity of each of the aforementioned 4 emission lines the contribution from AGN and LINER emission. 
The median values for the luminosities of the  $\rm{[O\textsc{iii}]}$,  $\rm{H\alpha}$,  $\rm{[N\textsc{ii}]}$ and $\rm{[S\textsc{ii}]}$ emission lines of all spaxels were then used as the new luminosities for the star formation  basis spectrum. The emission line luminosities for the AGN and LINER basis spectra were  the same ones as in the first iteration.  Then, we computed the  new values for m(j), n(j) and k(j) by conducting least-squares minimization on equation \ref{davies} with the new values for $\rm{L_{i}(SF)}$. The recovered model luminosities are in very good agreement to the observed ones.

\subsection{Determination of the electron density and temperature with \texttt{Pyneb} }
The electron temperature and density were computed by means of the \texttt{PyNeb} tool \citep{pyneb}, a \texttt{PYTHON} package for analysing emission lines.  \texttt{PyNeb}'s main functionality is to solve for line emissivities and to determine electron temperature and density given observed diagnostic line ratios. The tool works by solving the equilibrium equations for an n-level atom for collisionally excited lines and in the case of recombination lines, it works by interpolation in emissivity tables. 
Our main interest was the \texttt{getCrossTemDen} function of the tool,  a class which cross-converges the temperature and density derived from two sensitive line ratios, by inputting the quantity derived with one line ratio into the other and then iterating. When the iteration process ends, the two diagnostics are expected to give self-consistent results. The first line ratio provided by the user must be a temperature-sensitive one and the second a density-sensitive one. A large number of published diagnostic ratios are stored in PyNeb by default, and we have tested four of them: ($\rm{[O\textsc{iii}]} 4363/5007$,  $\rm{[S\textsc{ii}]}6731/6716$),  ($\rm{[N\textsc{ii}]}5755/6548$,  $\rm{[S\textsc{ii}]} 6731/6716$) , ($\rm{[N\textsc{ii}]} 5755/6584$,  $\rm{[S\textsc{ii}]} 6731/6716$) and  ($\rm{[N\textsc{ii}]} 5755/6584$,  $\rm{[Ar\textsc{iv}]} 4740/4711$). 
However, it is worth mentioning that we measure very weak $\rm{[O\textsc{iii}]}\:4363$,  $\rm{[Ar\textsc{iv}]}\:4740$ and $\rm{[Ar\textsc{iv}]}\:4711$ emission for M1931 BCG, making these diagnostics less robust.\\

\subsection{Determination of the star formation rate and colour excess E(B-V)}
The SFR was computed based on the extinction corrected luminosity of the $\rm{H\alpha}$ line, both for the integrated spectrum and for each spaxel of the MUSE cube subtending the BCG.
The $\rm{H\alpha}$ emission line is one of the most reliable SFR indicators, as this nebular emission arises directly from the recombination of HII gas ionised by the most massive O- and early B-type stars and, therefore, traces the star formation over the lifetimes of these stars. 
We make the simplifying assumption that the SFR is nearly constant over the past $\sim100$ Myr and that  case B recombination applies and therefore, the $\rm{H\alpha}$ luminosity can be used for estimating the SFR following the conversion proposed by  \citep{ken98} for solar metallicity and a Salpeter IMF: 
\begin{equation} \rm{SFR({M}_{_{\odot}} \cdot yr^{-1})=7.9 \cdot 10^{-42} L(H\alpha)}\:  (\rm{\frac{ergs}{s}})  \label{eq:5.3.1}\end{equation}
However, the intensities of emission lines arising from gas nebulae are affected by selectively absorbing material on the line of sight to the observer. Therefore, the luminosity of the $\rm{H\alpha}$  emission line was  corrected for extinction based on the Balmer decrement following the equations introduced by \cite{brock71}.  The colour excess was also calculated based on the Balmer decrement, following the same set of equations.\\
We also computed the SFR surface density  as $\rm{\Sigma SFR= SFR/area}$.  According to the cosmology, at  z=0.352 the observed  scale is 4.9 kpc/arcsec. As each spatial pixel is 0.2 arcsec,  this translates to each pixel having a side of 0.98 kpc.

\subsection{Determination of the oxygen abundance and ionisation parameter with  \texttt{HII-CH-mistry} }
The chemical abundances of galaxies are an important tool to study galaxy evolution, as they reflect the complex interplay between star formation, gas outflows through winds and supernovae, and galactic gas inflows.
The gas phase oxygen abundances for the M1931  BCG	 were computed by applying the direct $\rm{T_{e}}$  based methods described by \cite{pm17} (equations 38, 40 and 41), which use in addition to the emission line flux ratios, the temperature and density of the ionised gas. \\  As a consistency check, we have also used the  \texttt{HII-CH-mistry} pipeline \citep{pm14} for the computation of the gas phase metallicity.
Based on the diagnostic diagrams (see sect. \ref{BPTsect}), we have seen that the M1931 BCG does not have a strong optical AGN, and therefore, for the computation of (O/H)s, we have used  HII region grids calculated with \texttt{Cloudy v.17} \citep{cloudy} using the POPSTAR synthesis evolutionary models for  an instantaneous burst with an age of 1 Myr  and a Chabrier initial mass function \citep{chabrier03}. The grids of photoionization models  cover a wide range of input conditions of (O/H) and (N/O) abundances and ionisation parameter. 
The values offered by the code for the (O/H) of each spaxel are in very good agreement (within the errors) with the one obtained by us employing the  direct  $\rm{T_{e}}$  methods, giving us the confidence that the measurements are robust.\\
The ionisation parameter was also computed using  the  \texttt{HII-CH-mistry tool}  of \cite{pm14}.

\subsection{Determination of the stellar kinematics with \texttt{GIST}}
\label{starkinem}
The stellar kinematics were recovered using the  Galaxy IFU Spectroscopy Tool - \texttt{GIST}- \citep{gist}, a pipeline written in \texttt{PYTHON},  which extracts stellar kinematics,  performs emission-line analysis and derives stellar population properties from full spectral fitting by exploiting the well-known  penalized PiXel-Fitting -\texttt{pPXF}-  \citep{ppxf} and  Gas and Absorption Line Fitting -\texttt{GandALF}-  \citep{gandlaf} routines. This pipeline also implements the Voronoi binning \citep{vor} routine. We have used this tool only to recover the stellar kinematics via  \texttt{pPXF}. However,  it is worth mentioning that the fluxes recovered from  \texttt{GandALF}  are in good agreement to the ones recovered from \texttt{FADO}, \texttt{Porto3D} and \texttt{MPDAF}.\\
To spatially resolve the BCG stellar kinematics we probe each spaxel of the MUSE data-cube centred on the system, where the SNR of the stellar continuum is higher than 10. The SNR is computed in a 1000 $\AA$ window between 5250 $\AA$  and 6250 $\AA$, a region of the spectrum free of strong emission lines. The SNR of the stellar continuum is so low that we need spatial binning and therefore, we have applied the Voronoi tesselation technique. The MUSE cube was tessellated to achieve a SNR of 50 (per bin) in the emission line-free stellar continuum.
Therefore, we can measure the velocity fields only in the BCG core, corresponding to a  $2\times2$ arcsec region surrounding the supermassive black hole (SMBH).
We proceed with a  \texttt{pPXF} fit implementing the high resolution, UV-extended ELODIE models of \cite{maratson}. These SSP models are based on the template stellar library ELODIE \citep{prugniel07} merged with the theoretical spectral library UVBLUE \citep{meriono05}. The  SSPs have a Salpeter initial mass function \citep{salpeter}, a metallicity of $Z = 0.02\:Z_{\odot}$ and ages ranging from 3 Myr to 15 Gyr. The resolution is 0.55 $\AA$ (FWHM) and the  spectral sampling is  0.2 $\AA$, covering  the wavelength range 1000 - 6800 $\AA$.
In order to optimise the  \texttt{pPXF} absorption-line fits,  all the regions containing strong emission lines in all the spectra were masked ,  along with the telluric sky-lines at 5577 $\AA$, 5889 $\AA$, 6157 $\AA$, 6300 $\AA$ and 6363 $\AA$. During the  \texttt{pPXF} fitting routine, the Hermite coefficients were kept fixed. \\

\section{Results}
\label{results}
\subsection{Ionised gas flux maps}
Figure \ref{Ha} shows on the left hand side the spatially resolved map of the $\rm{H\alpha}$ emission in the M1931 BCG in units of $10^{-17} \rm{ergs\cdot s^{-1} \cdot cm^{-2}}$ for spaxels with a $\rm{SNR_{H\alpha}>10}$. The plot from the right hand side displays the $\rm{H\alpha}$ EW map of the BCG in units $\AA$, for which we have applied the same SNR criterium.  The spatial scale in (all) the plots corresponds to 90 by 90 kpc. The galaxies' intensity peak is coincident with the location of the AGN (as derived according to the different diagnostic diagrams, see sect. \ref{BPTsect}), and it shows an elongated $\rm{H\alpha}$ tail extending $\sim$ 30 kpc in N-E direction. The EW peak does not spatially coincide with the emission line brightness peak, but shows more enhanced values in the $\rm{H\alpha}$ arm.\\ We observe a similar distribution of fluxes and EWs for   $\rm{[O\textsc{ii}]}\:\lambda 3727$, $\rm{H\beta}$,  $\rm{[O\textsc{iii}]}\:\ \lambda 5007$,$\rm{[N\textsc{ii}]}\: \lambda 6584 $ and $\rm{[S\textsc{ii}]}\:\lambda 6718, 6732 $, see Fig. \ref{fluxtot}. The similarity between the flux maps of the different emission lines suggests that both recombination and forbidden lines probably originate from  the same clouds.\\  

\subsection{Comparison between ionised and molecular gas  fluxes}
We have created ratio maps between the line intensities of the ionised and molecular gas (see Fig. 2 from \citealt{fogarty19}),  after normalisation or rescaling by some  factor. Figure \ref{musealmaflux} displays in the panel from the left-hand side the ratio  between the $\rm{H\alpha}$ and CO(1-0)  flux and in the panel from the right-hand side the ratio between the $\rm{H\alpha}$ and CO(3-2) flux. The  $\rm{H\alpha}$ to CO flux ratios are close to unity and more or less constant all along the nebula and the peak of the CO flux intensity is located at the same position as the peak of the ionised gas flux intensity.  There are some regions that show more enhanced flux for $\rm{H\alpha}$ than CO, but there is an overall  close similarity between the line intensities of the ionised and molecular gas. The molecular gas is not as extended as the warm ionized gas, but this is likely due to the   sensitivity limit or due to the maximum resolution scale of ALMA. \\The molecular filaments are, thus, co-spatial in projection with the warm ionized gas, similar to what has been found in other cool core BCGs (\citealt{olivares19}, \citealt{tr18}, \citealt{Vantyghem16}).

\subsection{Ionised gas kinematics}
The gas kinematics were recovered from the Gaussian fits to the emission lines. We probe each  spaxel where the SNR>10 for the emission lines, and compute both the radial velocity and velocity dispersion of the gas from the  fits offered by \texttt{FADO} and \texttt{MPDAF}. We observe very good agreement between the results offered by the two different tools. It is worth mentioning that the kinematics of the $\rm{[O\textsc{ii}]}$ gas were recovered just from the fits offered  by \texttt{MPDAF}, making the derived kinematic parameters in this case  less robust than for the other emission lines. \\
 Figure \ref{vel} displays in the panel from the left  the  $\rm{H\alpha}$ radial velocity in km/s,  as recovered from \texttt{FADO}. The velocity  map is normalised  to the BCG's rest frame, i.e. to the velocity  obtained from the redshift of the central spaxel ($\rm{z_{center}}=0.3526$), whose location is coincident with that of the SMBH. We observe a clear gradient from negative ($\sim -300~\rm{km~s}^{-1}$) to positive velocities ( $\sim 300~\rm{km~s}^{-1}$) relative to the BCGs systemic velocity.  We recover a median value for the error of the velocity of $~\pm 30\: \rm{km~s}^{-1}$. 
 The core of the BCG shows mainly negative gas motions while the gas in the $\rm{H\alpha}$ tail shows positive velocities.  Such velocity profiles can be indicative of rotation but can also arise from coherently in- or outflowing material with an inclination to the plane of the sky. However,  the warm ionised nebula in the innermost 10 kpc of the galaxy is not in dynamical equilibrium, as there are no obvious signs of rotation in the core of the BCG.
 It is quite complicated to firmly establish whether the gas in the tail is inflowing or outflowing from the BCG core. The  $\rm{H\alpha}$ tail  does not seem to have  a bi-modal symmetry characteristic of jets, and it also does not lie along the axis connecting the cavities observed in the Chandra data by \cite{eh11}. Therefore, the most plausible explanation would be that the redshifted stream of gas in the $\rm{H\alpha}$ tail is radially in-falling towards the centre.  Such motions of  the gas have been observed in many BCGs (e.g. \citealt{hamer}, \citealt{olivares19}). \\
  The plot on the right hand  side shows the spatially resolved $\rm{H\alpha}$ velocity dispersion map. The measured velocity dispersions in all spaxels were corrected for instrumental broadening. The extended  gas has a consistently low velocity dispersion in the order of $\sim150-250\: \rm{km~s}^{-1}$, with a median value for the random error of the velocity dispersion of $\sim \pm 4\: \rm{km~s}^{-1}$. The ionised gas shows additional peaks in the line-width, suggesting the gas is more kinematically disturbed in these regions. Dispersions are lowest near the core and increase towards the northern and southern peripheries in the inner-most regions,  with  the largest dispersions being coincident with the $\rm{H\alpha}$ EW peak.  The rim with an enhanced dispersion along the NE-SW axis lies roughly along the line where the gas velocity changes from negative to positive values. The western side of the $\rm{H\alpha}$ tail shows the lowest velocity dispersion.\\
We measure  similar radial velocities and velocity dispersions for all other strong emission lines, see Fig. \ref{veltot}.\\

\subsection{Comparison between ionised and molecular gas  kinematics}
We then proceed to compare the recovered kinematics of the warm ionised gas to those of the cold, molecular gas (Fig. 3 from \citealt{fogarty19}). Figure \ref{veldif} shows the difference between the ionised and molecular gas kinematics. The panels on the left-hand side display the difference between the systemic velocity of the $\rm{H\alpha}$ gas and the velocity of the CO(1-0) gas (top panel) and CO(2-3) gas (bottom panel).  The velocity difference maps are predominantly filled with velocity offsets below $\pm 100\:\rm{km~s}^{-1}$, especially in the core of the BCG, where the velocity differences are in the order of $<\pm 50\: \rm{km~s}^{-1}$, indicating that the two gas phases are  likely co-moving.  The $ \sim \pm 100 \:\rm{km~s}^{-1}$ velocity differences might be explained by the different spatial resolutions of the MUSE and ALMA  data. \\ The two panels from the right-hand side of Fig. \ref{veldif} show the ratio maps between the velocity dispersion of the $\rm{H\alpha}$ and the CO(1-0) gas  on the top side and between   the velocity dispersion of the  $\rm{H\alpha}$ and the CO(3-2) gas on the bottom side. 
The maps show almost no structure and are close to unity along the nebula. We observe, thus,  very good agreement between the dispersions of the warm and cold  gas phases, with a median value for the ratio of $\sim 1.2$. It is worth noting that, some structures  with more enhanced velocity dispersions for $\rm{H\alpha}$ are visible in these maps (ratio $\sim 3$),  hinting to the fact that  the ionised gas is more kinematically disturbed in these regions than the molecular gas.  The velocity dispersion ratio map shows that, on average, the $\rm{H\alpha}$  velocity dispersion is by a factor of 1-2 times broader than that for CO(3-2) and CO(1-0). \cite{gaspari18} have shown that the warm ionised gas is  likely to be more turbulent compared to the cold molecular gas. On the other hand, \cite{tr18} suggested that lines-of-sight are more likely to intersect larger volumes of warm gas than cold one, which will lead to a broader velocity distribution in the ionised gas than in the molecular component. These two scenarios are, however, hard to distinguish. \\ To conclude, the comparison between the MUSE and ALMA data reveals evidence that the warm ionised and cold molecular nebulae are to some extent co-spatial and co-moving, consistent with the hypothesis that the optical nebula traces the warm envelopes of molecular clouds. The warm and cold gas are  "mixed" in the sense that that pockets of cold (and warm) neutral gas  are immersed within the warm  ionised gas.

\subsection{Ionisation sources}
\label{BPTsect}

\subsubsection{Diagnostic diagrams}
Figure \ref{BPT} shows on the  top left hand side the classical BPT diagram \citep{bald81}, which uses the ratios of   $ \rm{[O\textsc{iii}]/H\beta}$ vs $\rm{[N\textsc{ii}]/H\alpha}$. The blue solid curve represents the theoretical curve of \cite{kewley01}  and the green dashed one the empirical curve of \cite{kaufm03}, which separate SF  galaxies from AGNs. The orange solid curve of  \cite{sw17}  depicts  the separation line  between Seyfert II galaxies and LINERs.
Our results depict a system with mainly composite emission in the BPT, with contribution from both active star formation and an AGN, which is typical for cool-core BCGs (e.g. \citealt{l13}, \citealt{tr18}). \\The plots from the top middle and top right depict the BPT distribution on the sky, showing the Seyfert II emission in blue and LINER emission in green, respectively. We can identify the location of the SMBH (shown as a cross in all plots), according to the spaxels which fall in the Seyfert II region. It is also clear from these  plots that LINER emission can not be associated with the AGN, as this emission is mainly observed in the outskirts of the system.\\
Figure \ref{BPT} shows  in the middle left panel the $\rm{[O\textsc{iii}]/H\beta}$ vs $\rm{[S\textsc{ii}]/H\alpha}$  and in the lower left panel the  $\rm{[O\textsc{iii}]/H\beta}$ vs $\rm{[O\textsc{i}]/H\alpha}$ diagnostic diagram of \cite{vel87}. The blue curves in both panels represent  the separation  curve of \cite{kewley01}  which divides SF  galaxies from AGNs \& LINERS. The orange solid curve in both plots depicts  the separation curve of \cite{sw17}, which differentiates between Seyfert II  galaxies  and LINERs. The panels from the middle  side of both the middle and lower rows show the  $\rm{[S\textsc{ii}]}$ and  $\rm{[O\textsc{i}]}$-BPT distribution on the sky, displaying the Seyfert II emission in blue, while the panels from the right-hand side of both rows display the LINER emission in green.
The  $\rm{[O\textsc{iii}]/H\beta}$ vs $\rm{[S\textsc{ii}]/H\alpha}$ diagnostic depicts a system with both HII and LINER emission, while in the  $\rm{[O\textsc{iii}]/H\beta}$ vs $\rm{[O\textsc{i}]/H\alpha}$ plot, which is a sensitive diagnostic for shocks, the majority of spaxels fall in the LINER region.\\
We also observe considerable variation in the emission line ratio maps used for the computation of the diagnostic diagrams, suggesting that the source of the excitation is not localised at a specific region.\\
To conclude,   the three diagnostic diagrams reveal mainly composite-LINER emission for the M1931 system, hinting to the fact that there  are several mechanisms which ionise the gas. Based on this analysis, we can not draw any definitive conclusions, as the situation for cool-core BCGs is highly complex and likely represents a superposition of several different ionisation sources (e.g. \citealt{tr18}, \citealt{mc12}).\\

\subsubsection{Fully radiative shock models}

\begin{figure*}
\begin{subfigure}
    \centering
    \includegraphics[width=0.5\textwidth,angle=0,clip=true]{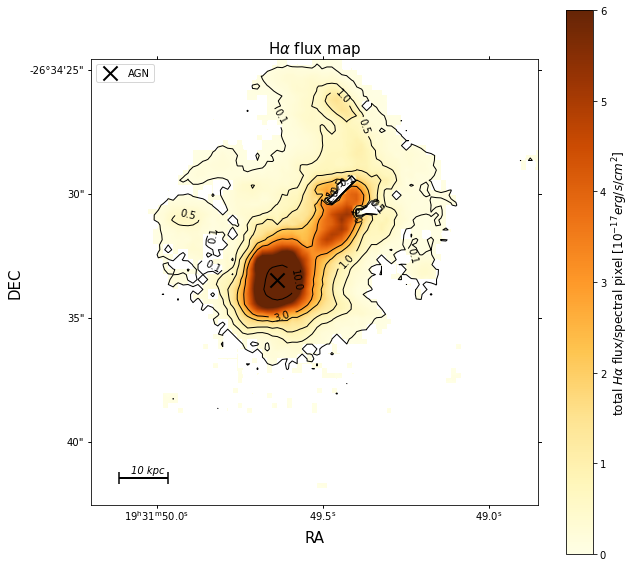}
    \centering
\end{subfigure} 
\begin{subfigure}
    \centering
    \includegraphics[width=0.5\textwidth,angle=0,clip=true]{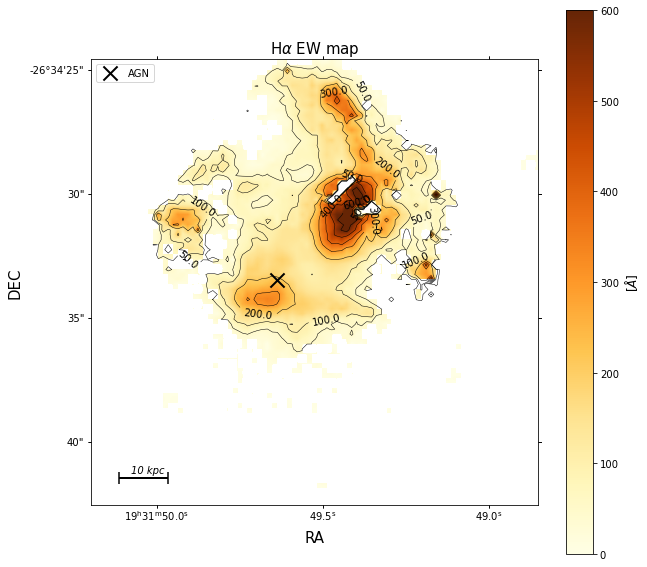}
    \centering
\end{subfigure} 
 \caption{ \textit{left}: Spatially resolved $\rm{H\alpha}$ emission line map for the BCG of the M1931 galaxy cluster. The colour-bar shows the flux of the $\rm{H\alpha}$ line in units of $10^{-17} \rm{ergs \cdot s^{-1} \cdot cm^{-2}}$.  \textit{right}: $\rm{H\alpha}$ equivalent width map measured in $\AA$. The contours correspond to different levels of flux and EW.
The white background in both plots corresponds to spaxels with $\rm{SNR_{H\alpha}<10}$. The cross shows the location of the AGN.}
\label{Ha}
\end{figure*}

\begin{figure*}
\begin{subfigure}
    \centering
    \includegraphics[width=0.5\textwidth,angle=0,clip=true]{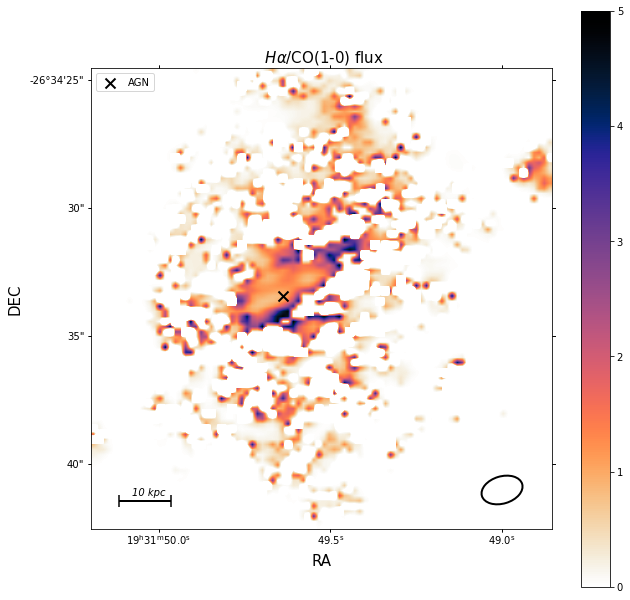}
    \centering
\end{subfigure} 
\begin{subfigure}
    \centering
    \includegraphics[width=0.5\textwidth,angle=0,clip=true]{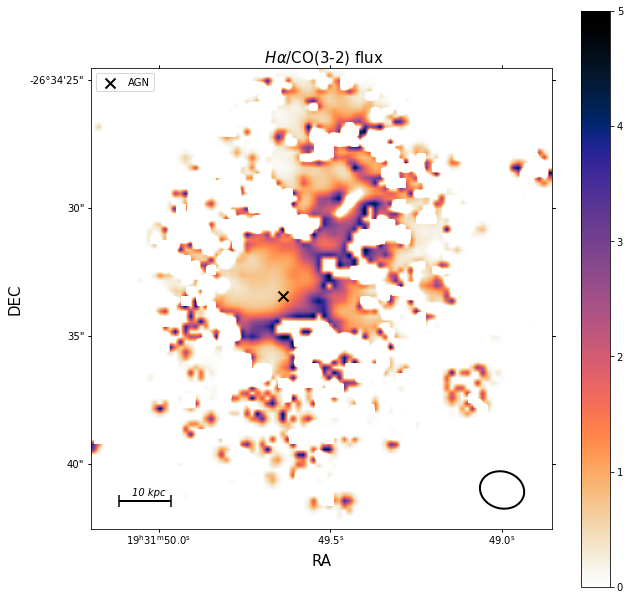}
    \centering
\end{subfigure} 
\caption{ Comparison between $\rm{H\alpha}$ and CO linear normalised flux. The panel from the \textit{left-hand side} displays the ratio between  $\rm{H\alpha}$ and CO(1-0) fluxes, while the panel from the \textit{right-hand side} shows the ratio between the fluxes of  $\rm{H\alpha}$ and CO(3-2). The cross shows the location of the AGN. The  ellipses in the lower right of both panels depict the beam sizes of the ALMA observations.} 
\label{musealmaflux}
\end{figure*}

Fig. \ref{3mdb} shows the fully radiative shock model grids, over-plotted  on the 3 diagnostic diagrams described above. Each black data point represents a spaxel of the MUSE cube with an SNR>10 in each emission line used for the diagnostic.  As can be seen from this plot, the different shock models partially reproduce the measured emission line ratios, mainly  for spaxels which fall in the Seyfert II region of the diagnostic diagrams. The only grids which partially seem to fit the data in all three diagnostic diagrams are the ones with a Solar (red lines) and twice Solar (blue lines) metallicity. We observe similar behaviour for all other shock model grids described in section \ref{ioni} as well. \\ We have also accounted for the contribution of star formation, AGN and LINER-like emission to the luminosity of each emission line in each spaxel of the data cube, allowing us to derive the luminosity of the $\rm{[O\textsc{iii}]}$, $\rm{[N\textsc{ii}]}$, $\rm{[S\textsc{ii}]}$, $\rm{[O\textsc{i}]}$, $\rm{H\alpha}$, and $\rm{H\beta}$ gas associated purely with LINER-like emission.   
LINER-like emission can arise due to large scale shocks, and if indeed shocks are responsible for ionising the gas in the M1931 BCG, then the fully-radiative shock models should reproduce the pure LINER fluxes of the emission lines in the three diagnostic diagrams.
In Fig. \ref{3mdb}, the grey data points represent the spaxels of the MUSE cube,  whose luminosities are associated with pure LINER-like emission. Even after removing the contribution from star formation and AGN emission from the luminosity of the emission lines, the different shock grids still do not fully reproduce this modelled emission. Therefore, we concluded that lower velocity shocks seem to play only a minor role as an ionising mechanism in the system. The  weakness of the $\rm{[O\textsc{iii}]}\: \lambda 4363$ emission also  suggests that shocks are not a major ionising mechanism in the M1931 BCG \citep{dh97}. This is in accordance with the findings of \cite{dh00}, who demonstrated, based on HST imaging of cool-core BCGs that, the radio sources in cool-core galaxy clusters are not injecting significant amounts of energy by strong shocks  to the emission-line gas and therefore, shocks cannot be a significant ionising source, nor a source of heating to counterbalance the cooling.\\
Regarding the AGN emission: we see from the diagnostic diagrams that M1931 does not have a strong optical AGN, as just a few spaxels fall in the Seyfert II region. The weakness of $\rm{He\textsc{ii}} \: \lambda\: 4686\:\AA$ also rules out such a hard ionising source. Therefore, AGN ionisation seems to influence only the most central regions of the BCG. This is in accordance with the findings of \cite{fogarty17}, who analysed the impact of AGN emission on the SED fit  from the UV through far-IR for M193, and concluded that the effect of AGN emission in M1931 is  marginal.\\ Hence, the dominant source of ionisation in M1931 BCG is a mix between star-formation  and other energetic processes which can “mimic” LINER emission. Several hypotheses have been proposed for the source of this extended LINER-like emission. Stellar populations may be responsible for this  emission, which could arise due to photoionisation of the gas by young starbursts,  or by old  pAGB  stars (e.g. \citealt{shields92}, \citealt{olsson10}, \citealt{l13} , \citealt{binette94}, \citealt{stasinska08}). For e.g. \cite{byler19} have studied the predictions of LINER-like emission from pAGB stars, based on fully self-consistent stellar models and photoionization modelling and they have demonstrated that  indeed,  post-AGB models produce line ratios in the LINER region of the $\rm{[S\textsc{ii}]/H\alpha}$ and $\rm{[O\textsc{i}]/H\alpha}$ diagrams, and in the “composite” region of the standard BPT diagram. This is exactly what we observe in the diagnostic diagrams for the M1931 BCG.  However, these post-AGB star models produce $\rm{H\alpha}$ EWs between 0.1 and 2.5 Å, while we observe by far higher  $\rm{H\alpha}$ EWs for M1931 ($\rm{EW_{H\alpha}}>50 \:\AA$ throughout the whole nebula, see the right-hand panel of Fig. \ref{Ha}). Therefore,  we can rule out post-AGB stars as the dominant source of LINER emission in the M1931 BCG. LINER-like emission may also be characteristic of starburst-driven outflows \citep{sharp10}, Lyman continuum photon escape through tenuous warm gas \citep{polis13}, Diffuse Ionized Gas (DIG) in galactic disks \citep{zhang16}, or gas heated by the surrounding medium. The latter mechanism includes  photoionisation by cosmic rays,  collisional heating by cosmic rays,  conduction from hot gas, suprathermal electron heating from the hot gas, X-ray photoionisation, turbulent mixing layers (e.g. \citealt{dh91}, \citealt{b91}, \citealt{dh11}, \citealt{mc12}, \citealt{sparks12}). \\ \cite{fogarty19} also studied the CO excitation mechanisms in the M1931 BCG, and their analysis of the CO spectral line energy distribution reveals evidence for multiple gas excitation mechanisms in the system, besides star formation. The CO(3-2) transition is highly excited, similar to Quasi Stellar Objects and Ultra Luminous Infrared galaxies, but the CO(4-3) is similar to that observed in normal, star forming galaxies. Therefore, there must be a mechanism not related to starburst which acts as an additional excitation mechanism. The authors conclude that the molecular gas excitation in the BCG is driven by  a mix of processes such as SF,  radiation from the SMBH and interaction between the cold gas and the ICM, in accordance with our findings.\\

\subsubsection{Spectral decomposition method}
\label{d}
To test the validity of the spectral decomposition method of \cite{davies17},  one should check  whether the fraction of emission attributed to each ionisation mechanism varies smoothly as a function of the diagnostic line ratios, peaking at the line ratios of the relevant basis spectrum and decreasing as the line ratios become closer to those of the other basis spectra. Figure  \ref{daviesbpt} demonstrates that this is exactly what we observe.
 The upper 3 panels depict the BPT diagram while the lower three ones show the diagnostic diagram of \cite{vel87}, with each data point colour-coded according to the fractional contribution of SF (left), AGN (middle) and LINER (right)  to the total $\rm{H\alpha}$ emission.\\
Figure \ref{daviesspatial} shows the emission line maps for $\rm{H\alpha}$ on the upper row, $\rm{[O\textsc{iii}]}$ on the second row, $\rm{[N\textsc{ii}]}$ on the third row and $\rm{[S\textsc{ii}]}$ on the fourth row colour coded according to  the fractional contribution of SF (left), AGN (middle) and  LINER-like emission (right) to the total line luminosity. \\
In the maps showing the fractional contribution of SF to the emission line luminosities (first column in Fig. \ref{daviesspatial}), some "structures" become visible, such as the clumps in the vicinity of the AGN (above and to the N-W) as well as some clumps in the tail,  which are most probably young, SF regions. These regions show in all the emission line maps more enhanced contribution from SF to the total line luminosity, probably making young stars the main ionising mechanism there. These "structures" are coincident with the compact knots seen in the naturally weighted CO(1-0) intensity image,  as well as with the UV knots present in HST  photometry. In the BCG core at the location coincident with that of the SMBH, the fractional contribution of SF to the total line-luminosity is the smallest.  In all the maps showing the AGN contribution to the total line luminosity (second column in Fig. \ref{daviesspatial}) we can observe an enhanced contribution of AGN emission to the total luminosity of the four emission lines in the most central regions of the system, corresponding to the location of the SMBH. In the maps showing the fractional contribution of LINER-like emission to the total line luminosity (third column in Fig. \ref{daviesspatial}), we observe again a low fraction in the central region of the system corresponding to the location of the AGN. Low LINER-like contribution can be found in the star-forming clumps as well. \\
Possible caveats regarding the spectral decomposition method are related to the fact that the relative luminosities of the emission lines  are primarily determined by the relative contributions of different ionisation mechanisms to the line emission, but they are also sensitive to the metallicity and ionisation parameter of the gas.\\
From the spectral decomposition method, we conclude that the  main source of ionisation in the M1931 BCG is a mix between SF and other energetic processes which give rise to LINER-like emission, while AGN ionisation is dominant only in the BCG core. Star formation  accounts for $\sim 50-60\%$  of the ionised gas emission, whereas AGN emission accounts only for about $\sim 10\%$. The rest of $\sim 30-40\%$ of the energetics needed to excite the gas  come from mechanisms which give rise to LINER-like emission. Star formation is, therefore,  the main ionising mechanism in the M1931 BCG and the main contribution to the luminosity of the emission lines.\\

\subsection{Electron Density and Electron Temperature}

Figure \ref{TD} shows in the upper left panel the  electron density vs electron temperature  diagnostic diagram for the most central regions of the M1931 BCG.  From the M1931 MUSE  data cube, we extracted a sub-cube corresponding to the core of the BCG ($2\farcs2\times2\farcs2$  circular aperture surrounding the AGN) and one corresponding to the $\rm{H\alpha}$ arm and provided the fluxes obtained from the integrated spectra of these two regions to \texttt{PyNeb}.
For the core,  we obtain degenerate values for the temperature and density when using different line diagnostics. The difference in the $\rm{n_e}$ and $\rm{T_e}$ obtained from different emission lines for the core  of M1931 BCG could be due to the superposition of gas components with different physical conditions and gas excitation mechanisms along the line of sight. As we have seen in sect. \ref{BPTsect}, the central region of the BCG is ionised by a mix of processes, and therefore, in this region, we likely see the luminosity-weighted average of emission lines arising from different volumes with different physical conditions, and eventually also different kinematics.
Also, different line ratios are sensitive to different ranges in $\rm{n_e}$ and $\rm{T_e}$, they therefore likely probe different components in the ionized gas.\\ The complexity of this issue may be demonstrated already on the example of a relatively 'simple' system, like a blue compact dwarf (BCD) galaxy:   \cite{james09} studied the BCD Mrk 996 based on VLT-VIMOS IFU observations and they have shown that the  ionized gas in the SF nucleus of that galaxy consists of a component with normal density $\sim 170\:\rm{cm^{-3}}$ and broad-line component with an electron density reaching $10^{7}\:\rm{cm^{-3}}$, similar to what we observe in the most central regions of the M1931 BCG.\\The plot from the right-hand side of  Fig. \ref{TD} also displays the  electron density vs electron temperature  diagnostic diagram, but for the $\rm{H\alpha}$ tail. The different diagnostics offer consistent values for the electron temperature and density, as can be observed  from the intersection of the different curves.\\The panel from the lower left hand side of Fig. \ref{TD} shows the spatially resolved  electron temperature map  measured in [K], as computed  from the  $\rm{[N\textsc{ii}]}$ 5755/6584  emission lines for individual spaxels of the M1931 MUSE cube. We measure a median electron temperature of $\rm{T_e}\sim11230$ K and a median random error for the temperature of  $\sim \pm 220$ K.  The errors for the electron temperature were  calculated with the Bootstrapping method, by assuming that the errors in the line fluxes are Gaussian and adding/subtracting them according to a random key from the measured flux values. These new flux measurements for each spaxel were provided to  \texttt{PyNeb} for a new computation of the electron temperature and density. This procedure was repeated a few tens of times to accurately recover the errors  for the electron temperature. It is worth mentioning that due to the weakness of the temperature-sensitive emission line $\rm{[N\textsc{ii}]}\:\lambda5755$ we can measure the electron temperature and density only in a small number of spaxels. The spatially resolved electron temperature map shows quite high variations, with the lowest temperatures of about $\rm{T_e}\sim 8 000$ K in the most central regions of the system. The temperature seems, thus, to  decrease  towards the core of the BCG. For e.g. \cite{v05} studied the  ICM temperature profiles in cluster cores and found that the temperature decreases in the cores of cool-core galaxy clusters, similar to what we observe for the interstellar medium (ISM) of the BCG.\\ The panel from the lower right-side of 
Fig. \ref{TD} displays the spatially resolved electron density map in units of $\rm{cm^{-3}}$ as computed  from the $\rm{[S\textsc{ii}]}\:\lambda 6718, \lambda6732$ emission lines  using the \texttt{PyNeb} tool. We measure a median electron density of $\rm{n_{e}=  361\:cm^{-3}}$, which is a value typical for BCGs, with a median  error of  $ \sim \pm  60\: \rm{cm^{-3}}$ (also computed through the Bootstrapping method). The recovered values for  $\rm{n_{e}}$   are below the critical density threshold of the $\rm{[S\textsc{ii}]}$ doublet, which is $\rm{n_e=3\cdot 10^{3}\: cm^{-3}}$ \citep{of06}, above which the lines become collisionally de-excited. Because of this, we can consider our derivation of the density reliable. \\

\subsection{ Color excess E(B-V)}
The panel on the  left-hand side of Fig. \ref{EU} shows the colour excess  map for the M1931 BCG. We recover a median value of E(B-V)=0.135 and median  random error of  $\pm0.002$ . The error was computed through error propagation by taking only the flux measurement errors and not the calibration uncertainties into account, meaning that the true errors are larger.  The highest extinction, in the order of  E(B-V)$\sim0.4 - 0.5$ is observed in the BCG core.  \\This is in accordance with the findings of \cite{fogarty19}, who also observed high dust continuum emission in  the core of the system, as well as in  the $\rm{H\alpha}$ tail. 
The regions 1, 2 and 3 marked by the black ellipses are the cold dust regions, for which \cite{fogarty19} measure a temperature of $\rm{T_{dust}}<10$ K. In the core of the BCG, the dust temperature $\rm{T_{dust,\: core}}> 11$ K. The dust temperature was estimated based on the  continuum emission in ALMA Bands 6 (336.0 GHz) and 7 (468.5 GHz), using the equations introduced by \cite{casey12} and a value of $\beta= 1.6\pm0.38$. The ionised gas and dust in the M1931 BCG are more or less co-spatial, except maybe for region 3, where we measure low emission line fluxes. Intriguingly, the ionised gas EW peak is coincident with the location of the cold dust region 2. The enhanced excitation of  the ionised gas in this region can come from collisional excitation of relativistic particles and cold dusty gas \citep{sparks12}. \\
It is  worth mentioning that \cite{fogarty17} also derived the extinction in the BCG of the M1931 cluster and they obtain a value of A(V) = $0.87 \pm 0.21$ assuming the Calzetti Law \citep{calz}, which corresponds to E(B-V) = $0.21\pm0.05$.  Therefore,  there is a small tension between the value inferred by us and that inferred by  \cite{fogarty17}  for the extinction.  A possible explanation for the small discrepancy  is that the spaxels with $ > 10 \sigma$ detections of $\rm{H\alpha}$ are systematically less reddened by dust than the broadband filters that  \cite{fogarty17} used to construct their SED fit from UV through far-IR.  However,  given the fact that our errors for the colour-excess are under-estimated as they are just the  statistical random errors, the two values for E(B-V) can be considered to be consistent. \\

\subsection{Ionisation parameter}
 The panel from the right-hand side of Fig. \ref{EU}  shows the spatially resolved map for the ionisation parameter log(U).  This parameter gives the ratio of ionizing photon density to hydrogen density and represents a  measure of the dimensionless intensity of ionizing radiation. We recover a median value for the ionisation parameter $log(U)=-2.9$ and a median error of $\pm0.3$. This value for the ionisation parameter is consistent with those observed for star forming galaxies (e.g. \citealt{cresci17}, \citealt{kewley19}). The highest log(U) values seem to be coincident with the location of the star forming structures which are visible in column 1 of Fig. \ref{daviesspatial}.

\subsection{Star formation rates}
\label{sf}
Figure \ref{SFROH} shows  the spatially resolved SFR map for the M1931 BCG in units of  $\rm{M_{\odot}/yr}$. 
 We observe regions with higher SFRs coincident with the BCG core, while in the tail, the SFR levels are lower. 
From  the spaxel by spaxel analysis, we measure a total $\rm{SFR\sim 144 \:M_{\odot}/yr}$ and a random error of $\rm{\pm0.33 \:M_{\odot}/yr}$ (computed through error propagation taking only the flux measurements into account). The recovered SFR surface density is $\rm{\Sigma SFR \sim 147  \:M_{\odot} \:yr^{-1}\: kpc^{-2}}$. 
However, these values for the SFR and $\rm{\Sigma SFR}$ are just  upper limits, because, as we have seen in section \ref{d},  SF accounts for $\sim 50-60\%$  of the ionised gas emission, AGN  ionisation accounts for about $\sim 10\%$, whereas  $\sim 30-40\%$ of the energetics needed to excite the gas  come from mechanisms which give rise to LINER-like emission. The  \cite{ken98} conversion for the estimation of the SFR was designed for pure star forming galaxies and therefore, we should exclude the contribution from AGN and LINER-like emission to the total luminosity of the $\rm{H\alpha}$ emission line to accurately estimate the SFR. 
By doing so, we recover a  $\rm{SFR\sim 97 \:M_{\odot}/yr}$ with a random error of  $\pm0.7 \:\rm{M_{\odot}/yr}$.  The integrated star formation rate surface density  that we recovered, after excluding the contribution of LINER-like and AGN emission to the total line luminosity of $\rm{H\alpha}$, is  $\rm{\Sigma SFR \sim 100  \:M_{\odot} \:yr^{-1}\: kpc^{-2}}$. However, LINER-like line ratios could also arise from an over-abundance of O-stars (but not indicating more star formation, \citealt{fogarty15}), meaning that the LINER-like excitation can also partially be due to star formation.  Therefore, it might be that the recovered SFR value of  $\sim 97 \:\rm{M_{\odot}/yr}$ is slightly under-estimated.  It is possibly the case that the "true" $\rm{H\alpha}$-based SFR is limited above by a maximum of 144 $\rm{M_{\odot}/yr}$ and limited below by minimum of 97 $\rm{M_{\odot}/yr}$. \\ Comparing to other works, \cite{dh15} reported an unobscured SFR$\sim 90 \:\rm{M_{\odot}/yr}$ for the M1931 BCG from  the CLASH HST rest-frame UV data,  without accounting for reddening. \cite{eh11} has reported a SFR$\sim 170 \:\rm{M_{\odot}/yr}$ from broadband Subaru $\rm{H\alpha}$ photometry for the same system. \cite{santos16} has computed a SFR$\sim 150\pm 15 \:\rm{M_{\odot}/yr}$  from a FIR SED fit using Herschel observations, after removing the AGN contamination. \cite{fogarty17} estimated a value for the SFR of $\sim 250 \pm 75 \:\rm{M_{\odot}/yr}$  from NUV-FIR SED fitting using photometry from the CLASH HST data set in combination with mid- and far-IR data from Spitzer and Herschel.  \cite{fogarty15}, on the other hand, derived a  reddening-corrected value for the UV-continuum estimate of the SFR for the M1931 BCG of $280 \pm 20\: \rm{M_{\odot}/yr}$ using CLASH HST ACS and WFC3 observations. Their derived SFR agrees with the value they obtained in  \cite{fogarty17} for the SED fit from the UV-FIR. They also estimated the  $\rm{H\alpha}$ luminosity for the system, for which they derived an SFR of $130 \pm40 \rm{M_{\odot}/yr}$.  All the $\rm{H\alpha}$ based SFR estimates are in accordance with both values derived by us, within the errors. The larger  SFR
$\sim250  \:\rm{M_{\odot}/yr}$ values reported by \cite{fogarty15} and \cite{fogarty17}  are heavily influenced by the UV continuum, and possibly reflect the SFR over a longer time period. The lower SFR values are most probably representative for the ongoing star-forming activity. \\
 According to \cite{fogarty19}, the M1931 BCG has a molecular gas reservoir with a mass of  $\rm{M_{H_2}=1.9 \pm 0.3\cdot 10^{10} M_{\odot}}$, and together with our inferred SFR, we can calculate the  gas depletion time as $\rm{t_{depl}=\frac{M_{gas}}{SFR}}$, and by doing so,  we recover a $\rm{t_{depl}\sim190}$ Myrs. \cite{dea08}  studied  BCGs based on imaging with the Spitzer Space Telescope and inferred a typical gas depletion timescale for such systems of about 1 Gyr. \cite{ken12} have demonstrated that when the gas depletion time is shorter the SFR efficiency is higher.  This might imply that the M1931 BCG has a high star formation efficiency and that currently, the system is consuming the gas at a higher rate than typical.
On the other hand, \cite{voit11} have  demonstrated that BCGs with large SFRs ($>10 \:\rm{M_{\odot}/yr}$) have depletion times less than 1 Gyr. These depletion times are shorter than those of BCGs with lower SFR levels. The depletion time we infer for the M1931 BCG is thus, not so different from that other BCGs with large SFRs and is also similar to that of other star forming galaxies, especially to those with starburst rates \citep{dk20}. Hence, the global Kennicutt star formation relation \citep{ken98a} relating molecular gas quantity to the SFR is not so different in cool-core, multiphase BCGs and disk galaxies. \\ To conclude, the M1931 BCG shows elevated levels of SF in the order of $\sim 97 \:\rm{M_{\odot}/yr}$, with the largest values in the core of the system,  in accordance with previous studies.

\subsection{Oxygen abundances}
\label{oh}
Figure \ref{OH} displays the spatially resolved oxygen abundance map for the BCG of the M1931 cluster. 
We measure a median gas-phase metallicity $\rm{12+log(O/H)}\sim8$ and an error of $\pm0.35$, with the lowest (O/H) values in the BCG core, and slightly more enhanced values in the outskirts.\\
For example, \cite{km15} studied the hot gas-phase  metallicity distribution in 29 galaxy cluster cores based on Chandra X-ray data and found that over the life cycle of AGN activity, hot outflows  can be responsible for the broadening of abundance peaks in cool-core clusters, and effectively transport metal-enriched gas from the BCG to great distances in the cluster atmospheres. Similarly, \cite{eh11} studied the ICM metallicity based on Chandra X-ray data in the M1931 cluster and concluded that this cluster core is missing the central metallicity peak which is normally measured in  cool core  clusters, thus suggesting bulk transport of hot X-ray gas out to large distances from the centre due to  AGN outburst. This is similar to what we observe for the warm gas in  the ISM of this system, with the lowest metallicity values inferred in the core and slightly more enhanced values in the tail. \cite{eh11} measure a more or less constant ICM metallicity of $\rm{Z = 0.36\:Z_{\odot}}$ out to distances as large as 400 kpc from the M1931 cluster core, which translates to a 12 + log(O/H) = 8.25. This value is consistent to the gas-phase metallicity we measure in the ISM of the BCG. The similarity may hint to the fact that the gas we observe in the ISM of the galaxy has condensed from the ICM. \\

\subsection{Star formation history}
Besides emission line fluxes, equivalent widths and stellar parameters, \texttt{FADO} can also recover the  SFH of a galaxy, ensuring consistency between the best-fitting  SFH and the observed nebular emission. As a consistency check, we have compared the SFH recovered from \texttt{FADO} to the one recovered from  \texttt{Porto3D}  and we observe a very good agreement between the two.\\
The SFH module of FADO allows us to  recover  the stellar mass ever formed within the system: 
$\rm{M^*_{ever\:\:formed}=5.9\cdot 10^{11}M_{\odot}}$.  
This value is in perfect agreement to the one inferred by \cite{bell16} for the mass of the M1931 BCG   using only K-band photometry:  $\rm{M^{*}=5.9\pm1.1\cdot 10^{11}M_{\odot}}$.\\
Figure \ref{sfh} shows the discretised approximation to the best-fitting SFH for the integrated spectrum of the M1931 BCG, as recovered from \texttt{FADO}. The upper panel shows the  contribution of  the individual SSPs in the best-fitting population vector  to the monochromatic luminosity at 6150 $\AA$  and the lower one the contribution to the stellar mass as a function of age. 
It is clear from these two plots that the M1931 BCG had a complex SFH. Approximately  $50\%$  of  the systems'  stellar mass, i.e. $\rm{M^*\sim 3\cdot 10^{11} \: M_{\odot}}$, was in place at an early epoch, $t_{\frac{1}{2}}\sim 6.98$ Gyr ago (z$\gtrsim1.5$), followed shortly by  subsequent mass build-up episodes, but less strong. This main episode of SF also contributes the most to the luminosity of the galaxy ($\sim 30\%$).  This mass build-up episode is shortly followed by another episode, which occurred 6.3 Gyr ago and leads to the formation of $30\%$ of the mass of the system, i.e. to the formation of  $\rm{M^*\sim 1.7\cdot 10^{11} \: M_{\odot}}$. This second strongest and oldest mass build-up event might be associated with a wet merger (given that there are SSPs of different ages and metallicities  contributing to the  mass build-up).  Hence, most of the mass of the BCG ($\sim 80\%$) was built up  at an early epoch, more than 6 Gyr ago, in accordance with theoretical models, which favour the scenario in which BCG  growth follows two–phase hierarchical formation, with  rapid cooling and in-situ star formation at high redshifts followed by a subsequent growth through repeated mergers, (e.g. \citealt{d07}). Our  results are also in accordance with the findings of  \cite{collins09}, who demonstrate  that $\sim 90 \%$ of BCGs stellar masses are in place by z = 1.\\
Many other mass build-up events occur for the M1931 BCG over its SFH, but they are  less strong than the two initial episodes,  and only lead, combined, to the build-up of  $\sim 20\%$ of the mass of the system. The second two strongest mass build-up events occur at a redshift $\rm{z\lesssim0.5}$: one $\sim 1$ Gyr ago - contributing  $\sim 10 \%$ to the  stellar mass ever formed (leading to a mass build-up of $\rm{M^*\sim 5.3\cdot10^{10}\: M_{\odot}}$ ) and  $\sim 13\%$ to  the luminosity  and the other  $\sim   5.6\cdot 10^7$ yrs ago - contributing $\sim 2 \%$  to today's stellar mass  (leading to a mass build-up of $\rm{M^*\sim 1.1\cdot10^{10}\: M_{\odot}}$ ) and $\sim 20\%$ to the luminosity. The ages of these two "post-assembly" mass build-up episodes are  a good match to the starburst age range \cite{fogarty17} derive for the M1931 BCG using a Bayesian photometry-fitting technique that accounts for both stellar and dust emission from the UV through far-IR. Using a simpler single-burst SFH parameterization, \cite{fogarty17} estimate an age range for a "post-assembly" star formation episode of $\sim 4.5 \times 10^{7} - 2.2 \times 10^{8}$, in pretty good agreement with the ages we recover.\\
Other weak mass build-up events probably occurred for the M1931 BCG, leading to the build-up of less than $1\%$ of the stellar  mass and  less than $10\%$ of the total luminosity of the system. However, robust conclusions of such minor SF episodes are difficult due to the probabilistic nature of PSS.
 The most recent mass build-up event occurred 1 Myr ago and lead to the formation of  $0.2\%$ of the total mass, i.e.  to the formation of $\rm{M^*=1.1\cdot 10^9 M_{\odot}}$. \\
All the weaker episodes of mass build-up, taken at face value, might be due to minor dry mergers  or due to "cooling-flow" induced star formation episodes (or  a combination of the two).  If we consider all the mass build-up events that happened during the last one Gyr,  the redshift of the universe at that time was z$\sim0.5$, an epoch  at which wet mergers have mostly ended in galaxy clusters, which are more or less fully assembled. Therefore,  the mass build-up episodes which occurred less  than 1 Gyr ago can, most probably, not be attributed to wet mergers and associated minor in situ SF episodes, although a contribution from dry mergers cannot be ruled out. For e.g. \cite{burke15} examine the stellar mass assembly in galaxy cluster cores using CLASH data and find that BCGs grow in stellar mass by a factor of 1.4 on average from the accretion of their companions and find major merging to be very rare for  their sample. They conclude that minor mergers constitute  the dominant process for stellar mass assembly of BCGs at low redshifts, but with the majority of the stellar mass from interactions ending up contributing to the ICL rather than building up the system. Similarly, \cite{webb15} studied the SFH of BCGs to z = 1.8 from the  SPARCS/SWIRE survey and concluded that,  star formation episodes  below $z\sim1$ do not contribute more than 10\% to the final total stellar mass of the BCGs and that BCG growth is likely dominated by dry mergers at low z. This is consistent with the recent mass build-up episodes that we infer for the M1931 BCG, which contribute to the build-up of less than 20\% of its  stellar mass. \\
Another interpretation  could be that these weaker mass build-up events are related to "cooling flow" induced SF episodes, as we know that the M1931 BCG resides in a cool-core cluster. Thermally unstable cooling of the ICM into cold clouds which start sinking towards the SMBH was shown to explain the on-going star formation as well as the AGN activity of BCGs, (e.g.\citealt{voit15b}). This scenario is supported by the high SFR levels we measure for the M1931 BCG, which can probably  not be attributed to wet  mergers with satellite galaxies, as we know that satellite cluster galaxies approaching the cluster core should be devoid of gas due to ram pressure stripping and starvation. The co-spatiality and kinematics of the ionised and cold gas component, as well as the metallicity of the ISM, also point to a common origin for  the two gas-phases, such as  cooling from the ICM.\\
 Making a simplified assumption that the SFR was more or less constant for the last 100 Myrs, then  a SFR =97\:$\rm{M_{\odot}/yr}$ yields a stellar mass of  $\sim 10^{10} \rm{M_{\odot}}$ in this time span (and similarly a SFR =144\:$\rm{M_{\odot}/yr}$ yields a mass of $\sim 1.4 \cdot10^{10} \rm{M_{\odot}}$ in the last 100 Myrs).
According to Fig.   \ref{sfh}, in the last 100 Myrs, $\sim 3\%$ from of the total mass of the system was formed, yielding a value of about $\sim 1.7 \cdot 10^{10}\rm{M_{\odot}}$. This value is in good agreement to the mass that would have formed in-situ in the last 100 Myrs, given the star formation levels inferred from the $\rm{H\alpha}$ emission line, making, thus, "cooling-flows" induced SF-episodes a good candidate to explain the recent SFH of the M1931 BCG. \\
The SFH module of \texttt{FADO} also allows us to put some rough constraints on the stellar parameters, such as the SSP ages (and metallicities). We have investigated the stellar age gradients in a region corresponding to the core of the galaxy (i.e. $2\farcs2\times2\farcs2$  circular aperture around the SMBH) and in a region encompassing the whole system, and we recover mainly positive (to flat) age gradients. Stellar ages are lowest in the core of the system and seem to increase towards the outskirts of the galaxy.   These findings  are also supported by our spatially resolved SFR map, where we  demonstrate that the highest SFR levels  are confined to the core of the galaxy.  Hence, it is expected that the core of the BCG contains a  young population of stars.   \cite{l09} has demonstrated the youngest BCG cores are mostly line-emitting galaxies in cool-core clusters, similarly to what we observe in the case of the M1931 BCG. Likewise, \cite{edwards20} investigated age and metallicity gradients in a sample of 23 low redshift BCGs (z<0.7) and observed positive age gradients for 6 of their systems, and negative gradients for the rest (which is often seen for massive elliptical galaxies). All of the BCGs showing positive age gradients are star forming multi-phase systems, similar to the BCG of the M1931 cluster. \\ To conclude, both dry mergers  and/or  "cooling-flow" induced in-situ star formation episodes  dominate  the mass build-up of M1931 BCGs at late epochs, but they account for only a  small fraction  of less than $ 20\%$ of the total mass of the system, while probably in-situ star formation  (either due to rapid cooling or wet mergers) at early epochs, more than 6 Gyr ago, lead to the formation of up to  $\sim 80\%$ of the stellar mass of the system.

\subsection{ Stellar kinematics}
\label{starkin}

Figure \ref{stvel} shows in the panel on the left-hand side the stellar radial velocity and in the  panel on the right hand side the stellar velocity dispersion in the BCG core, as recovered from the  \texttt{GIST} pipeline.  The stellar velocity map is normalised to the BCGs' rest frame, i.e. to the velocity of the central bin, whose position is coincident with that of the SMBH. The recovered stellar velocity dispersions are corrected for instrumental broadening. 
While the stellar radial velocity shows a gradient from -200 to 200 $\rm{km~s}^{-1}$ (with a median error for the radial velocity of $\pm12\: \rm{km~s}^{-1}$) with respect to the systemic velocity of the BCG, being indicative of rotation, the values for the stellar velocity dispersion are by far higher, ranging from 300 $\rm{km~s}^{-1}$ up to $\sim 650\: \rm{km~s}^{-1}$ (with a median error for the velocity dispersion of $\pm12\: \rm{km~s}^{-1}$). This means that the BCG core is  dispersion dominated, i.e. a slow rotator, which is expected for massive, elliptical galaxies. Since this high stellar velocity dispersion is going to be preserved after the gas was consumed by SF and/or ejected by the AGN, our data shows an intermediate-late stage in the assembly history of present-day massive early-type galaxies  (or central dominant galaxies) at the centre of galaxy clusters.\\
\subsection{Comparison stellar and gaseous kinematics}
Figure \ref{stveldif} displays the difference between the systemic velocity of the stars and the $\rm{H\alpha}$ gas on the left-hand side and on the right-hand side, the ratio between the velocity dispersions of the stellar and ionised gaseous component. For this comparison, we have used the stellar kinematics recovered by the  \texttt{GIST} pipeline for the un-binned data.  We measure  velocity differences between the stellar and  gaseous component  as high as $\sim \pm200 \: \rm{km~s}^{-1}$, and a median value for the  ratio between the velocity dispersion of stars and gas  of 2.4. Hence, the stars have much higher velocity dispersions than the gas.\\
The  gaseous and stellar kinematics  in the M1931 BCG core are, thus, decoupled, with the gaseous component not following the potential well of the stars. Such decoupled kinematics have been observed in other  BCGs (e.g. \citealt{hamer}, \citealt{l13}). The kinematics of the gas are  more closely related to the turbulence and bulk motions in the hot ICM than they are related to the motions of the stars. The high velocity dispersion of the stars argues  strongly that the gas  has originated from the more quiescent circum-galactic medium, and not from the stars.  These findings are in accordance with the predictions of the precipitation model (\citealt{voit17}) as well as of the Chaotic cold Accretion model (\citealt{gaspari18}),  which state that  the gas motions will track the much larger X-ray gas reservoir rather than the stellar component. The precipitation  models also predict that  $\sigma_{gas} <  0.5 \cdot \sigma_{stars}$,  in accordance with what we observe for the M1931 BCG.

\section{Discussion}
\label{discussion}

\begin{figure*}
\begin{subfigure}
    \centering
    \includegraphics[width=0.51\textwidth,angle=0,clip=true]{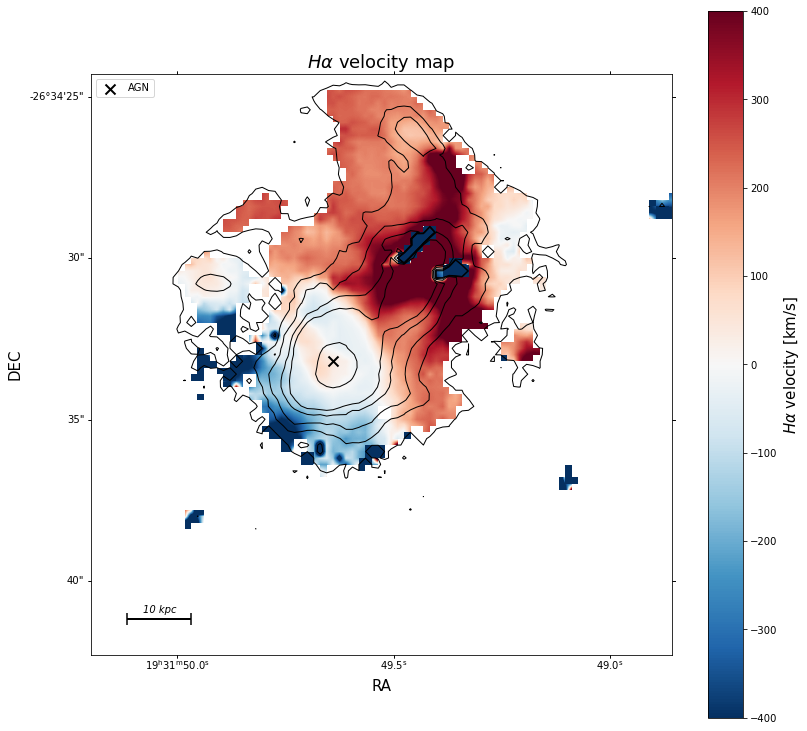}
    \centering
\end{subfigure} 
\begin{subfigure}
    \centering
    \includegraphics[width=0.5\textwidth,angle=0,clip=true]{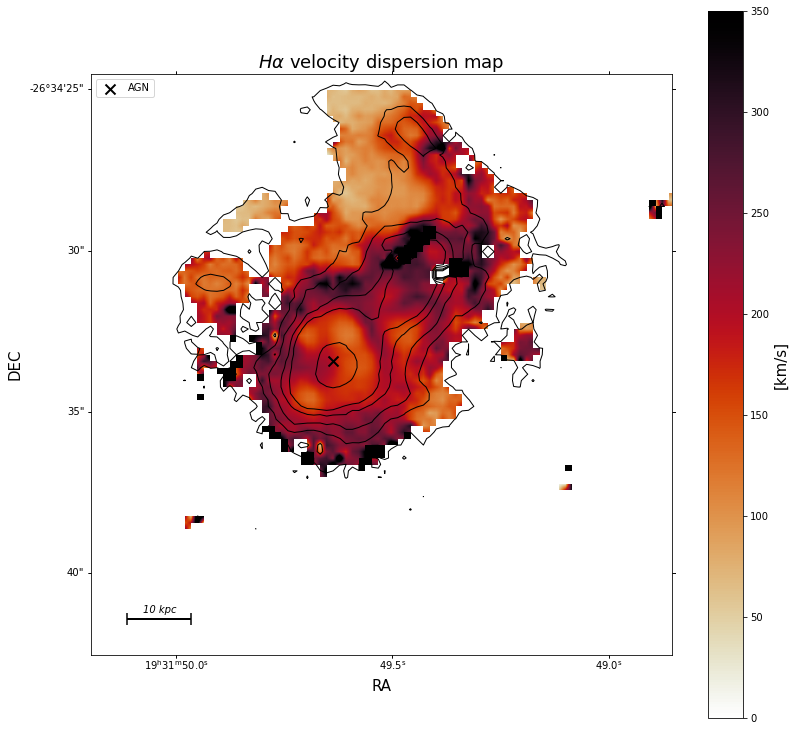}
    \centering
\end{subfigure} 
 \caption{\textit{left}: Spatially resolved $\rm{H\alpha}$ radial velocity map for the BCG of the M1931 galaxy cluster. The colour-bar displays the radial velocity  of the $\rm{H\alpha}$ gas with respect to the BCG rest frame, measured in [km/s]. \textit{right}: Spatially resolved $\rm{H\alpha}$ velocity dispersion  map measured in [km/s]. The white background in both plots  corresponds to spaxels with a $\rm{SNR_{H\alpha}<10}$. The cross shows the location of the AGN.  The contours show the $\rm{H\alpha}$ flux intensity.}  
\label{vel}
\end{figure*}

\begin{figure*}
\begin{subfigure}
    \centering
    \includegraphics[width=0.5\textwidth,angle=0,clip=true]{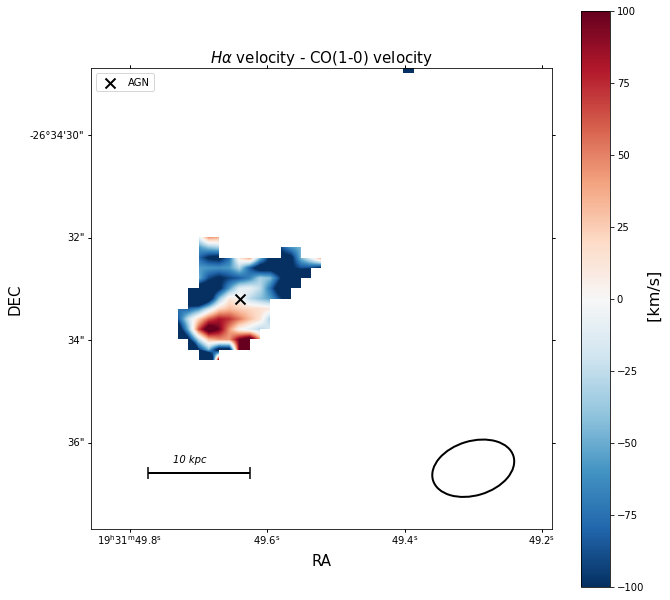}
    \centering
\end{subfigure} 
\begin{subfigure}
    \centering
    \includegraphics[width=0.48\textwidth,angle=0,clip=true]{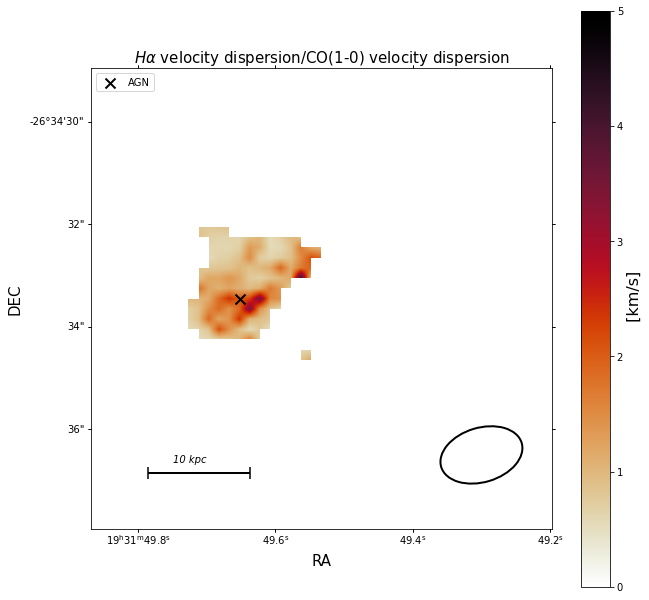}
    \centering
\end{subfigure} 
\begin{subfigure}
    \centering
    \includegraphics[width=0.5\textwidth,angle=0,clip=true]{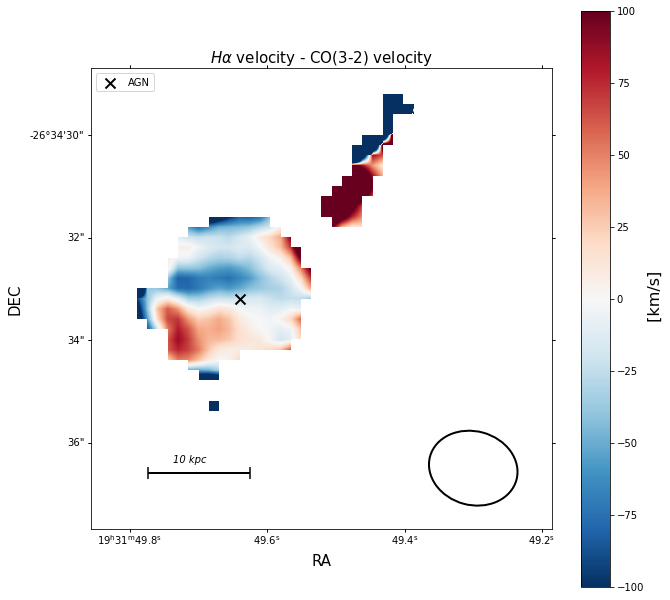}
    \centering
\end{subfigure} 
\begin{subfigure}
    \centering
    \includegraphics[width=0.48\textwidth,angle=0,clip=true]{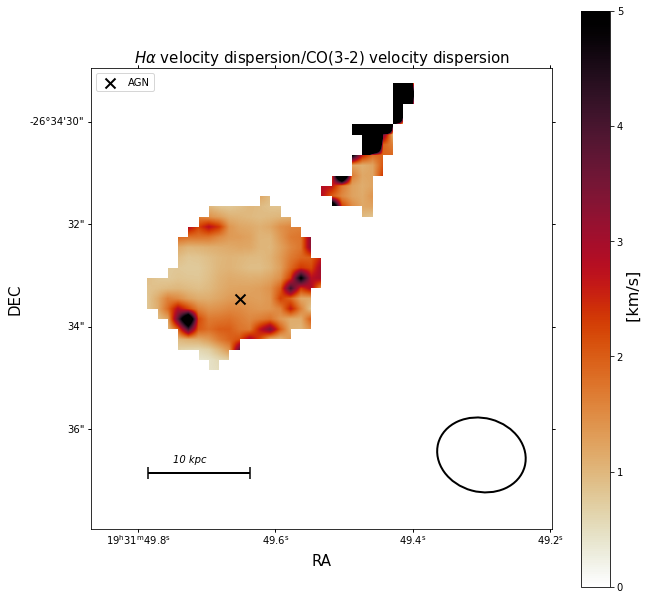}
    \centering
\end{subfigure} 
 \caption{Difference between the kinematics of the ionised gas and  those of the cold molecular gas. For a better visualisation, we present the zoomed in plots corresponding to a spatial scale of 45 by 45 kpc.
 \textit{left-hand side}: Difference between the radial velocity of the $\rm{H\alpha}$ gas and the velocity of the CO(1-0) gas (\textit{top}) and  difference between the  velocity of the $\rm{H\alpha}$ gas and the velocity of the CO(3-2) gas (\textit{bottom}). \textit{right-hand side}: Ratio between the $\rm{H\alpha}$  velocity dispersion and the CO(1-0) velocity dispersion (\textit{top}) and ratio between the  velocity  dispersion of the $\rm{H\alpha}$ gas and the velocity dispersion of the CO(3-2) gas (\textit{bottom}). 
 The cross in all diagrams shows the location of the AGN, as identified according to the different diagnostic diagrams. The  ellipses in the lower right of all panels depict the beam sizes of the ALMA observations.}
\label{veldif}
\end{figure*}

\begin{figure*}
\begin{subfigure}
    \centering
    \includegraphics[width=0.32\textwidth,angle=0,clip=true]{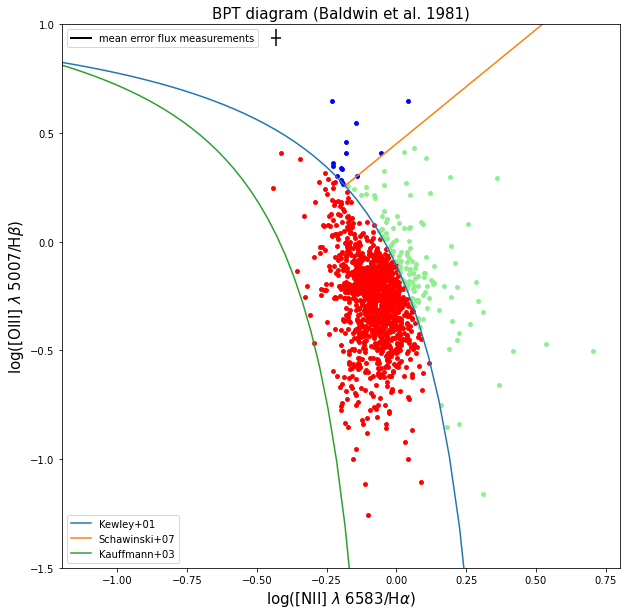}
    \centering
\end{subfigure} 
\begin{subfigure}
    \centering
    \includegraphics[width=0.32\textwidth,angle=0,clip=true]{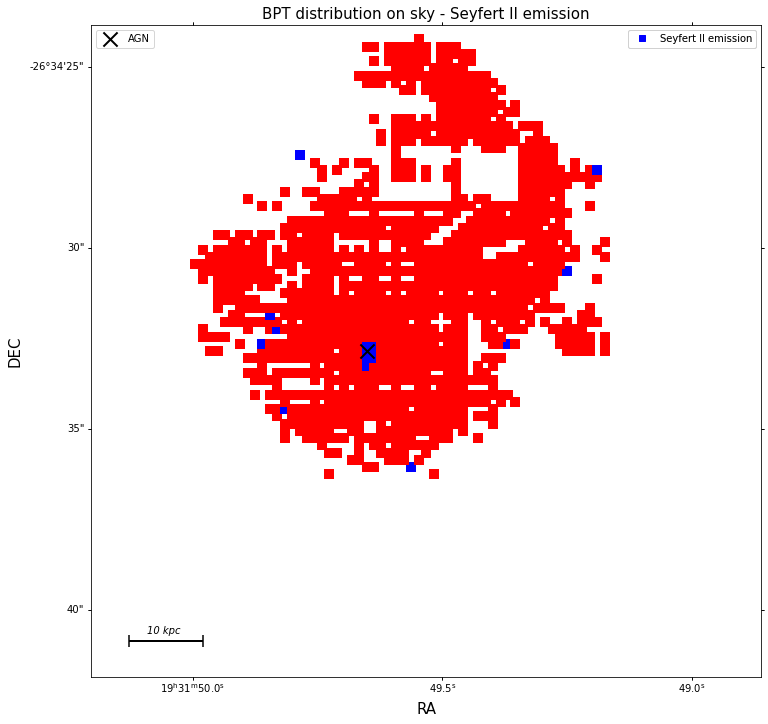}
    \centering
\end{subfigure} 
\begin{subfigure}
    \centering
    \includegraphics[width=0.32\textwidth,angle=0,clip=true]{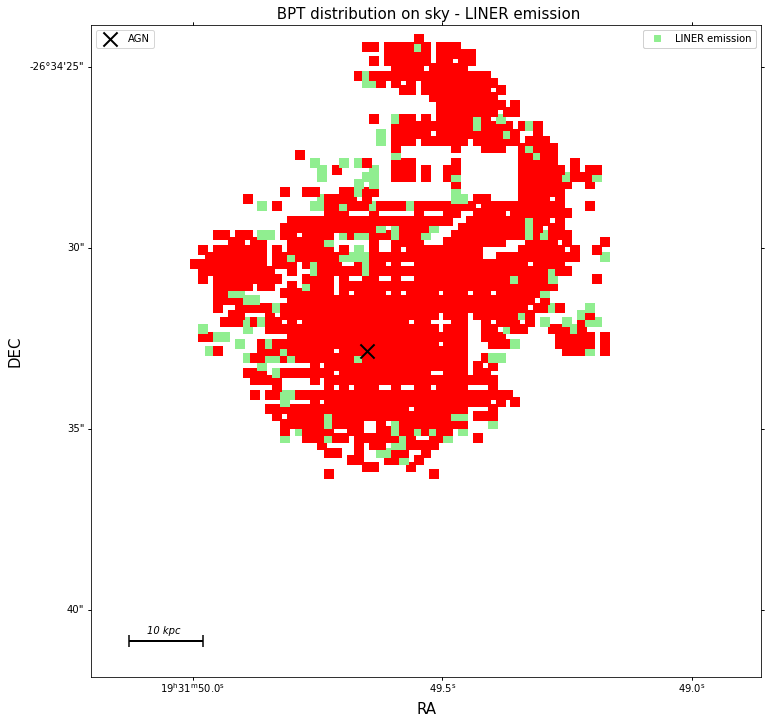}
    \centering
\end{subfigure} 
\begin{subfigure}
    \centering
    \includegraphics[width=0.32\textwidth,angle=0,clip=true]{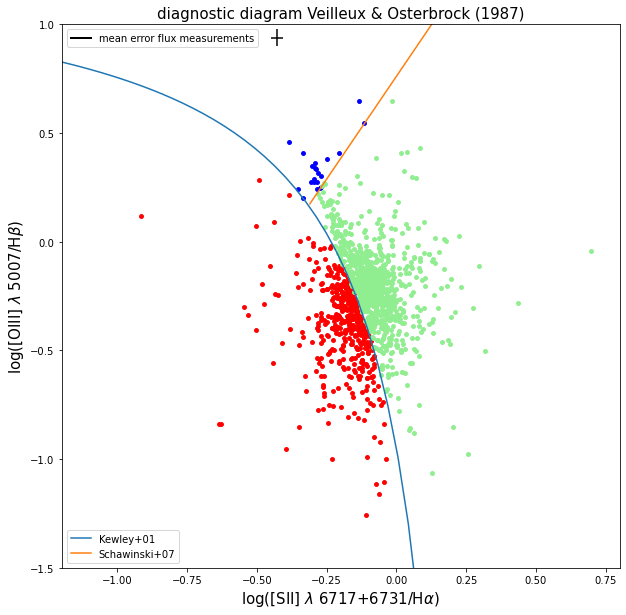}
    \centering
\end{subfigure} 
\begin{subfigure}
    \centering
    \includegraphics[width=0.32\textwidth,angle=0,clip=true]{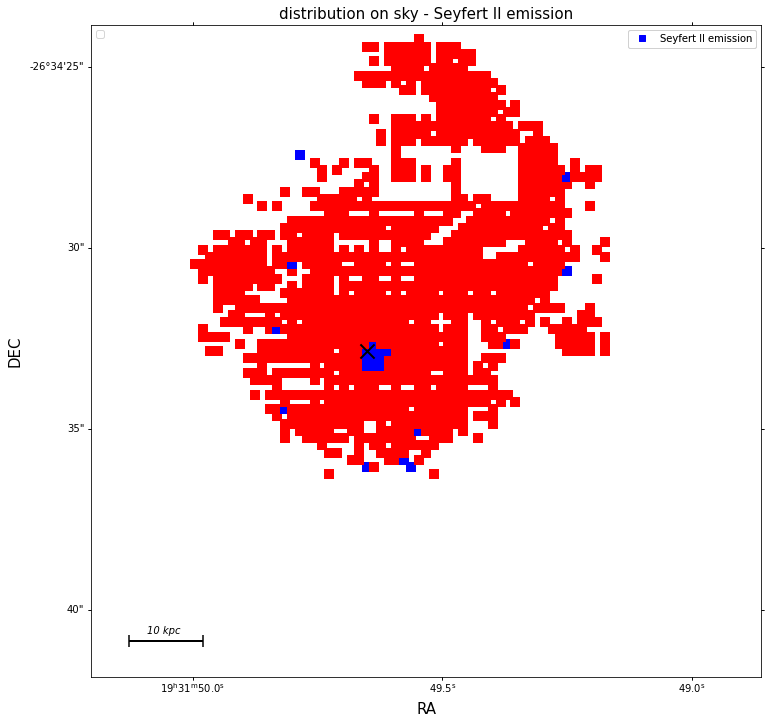}
    \centering
\end{subfigure} 
\begin{subfigure}
    \centering
    \includegraphics[width=0.32\textwidth,angle=0,clip=true]{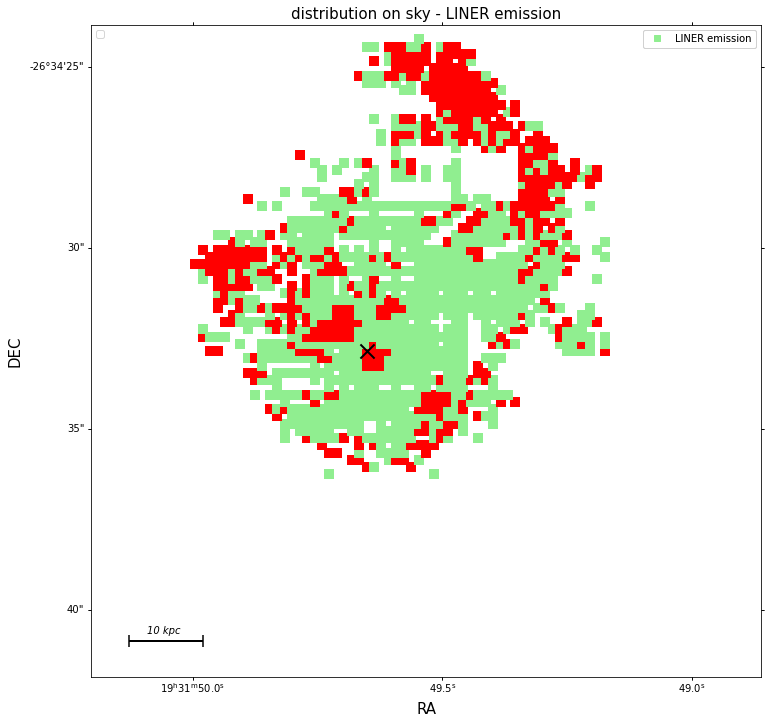}
    \centering
\end{subfigure} 
\begin{subfigure}
    \centering
    \includegraphics[width=0.32\textwidth,angle=0,clip=true]{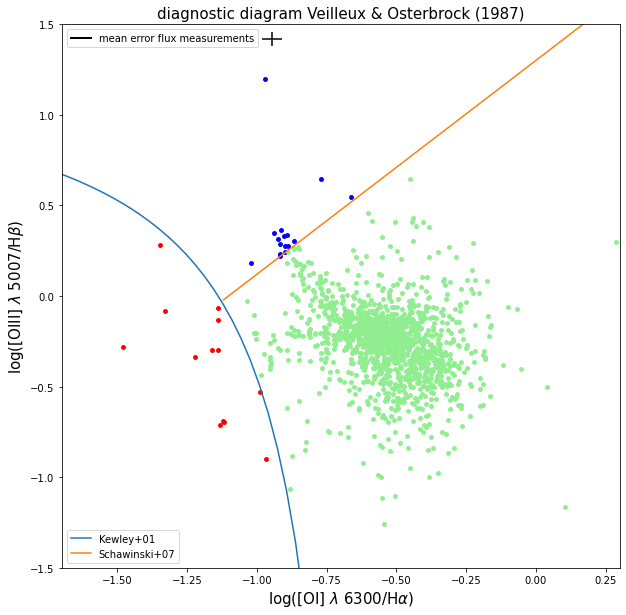}
    \centering
\end{subfigure} 
\begin{subfigure}
    \centering
    \includegraphics[width=0.32\textwidth,angle=0,clip=true]{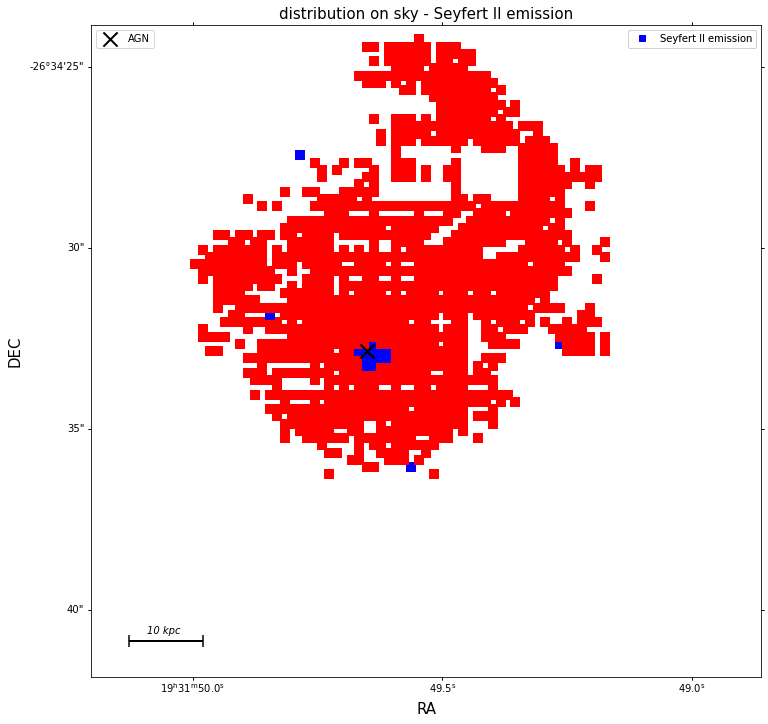}
    \centering
\end{subfigure} 
\begin{subfigure}
    \centering
    \includegraphics[width=0.32\textwidth,angle=0,clip=true]{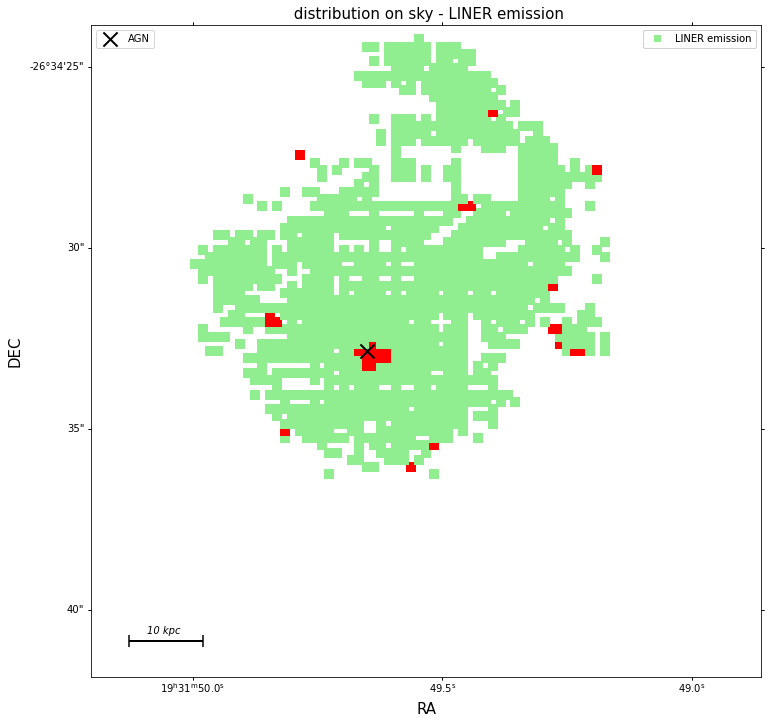}
    \centering
\end{subfigure} 
  \caption{ Diagnostic diagrams  to distinguish the ionization mechanism of the nebular gas.  \textit{first row-left}:   BPT diagram \citep{bald81} for all spaxels of the MUSE data cube with SNR>10 in the emission lines used for the diagnostic. \textit{firs row-middle}: BPT distribution on the sky showing the  spaxels with Seyfert II emission in blue.     \textit{first row-right}: BPT distribution on the sky showing the spaxels which fall in the LINER region of the diagnostic diagram in green.  \textit{second row-left}: Diagnostic diagram of \cite{vel87} using the  $\rm{[O\textsc{iii}]/{H\beta}}$ vs $\rm{[S\textsc{ii}]/{H\alpha}}$ emission line ratios. \textit{second row-middle}: Distribution on the sky showing the Seyfert II emission in blue. \textit{second row-right}: Distribution on the sky showing the LINER emission in green. \textit{third row-left}: $\rm{[O\textsc{iii}]/{H\beta}}$ vs $\rm{[O\textsc{i}]/{H\alpha}}$  diagnostic diagram of \cite{vel87}. \textit{third row-middle}: Distribution on the sky showing the Seyfert II emission in blue. \textit{third row-right}: Distribution on the sky showing the LINER emission in green. In the latter 2 diagrams, each data point represents a spaxel of the MUSE cube with a SNR>10 in each emission line used for the diagnostic. The cross from the upper left corner of all three diagnostic diagrams shows  the mean error of the flux measurements. The cross in the neighbouring maps show the location of the AGN.}
  \label{BPT}
    \centering
  \end{figure*}

\begin{figure*}
    \centering
\begin{subfigure}
    \centering
    \includegraphics[width=0.32\textwidth,angle=0,clip=true]{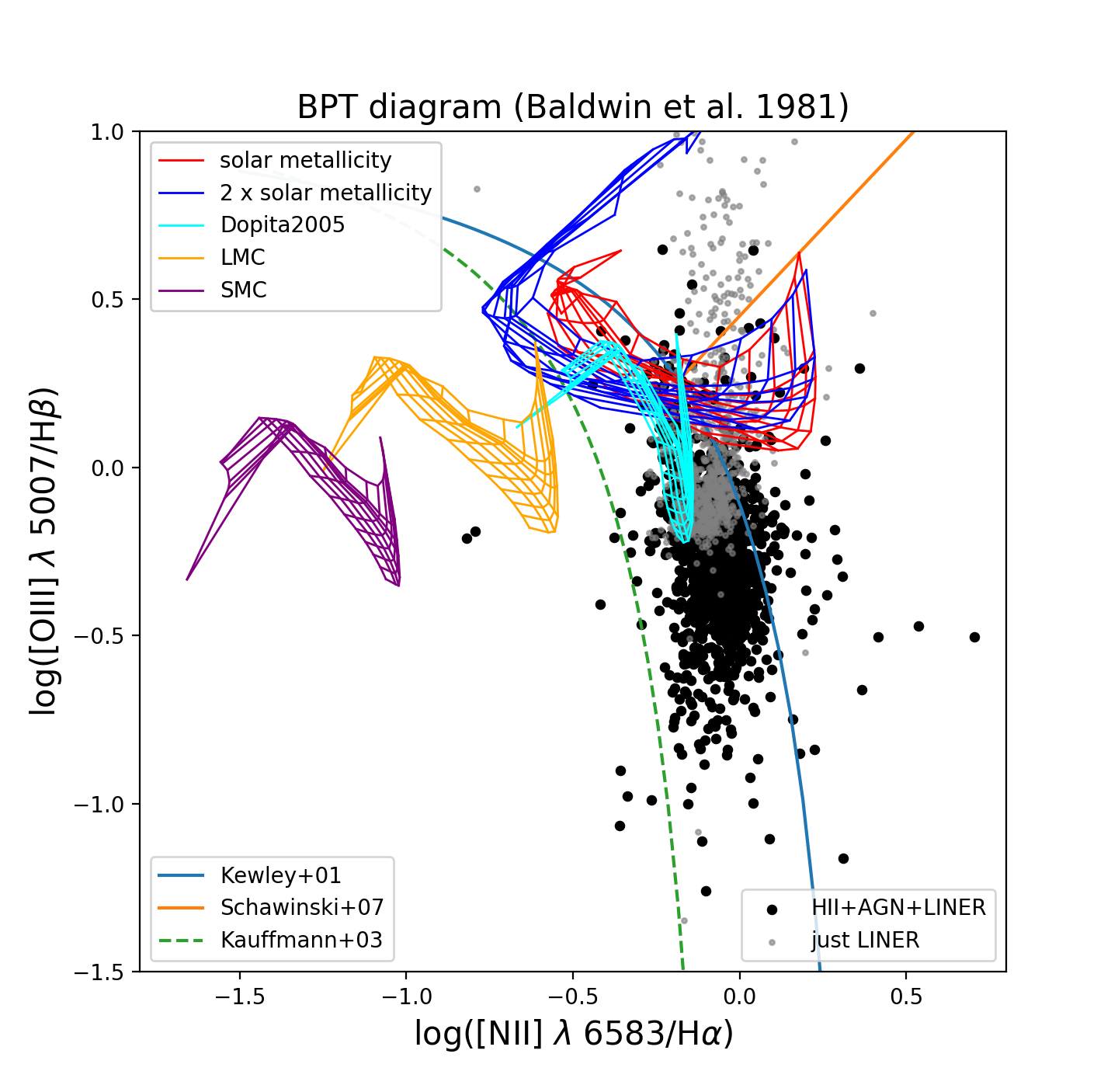}
    \centering
\end{subfigure} 
\begin{subfigure}
    \centering
    \includegraphics[width=0.31\textwidth,angle=0,clip=true]{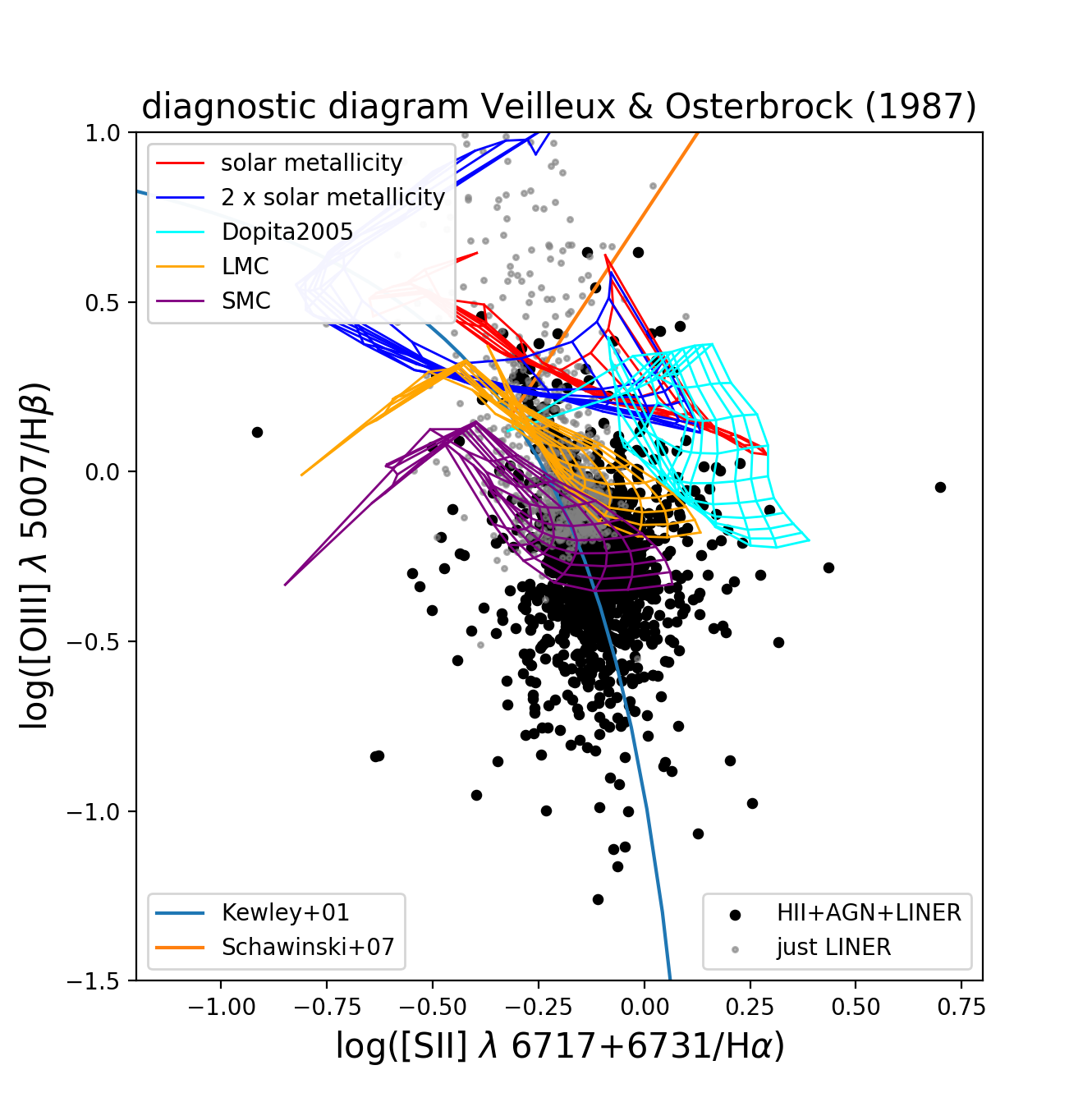}
    \centering
\end{subfigure} 
\begin{subfigure}
    \centering
    \includegraphics[width=0.32\textwidth,angle=0,clip=true]{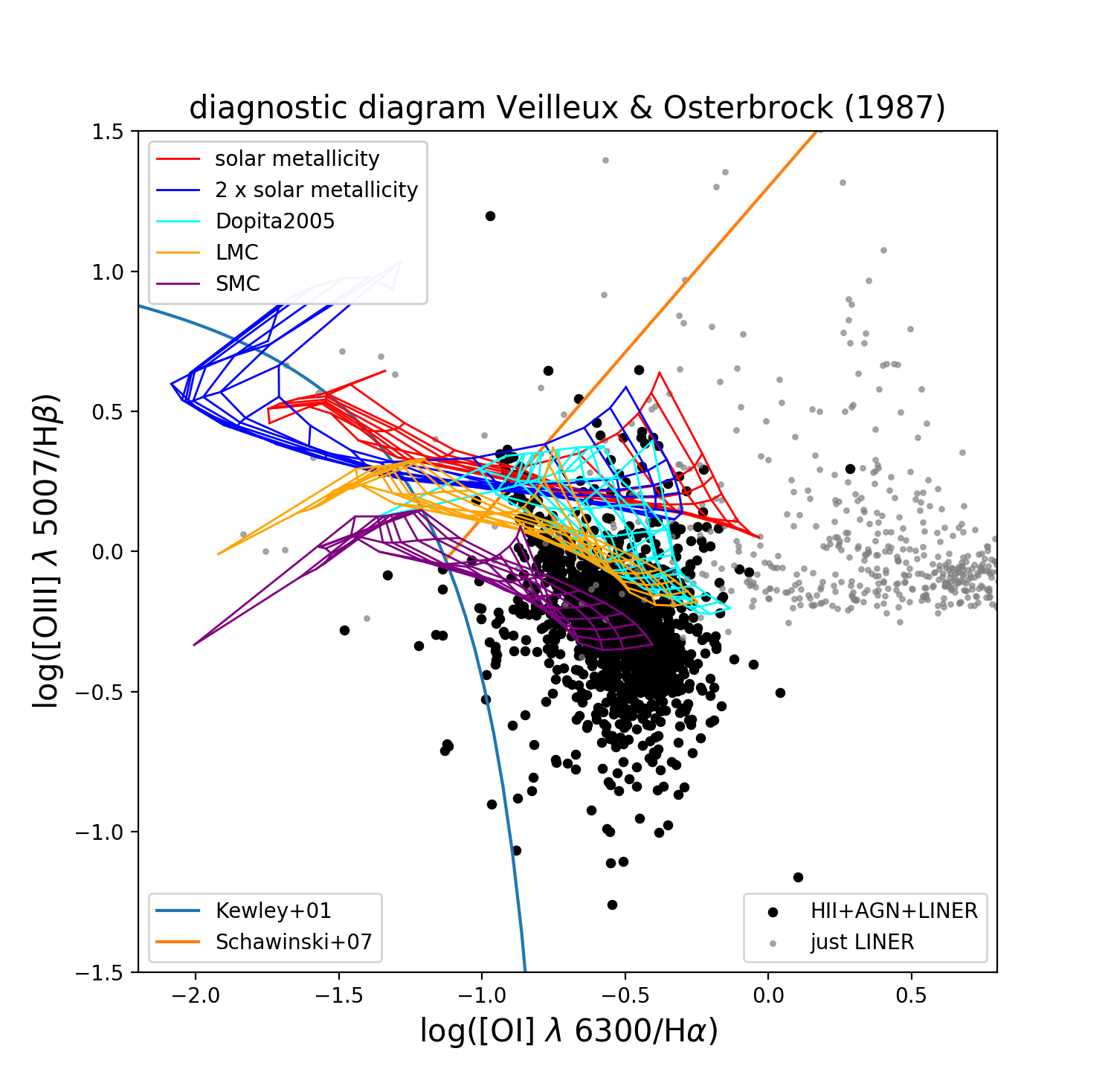}
    \centering
\end{subfigure} 
  \caption{Same as Figure \ref{BPT}, but showing in addition the predictions from the  fully radiative shock models  of  \cite{allen08} computed with MAPPINGS V. We consider a pre-shock density of 1 $\rm{cm^{-3}}$, shock velocities ranging from 10-350 $\: \rm{km~s}^{-1} $, and different metallicities.   The choice of these shock velocities are motivated by the observed velocity dispersion of the nebular gas. The red lines depict a grid with solar metallicity, the blue ones a grid with twice solar metallicity, the cyan lines one with metallicities from \cite{dopita05}, the orange ones a grid with Large Magellanic Cloud metallicities and the purple ones a grid with Small Magellanic Cloud metallicities. The black data-points represent the spaxels of the MUSE cube, with a SNR>10 in each emission line used for the diagnostic. The grey data-points represent the spaxels of the MUSE cube, from which we have subtracted the contribution from star formation and AGN emission to the total luminosity of each emission line, after applying the decomposition method of \cite{davies17}. These data-points are representative for pure LINER-like emission. }
  \label{3mdb}
    \centering
  \end{figure*}

\begin{figure*}
  \centering
\begin{subfigure}
    \centering
    \includegraphics[width=0.3\textwidth,angle=0,clip=true]{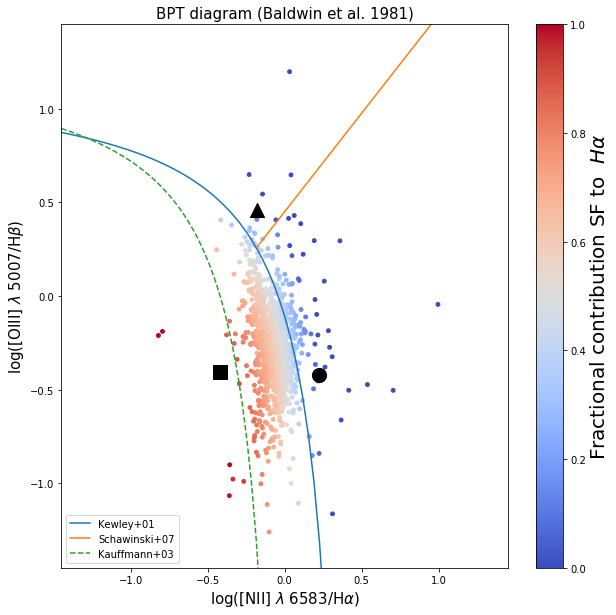}
    \centering
\end{subfigure} 
\begin{subfigure}
    \centering
    \includegraphics[width=0.3\textwidth,angle=0,clip=true]{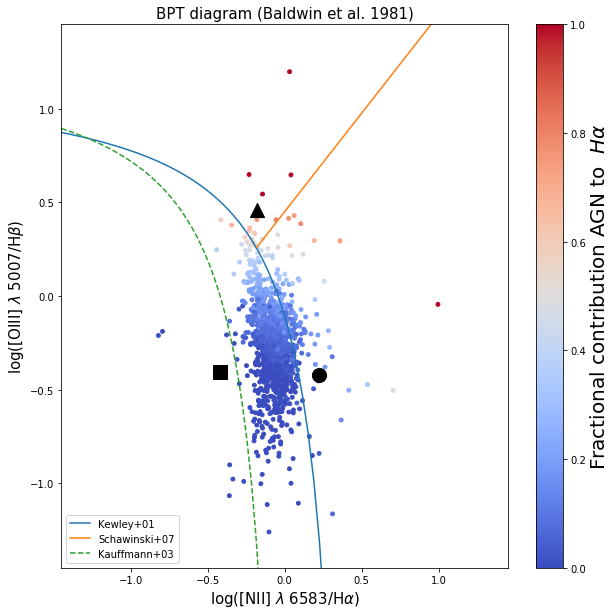}
    \centering
\end{subfigure} 
\begin{subfigure}
    \centering
    \includegraphics[width=0.3\textwidth,angle=0,clip=true]{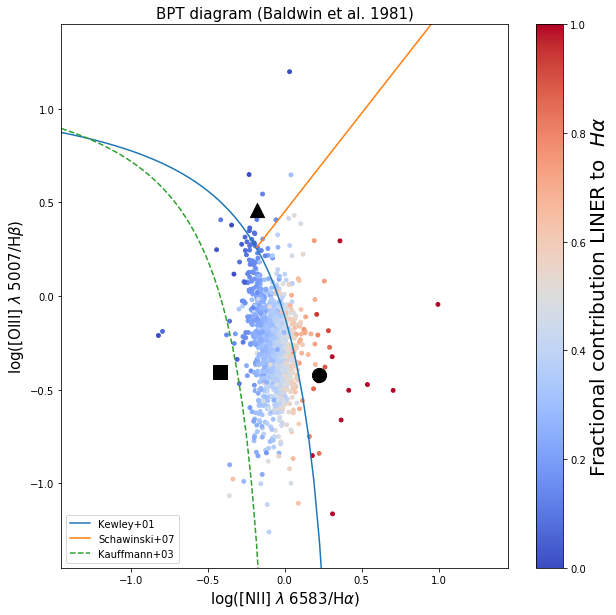}
    \centering
\end{subfigure} 
\begin{subfigure}
    \centering
    \includegraphics[width=0.3\textwidth,angle=0,clip=true]{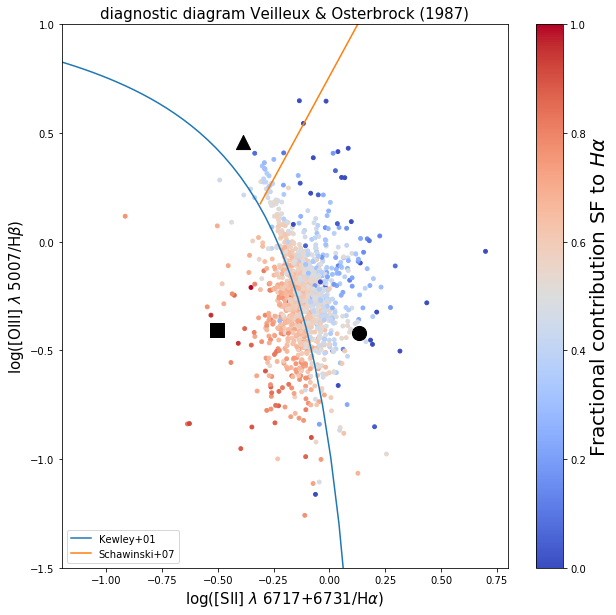}
    \centering
\end{subfigure} 
\begin{subfigure}
    \centering
    \includegraphics[width=0.3\textwidth,angle=0,clip=true]{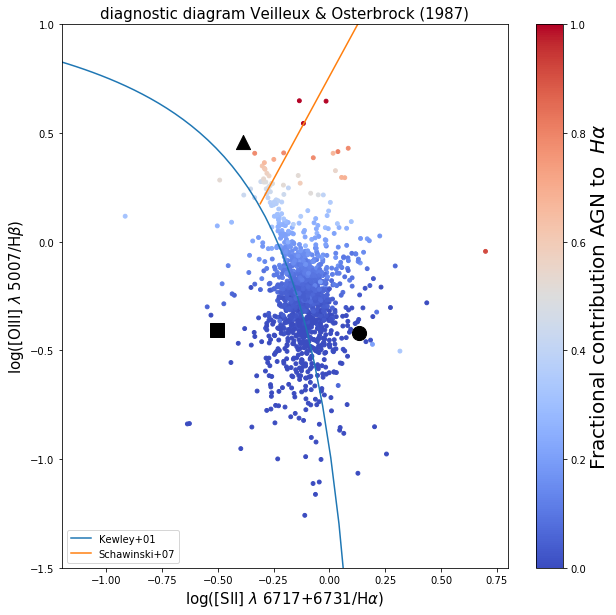}
    \centering
\end{subfigure} 
\begin{subfigure}
    \centering
    \includegraphics[width=0.3\textwidth,angle=0,clip=true]{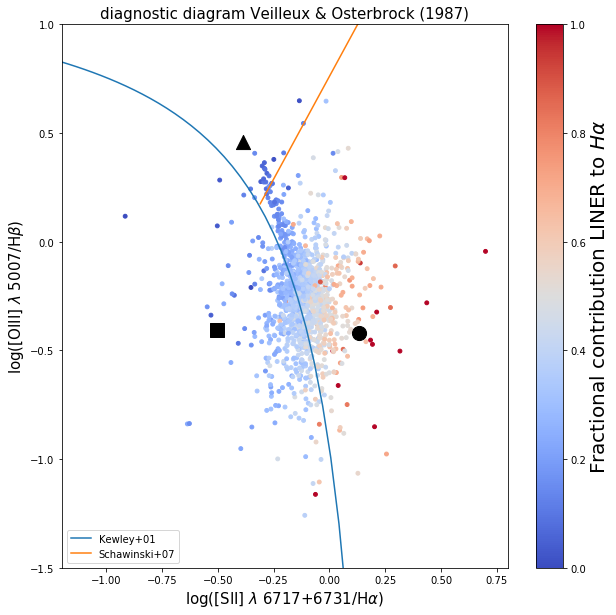}
    \centering
\end{subfigure}  
\centering
\caption{The \textit{upper three panels} show the BPT diagram, while the \textit{lower three panels} show the diagnostic diagram of \cite{vel87}, with the line ratios extracted from the total emission line fluxes of the individual spaxels. The SF, AGN and LINER base spectra are depicted as the black  square, triangle and circle, respectively. The data points are  colour coded according to the fraction of $\rm{H\alpha}$ emission attributable to SF (\textit{left-hand panels}), AGN  (\textit{middle panels}) and "LINER" (\textit{right-hand panels}), as computed  from the spectral decomposition method described in \cite{davies17}. }
\label{daviesbpt}
\end{figure*}

\begin{figure*}
  \centering
\begin{subfigure}
    \centering
    \includegraphics[width=0.3\textwidth,angle=0,clip=true]{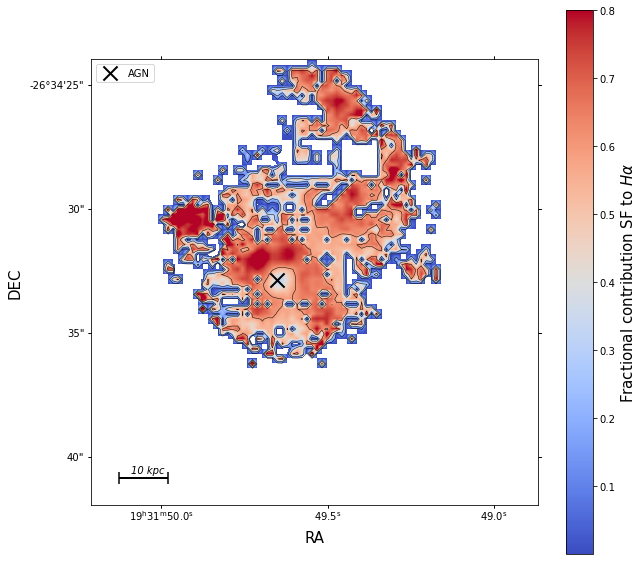}
    \centering
\end{subfigure} 
\begin{subfigure}
    \centering
    \includegraphics[width=0.3\textwidth,angle=0,clip=true]{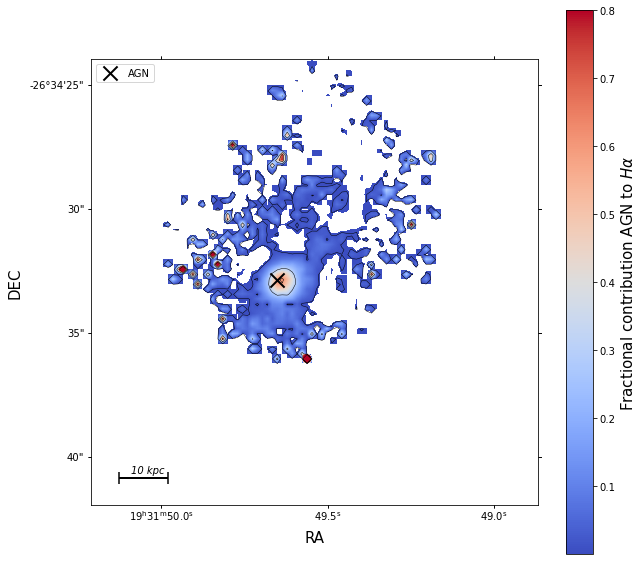}
    \centering
\end{subfigure} 
\begin{subfigure}
    \centering
    \includegraphics[width=0.3\textwidth,angle=0,clip=true]{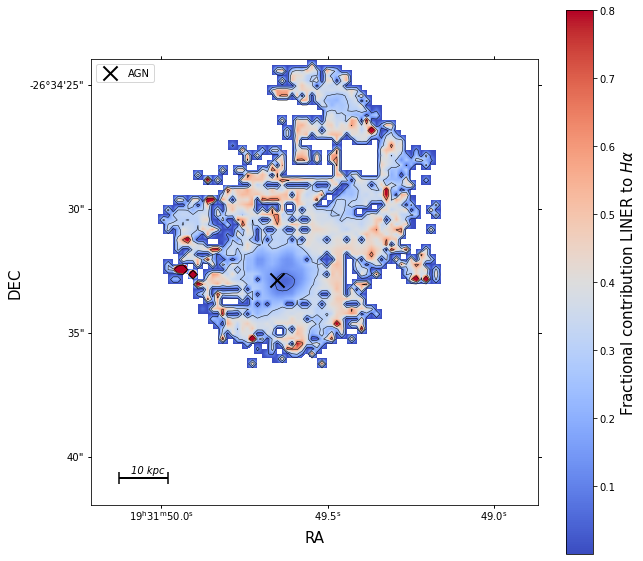}
    \centering
\end{subfigure} 
\begin{subfigure}
    \centering
    \includegraphics[width=0.3\textwidth,angle=0,clip=true]{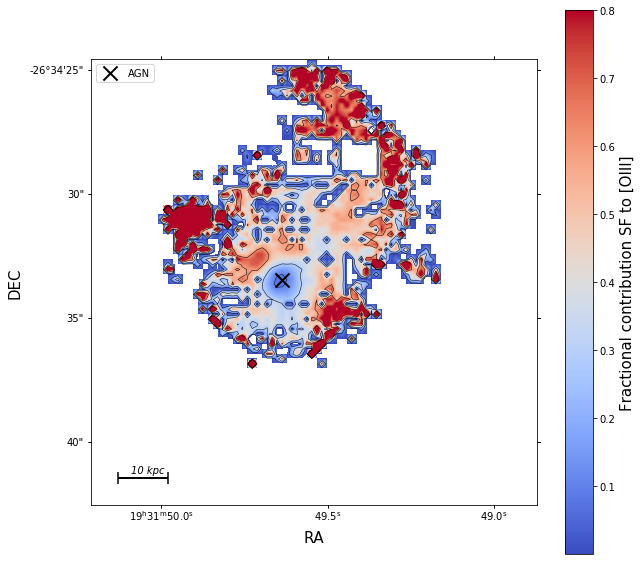}
    \centering
\end{subfigure} 
\begin{subfigure}
    \centering
    \includegraphics[width=0.3\textwidth,angle=0,clip=true]{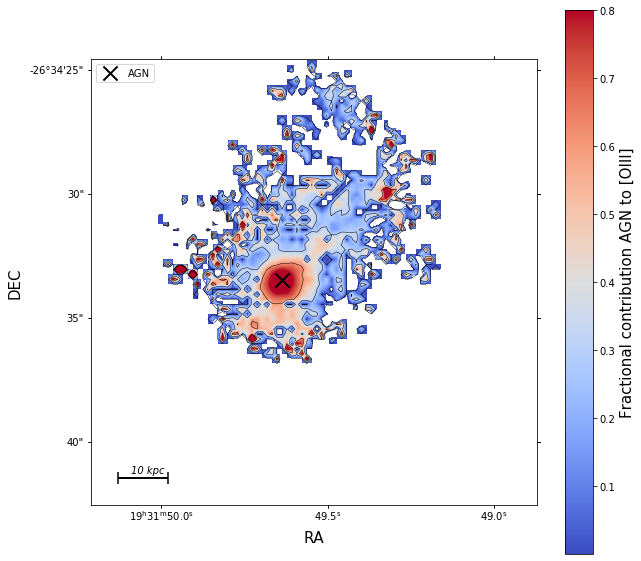}
    \centering
\end{subfigure} 
\begin{subfigure}
    \centering
    \includegraphics[width=0.3\textwidth,angle=0,clip=true]{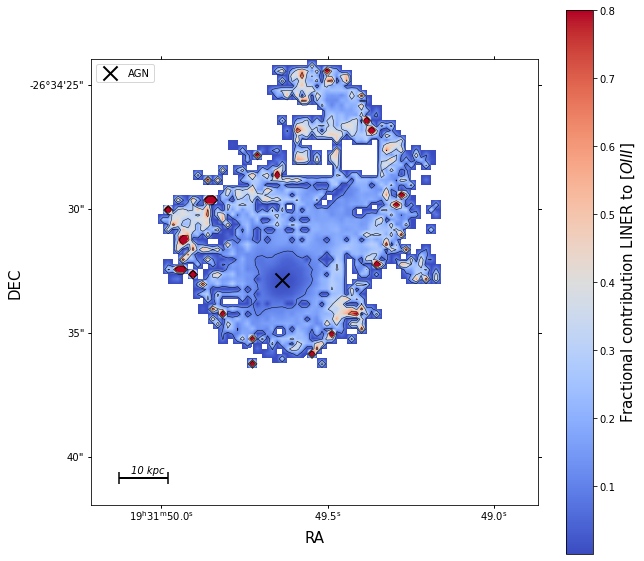}
    \centering
\end{subfigure}  
\begin{subfigure}
    \centering
    \includegraphics[width=0.3\textwidth,angle=0,clip=true]{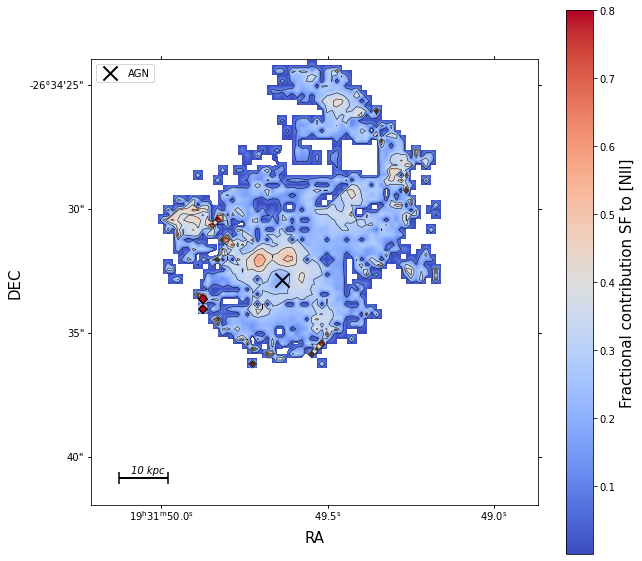}
    \centering
\end{subfigure} 
\begin{subfigure}
    \centering
    \includegraphics[width=0.3\textwidth,angle=0,clip=true]{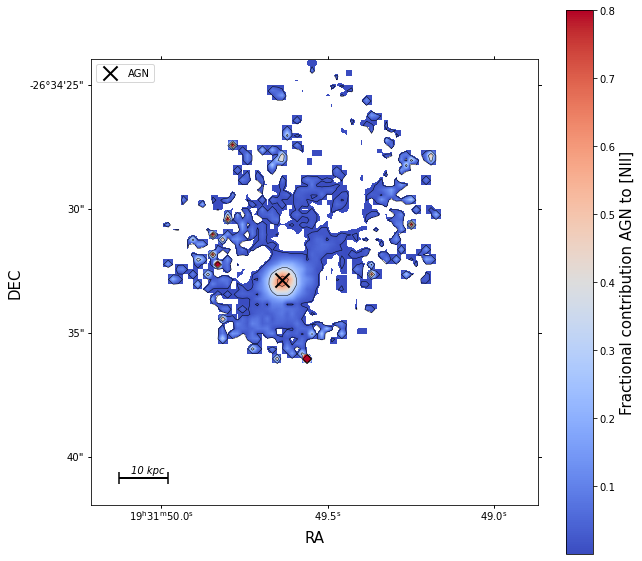}
    \centering
\end{subfigure} 
\begin{subfigure}
    \centering
    \includegraphics[width=0.3\textwidth,angle=0,clip=true]{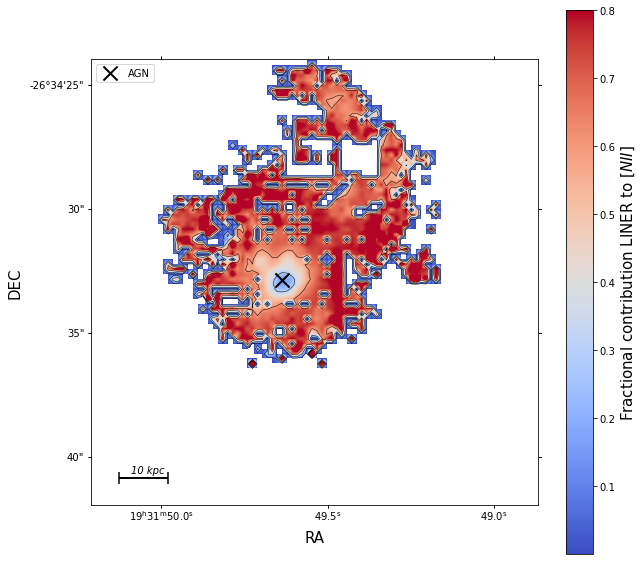}
    \centering
\end{subfigure} 
\begin{subfigure}
    \centering
    \includegraphics[width=0.3\textwidth,angle=0,clip=true]{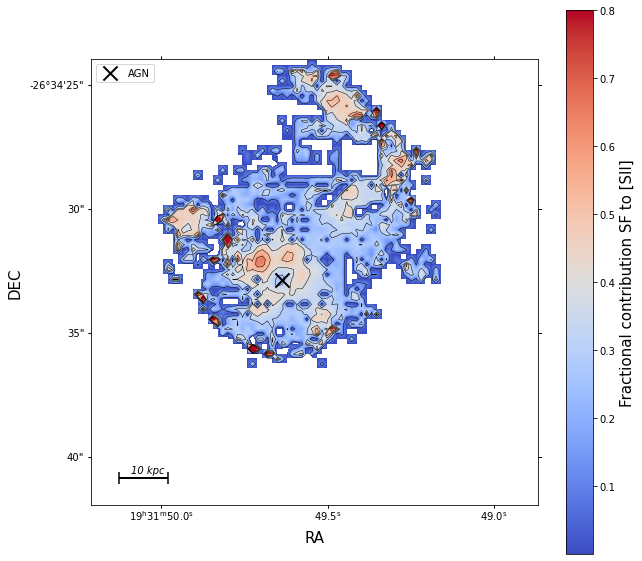}
    \centering
\end{subfigure} 
\begin{subfigure}
    \centering
    \includegraphics[width=0.3\textwidth,angle=0,clip=true]{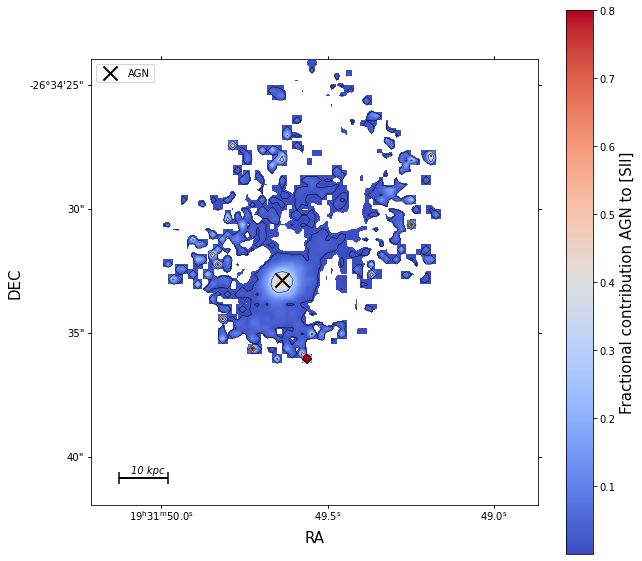}
    \centering
\end{subfigure} 
\begin{subfigure}
    \centering
    \includegraphics[width=0.3\textwidth,angle=0,clip=true]{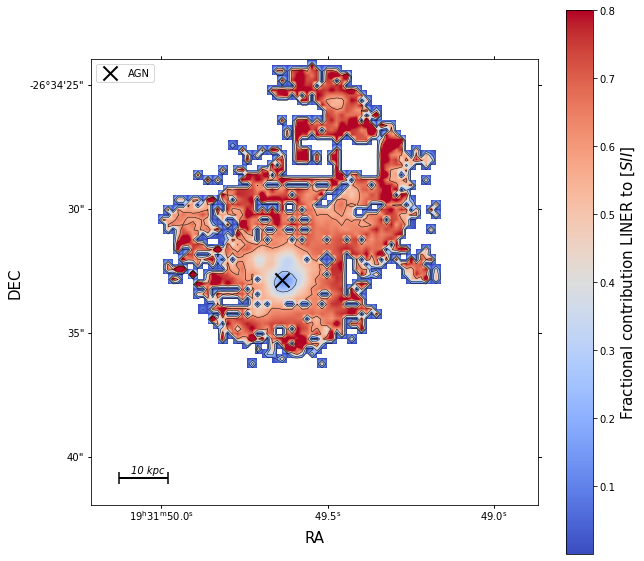}
    \centering
\end{subfigure}  
\centering
\caption{Maps depicting  the fractional contribution of SF (\textit{left}), AGN (\textit{middle}), "LINER" (\textit{right}) to the $\rm{H\alpha}$ (\textit{first row}), $\rm{[O\textsc{iii}]} \: \lambda5007$ (\textit{second row}), $\rm{[N\textsc{ii}]}\:\lambda 6584 $ (\textit{third row}) and $\rm{[S\textsc{ii}]}$ (\textit{last row)} emission lines, as calculated based on the spectral decomposition method. The colour-bar limit displaying the fractional contribution of each ionising mechanism to the total luminosity of the emission lines  was set to 0.8, and not the nominal value of 1, for better visualisation purposes. The cross in all diagrams shows the location of the AGN. }
\label{daviesspatial}
\end{figure*}

\begin{figure*}
\begin{subfigure}
    \centering
    \includegraphics[width=0.51\textwidth,angle=0,clip=true]{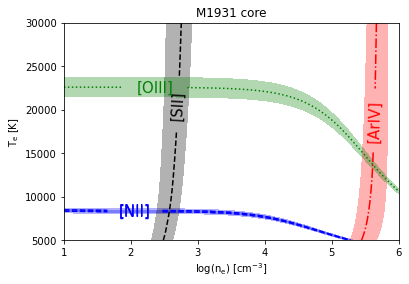}
    \centering
\end{subfigure} 
\begin{subfigure}
    \centering
    \includegraphics[width=0.5\textwidth,angle=0,clip=true]{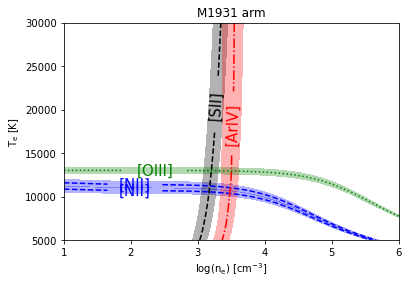}
    \centering
\end{subfigure} 
\begin{subfigure}
    \centering
    \includegraphics[width=0.51\textwidth,angle=0,clip=true]{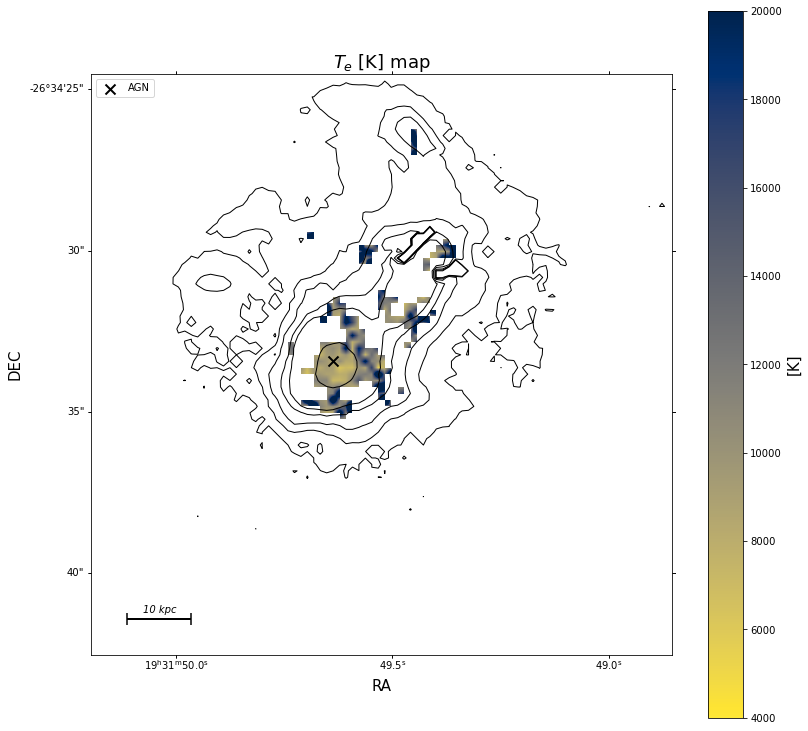}
    \centering
\end{subfigure} 
\begin{subfigure}
    \centering
    \includegraphics[width=0.5\textwidth,angle=0,clip=true]{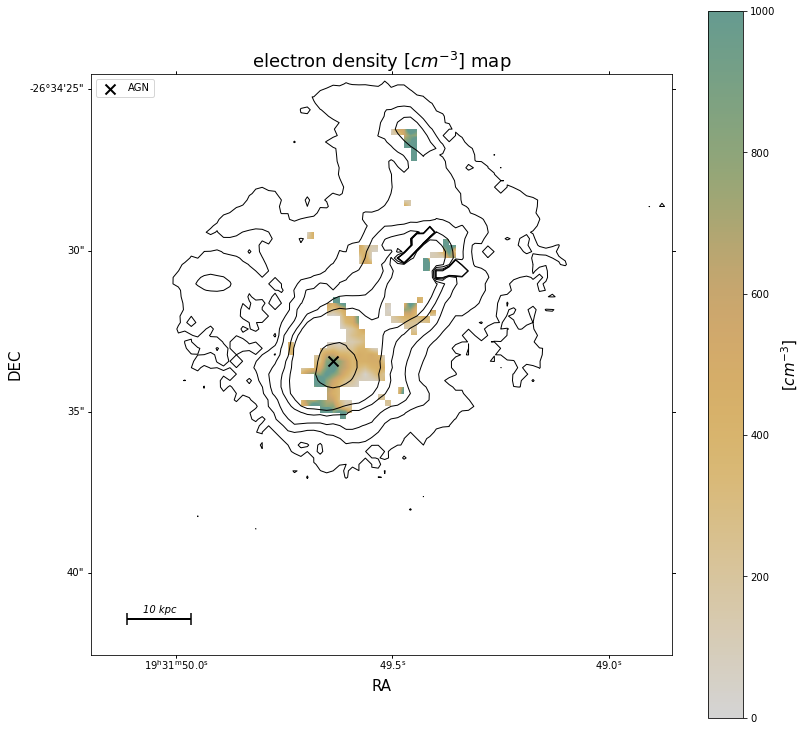}
    \centering
\end{subfigure} 
 \caption{\textit{top-left}: Electron density vs electron temperature  diagnostic diagram for the core of the M1931 BCG.  For this plot, we have used the diagnostics:  $\rm{[N\textsc{ii}]}\:\lambda\:5755/6548$,  $\rm{[N\textsc{ii}]}\:\lambda\:5755/6584$, $\rm{[S\textsc{ii}]}\:\lambda\:6731/6716$, $ \rm{[Ar\textsc{iv}]}\:\lambda\:4740/4711$ and  $\rm{[O\textsc{iii}]}\:\lambda\:4363/5007$. The intersection of the curves gives the best fit electron temperature and density.    \textit{top-right}: Same as \textit{top-left}, but for the $\rm{H\alpha}$ tail of the BCG. \textit{bottom-left}: Spatially resolved electron temperature map for the BCG of the M1931 cluster. The colour bar shows the temperature measured in [K], as computed using the \texttt{PyNeb} tool, from the  $\rm{[N\textsc{ii}]}\: \lambda5755/6584 $ emission lines.   \textit{bottom-right}: Electron density map as computed using the \texttt{PyNeb} tool, from the  $\rm{[S\textsc{ii}]}\:\lambda6731/6716$ doublet. The colour-bar depicts the density in units of $\rm{cm^{-3}}$. The contours in both lower panels show the $\rm{H\alpha}$ flux intensity, while the cross shows the location of the AGN. }  
\label{TD}
\end{figure*}

\begin{figure*}
\begin{subfigure}
    \centering
    \includegraphics[width=0.5\textwidth,angle=0,clip=true]{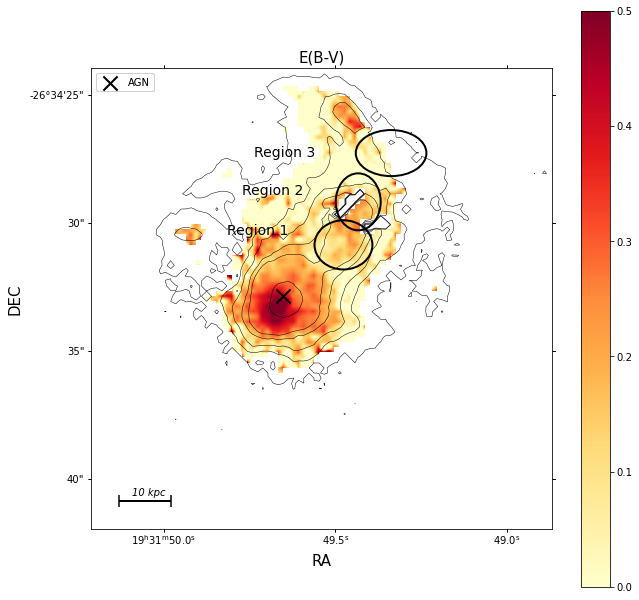}
    \centering
\end{subfigure} 
\begin{subfigure}
    \centering
    \includegraphics[width=0.5\textwidth,angle=0,clip=true]{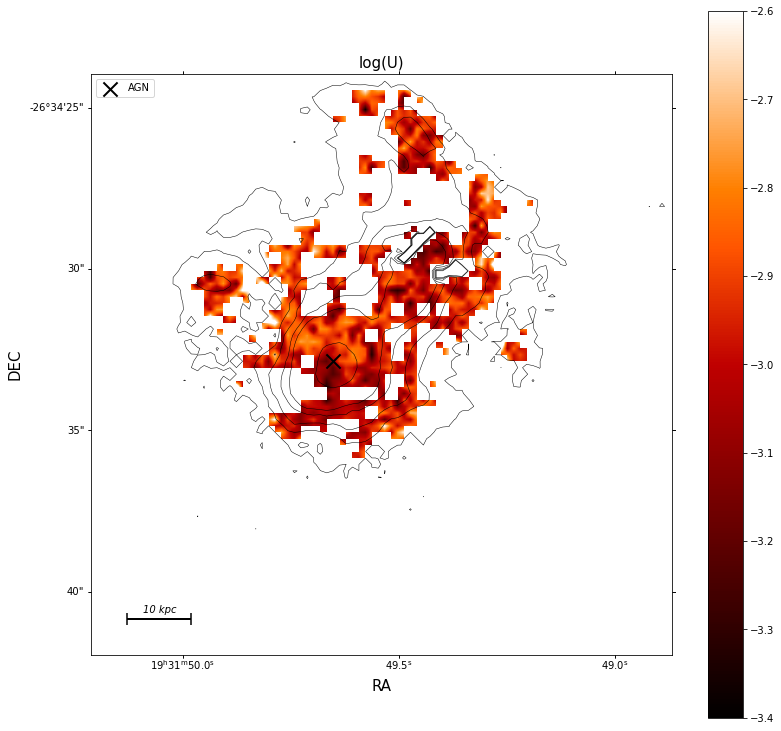}
    \centering
\end{subfigure} 
 \caption{ \textit{left}: Colour excess map for the M1931 BCG as computed from the Balmer decrement. \textit{right}: Ionisation parameter map, as computed from the the  \texttt{HII-CH-mistry tool} \citep{pm14}.  The cross in all diagrams shows the location of the AGN. The contours show the $\rm{H\alpha}$ flux intensity. The white background corresponds to the spaxels with a SNR<10 in the emission lines of interest.}  
\label{EU}
\end{figure*}

\begin{figure*}
    \centering

\begin{subfigure}
    \centering
    \includegraphics[width=0.45\textwidth,angle=0,clip=true]{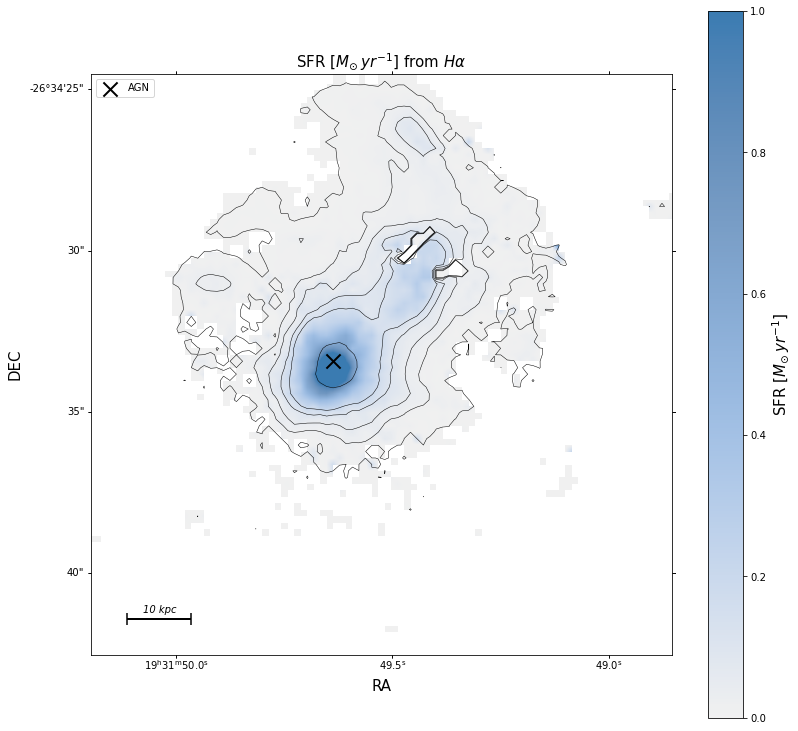}
    \centering
\end{subfigure}

  \caption{Spatially resolved SFR map calculated from the extinction corrected $\rm{H\alpha}$ emission line for each spaxel of the MUSE sub-cube. The colour-bar shows the SFR  in units of $\rm{M_{\odot}/yr}$.
The cross shows the location of the AGN, while the contours display the $\rm{H\alpha}$ flux intensity. The white background corresponds to the spaxels, which have a $\rm{SNR_{H\alpha}<10}$.}
  \label{SFROH}
    \centering
  \end{figure*}

\begin{figure*}
    \centering
\begin{subfigure}
    \centering
    \includegraphics[width=0.45\textwidth,angle=0,clip=true]{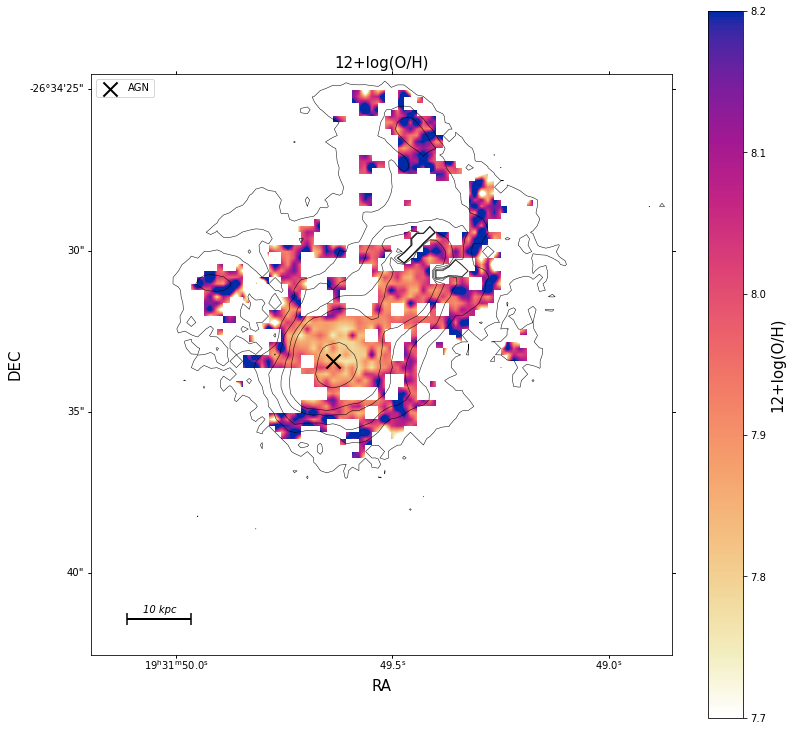}
    \centering
\end{subfigure} 
 \caption{ Gas phase metallicity (O/H), as computed by the \texttt{HII-CH-mistry tool}. The colour-bar shows the oxygen abundance in units of 12+log(O/H). The cross shows the location of the AGN, while the contours display the $\rm{H\alpha}$ flux intensity.  The white background corresponds to the spaxels with a SNR<10 in the emission lines of interest.}
  \label{OH}
    \centering
  \end{figure*}

\begin{figure*}
    \centering
    \includegraphics[width=0.45\textwidth,angle=0,clip=true]{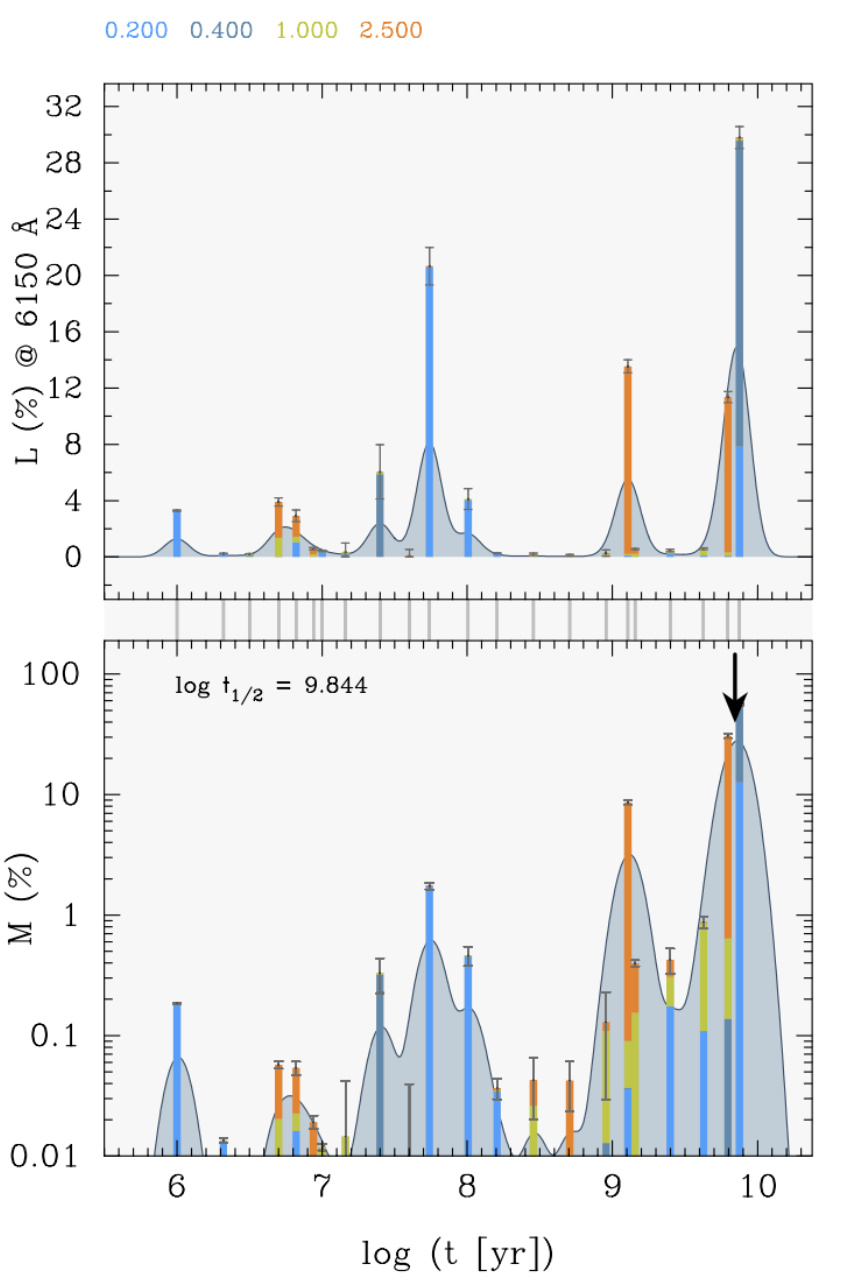}
    \centering
\caption{ Star formation history of the M1931 BCG. The \textit{upper panel} shows the contribution of  the individual SSPs in the best-fitting population vector to the monochromatic luminosity at 6150 $\AA$ as a function of age. The \textit{lower panel}  displays the mass contribution of the SSPs to the total mass of the system as a function of age. The vertical arrow marks the age  when $50\%$ of the present-day stellar mass has been in place.  The colour coding in both panels display the metallicities of the SSPs, whose values can be found above the upper plot.  The vertical bars represent the $1\sigma$ uncertainties. The grey vertical lines connecting the two panels mark the ages of the SSPs.  The light-blue shaded area in both panels shows an Akima-smoothed \citep{Akima70} version of the SSP contributions.}  
\label{sfh}
\end{figure*}

\begin{figure*}
    \centering
\begin{subfigure}
    \centering
    \includegraphics[width=0.45\textwidth,angle=0,clip=true]{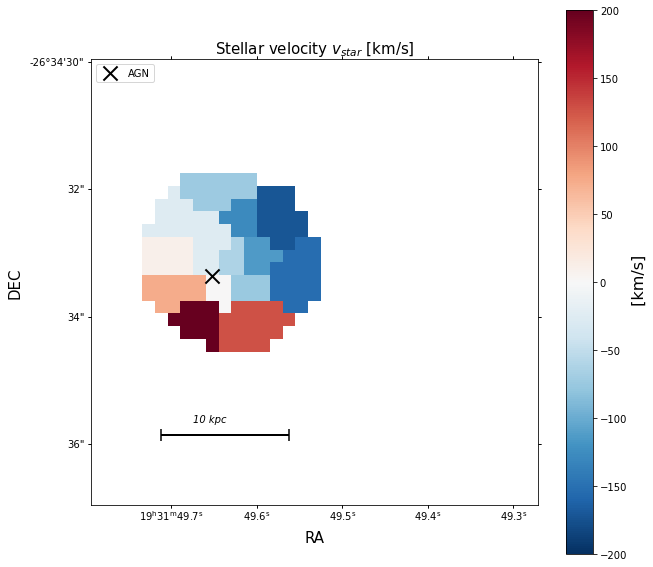}
    \centering
\end{subfigure} 
\begin{subfigure}
    \centering
    \includegraphics[width=0.45\textwidth,angle=0,clip=true]{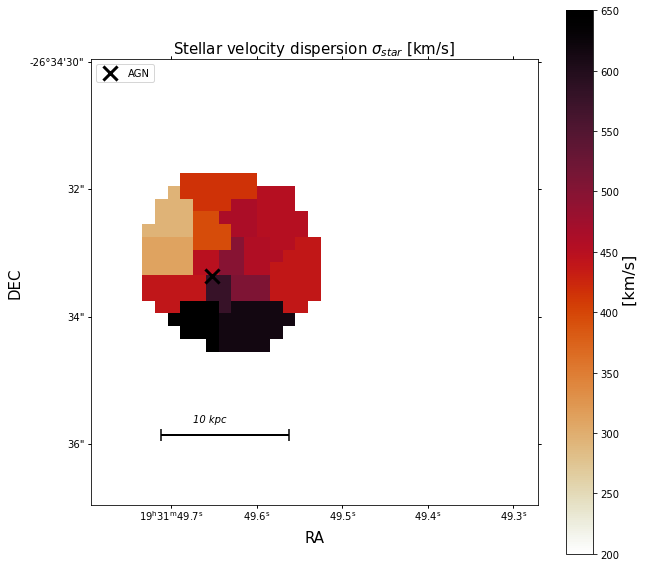}
    \centering
\end{subfigure}    
 \centering
\caption{\textit{left:} Radial velocity of the stars with respect to the systemic velocity of the most central region of the system. \textit{right:} Velocity dispersion of the stars in the  BCG. The colour-bar in both plots depicts the radial velocity and velocity dispersion in units of  [km/s]. Due to the low signal to noise in the stellar continuum, the data needs binning, and we can, therefore, measure the stellar kinematics only in the core of the system. To better visualise the results, we present maps showing just the most central regions of the BCG, corresponding to a spatial scale of 45x45 kpc. The white background corresponds to the spaxels, where the $\rm{SNR_{stellar\:continuum}< 10}$.   The cross shows the location of the AGN. }  
\label{stvel}
\end{figure*}

\begin{figure*}
    \centering
\begin{subfigure}
    \centering
    \includegraphics[width=0.45\textwidth,angle=0,clip=true]{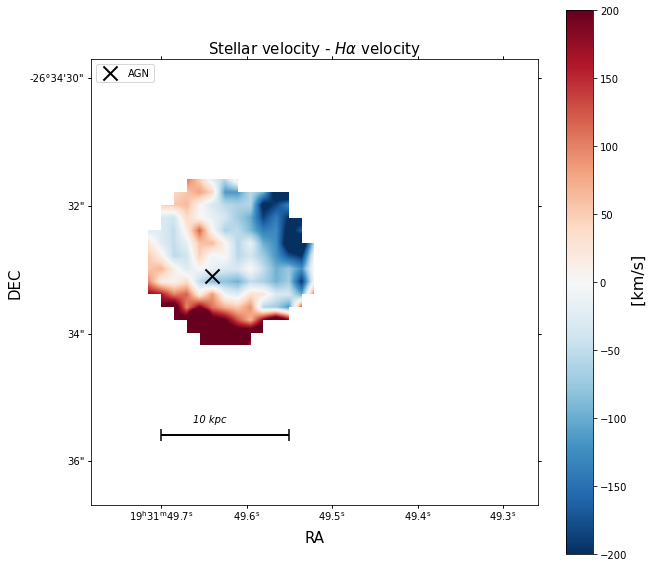}
    \centering
\end{subfigure} 
\begin{subfigure}
    \centering
    \includegraphics[width=0.45\textwidth,angle=0,clip=true]{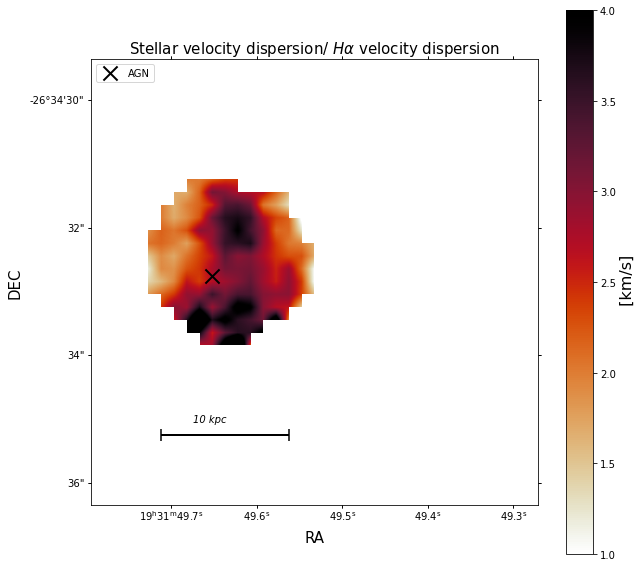}
    \centering
\end{subfigure}     \centering
\caption{\textit{left:} Difference between the radial velocities of the stars and the $\rm{H\alpha}$ gas in the BCG core.  \textit{right:} Ratio between the velocity dispersion of the stars and the $\rm{H\alpha}$ gas in the core of the system. To better visualise the results, we present maps showing just the most central regions of the BCG, corresponding to a spatial scale of 45x45 kpc. The white background corresponds to the spaxels, where the $\rm{SNR_{stellar\:continuum}< 10}$.   The cross shows the location of the AGN.}  
\label{stveldif}
\end{figure*}

The archival MUSE and ALMA observations used in our analysis of the M1931 BCG reveal an extended, filamentary system,  displaying dynamically unrelaxed structures with disturbed motions along the filaments. \\
Both the molecular and ionised gas distribution  shows a nuclear emission component closely related to the BCGs' core, as well as a set of clumpy filaments, which form an elongated $\rm{H\alpha}$ arm, extending $\sim 30$ kpc in NE direction. Both gas phases are co-spatial and co-moving, with the  molecular gas being spatially distributed along the brightest emission from the warm ionised nebula.\\
These findings are  in accordance with the hypothesis that the optical nebula traces the envelopes of the cold molecular clouds. As we have seen in sect. \ref{BPTsect},  the cloud surfaces are excited by a manifold of  mechanisms and are bright in Balmer and forbidden line emission.  
The pockets of cold gas are immersed within ionised gas, and the in situ SF  within the molecular clouds as well as interactions with the ICM are the most plausible mechanisms that give rise to extended $\rm{H\alpha}$ emission all over the core and NE filament. \\Such correlations between the two gas phases can  be  interpreted as the manifestation of a common origin, like the condensation of low-entropy ICM gas  through thermal instabilities. The  ICM surrounding the $\rm{H\alpha}$ tail has the
 lowest entropy according to \cite{eh11}, therefore, being a prime candidate for the reservoir of gas that cools to become star-forming molecular gas. Moreover, the recovered values for the gas phase metallicity in the BCG are consistent to the ones recovered by \cite{eh11} for the ICM, within the errors.  These findings support the scenario, that the warm gas  has condensed from the ICM, without being further polluted by the stars in the system. \\For example, \cite{fogarty17} studied the nature of feedback mechanisms in the 11 CLASH  BCGs and their results strongly suggest that thermally unstable ICM plasma with a low cooling time is the source of material that forms the reservoir of cool gas fuelling star formation in the BCGs and that BCG star formation and feedback either exhausts the supply of this material on gigayear timescales or settles into a state with relatively modest and continuous star formation.\\
The role of AGN feedback on the turbulent and chaotic behaviour of the gas was emphasised by \cite{voit17} who demonstrated that the AGN jets and bubbles can promote condensation of the hot ambient medium by raising some of it to greater altitudes and thus lowering its $\rm{(t_{cool} /t_{ff})} $ ratio.
After the condensates form, they start raining down towards the SMBH, representing  additional fuel for  both star formation and AGN activity.  In simulations, complex arcs of gas fall back into the centres of BCGs after initially being propelled outwards by AGN jets and cavities (\citealt{gaspari13}, \citealt{li15}), similar to what we observe in the M1931 BCG. The significance of uplift and radial infall  is now an integral aspect of the precipitation model (\citealt{voit15b}, \citealt{voit17}) as well as of the Chaotic cold Accretion model (\citealt{gaspari18},  \citealt{tr18}).\\
The kinematics of the gas in the M1931 BCG  point out to such a scenario. According to \cite{fogarty19} and to our observations, the molecular gas in the $\rm{H\alpha}$ tail is probably falling inward at $\sim300\: \rm{km~s}^{-1}$ , with the ionised gas component closely following these motions, as can be seen from the left-hand panel of Fig. \ref{vel}. The redshifted stream of gas we observe would, thus,  relate to the material that is radially in-falling towards the centre of the system.  
These infalling clouds can provide a substantial component of the mass flux toward the SMBH accretion reservoir,  while the physical conditions of the gas within these clouds could satisfy the criteria for the ignition of star formation. \\Moreover, the gaseous and stellar kinematics are decoupled, as they do not share a common velocity field, with the stars showing larger velocities than the gas. These distinct properties of the gas velocities  and the stars' velocities  suggest that the motion of the gas is more closely related to the turbulence and bulk motions within the ICM than it is related to the motion of the stars in the system.\\
 The kinematics of the warm and cold gas phases lead us to conclude that the AGN in the system might have recently experienced an energetic outburst. This  AGN outburst temporarily led to the condensation of the uplifted gas, which is now probably cycling back down towards the AGN, promoting an elevated star formation rate of $\sim 100 \:\rm{M_{\odot}/yr}$. The observed distribution of  dust  in the system  \citep{fogarty19}  points to an  uplift mechanism, with dust emission most prominent in the core and at the furthest extremity of the tail. The observed distribution of the  gas phase metallicity  can also hint to such an uplift mechanism. We observe the lowest (O/H) values of $\sim$ 7.9 in the most central regions of the systems, while in the $\rm{H\alpha}$ tail, the metallicity is slightly higher. This might mean that the metal enriched gas has been expelled outwards from the centre due to AGN outflows. \\
However, the infalling multiphase tail is not aligned with the jet axis implied by features consistent with E-W oriented X-ray cavities observed in the Chandra data \citep{eh11}. This indicates that the gaseous component either might have originally formed at the interface between the jet-inflated cavities and the ICM and then migrated away from these positions, or it might have formed at an earlier epoch, sufficiently long ago for the X-ray cavities to have dissipated \citep{fogarty19}. \\
The cloud envelopes in this system are ionised by a superposition of many physical processes, including photoionisation from young stars  and other mechanisms which give rise to  LINER-like emission, the best candidates being photoionisation by cosmic rays, conduction from the hot ICM, X-ray photoionisation, turbulent mixing layers or collisional heating.  AGN emission dominates only the BCG core, see Figs. \ref{BPT} and \ref{daviesspatial}. This is in accordance with the findings of \cite{fogarty19}, whose analysis of the CO spectral line energy distribution  in the M1931 BCG also reveals evidence for multiple gas excitation mechanisms including SF, interaction between the molecular gas and the ICM and AGN emission. According to \cite{dh11}, such multiphase-star forming BCGs also tend to have unusually luminous vibrational and rotational $\rm{H_2}$ lines, which are by far more luminous  than in normal, star formation objects. The emission in such galaxies is not fit by $\rm{H_2}$ gas at a single excitation temperature, hinting to the fact that there must be a mix of excitation mechanisms at play, in accordance with our findings. The spectral decomposition method of \cite{davies17} allows us to infer the fractional contribution of each mechanism to the total luminosity of the emission lines.  Ionisation from star formation  has a fractional contribution to the luminosity of the emission lines of about 50-60\%, whereas  AGN emission accounts only for $\sim$ 10\%. LINER-like emission accounts for  30-40\% of the energetics needed to ionise the gas. The occurrence of  such "composite" emission is typical for cool-core BCG. For example, \cite{tr18}, \cite{iani19}, \cite{hamer} also studied (cool-core) BCGs based on  IFU data and infer composite emission for their systems in the BPT diagram and  concluded that the systems are ionised by a superposition of many physical processes. \\ The spectral decomposition method also allows us to recover a more accurate estimate for the SFRs, after excluding the fractional contribution of AGN and LINER emission to the luminosity of the $\rm{H\alpha}$ line, inferring a  lower limit of  $\rm{SFR\sim 97\:M_{\odot}/yr}$.  This value for the SFR, as well as the one we inferred without removing the contribution from both AGN and LINER to the luminosity of the $\rm{H\alpha}$ line, i.e. the   SFR value of $\sim 144\:\rm{M_{\odot}/yr}$ , are in accordance with literature $\rm{H\alpha}$ based SFRs for the M1931 BCG, within the errors (\citealt{eh11}, \citealt{dh15},  \citealt{fogarty15},  \citealt{fogarty17}). \\
The properties of the ionised gas that we infer by analysing the spectrum of each spectral pixel of the MUSE cube, such as the electron temperature, density, colour excess and  ionisation parameter are typical for star forming systems.  We have also computed the physical properties of integrated regions in the system, see  appendix \ref{appendix3} for more details. \\The temperature that we infer from the  $\rm{[N\textsc{ii}]}$ 5755/6584 emission lines is in the order of $\rm{T_{e}}=11000$ K, with the lowest values observed in the most central regions of the system. These values are in accordance with the ones inferred for the BCG of the cool-core cluster Abell 2597 by \cite{dh97}.\\  For example \cite{hamer} and \cite{iani19} determined the electron density for their sample of BCGs based on the $\rm{[S\textsc{ii}]}$ 6731/6716 emission line ratio and their inferred values in order of a few hundred $\rm{cm^{-3}}$ are in very good agreement to the value $\rm{n_e=361 cm^{-3}}$ that we determine for the M1931 BCG. \\  We observe the highest density and extinction ($E(B-V)\sim0.5$)  in the most central regions of the system, in accordance with the findings of \cite{fogarty19}, who also measures the highest dust continuum emission  in the core of the system.  Thus, the densest, and perhaps dustiest gas is found to be coincident (and to the South) with the nucleus, where the radio source is also detected.\\
The measured ionisation parameter is in the order of  log(U) = -2.9, a value typical for star forming galaxies according to \cite{kewley19}, and it shows some variation within the nebula, hinting to the fact that the source of excitation  is not confined within a specific region.\\
To conclude, it is possible that the molecular gas and ionized nebula at the centre of the M1931 cluster is a galaxy-scale “fountain” similar to what was observed by \cite{tr18} and \cite{olivares19} for their sample of BCGs: the AGN feedback has uplifted the low- entropy gas,  that ultimately condensed  and  began raining back toward the galaxy centre from which it came. This rain of  gas  back to the SMBH accretion reservoir is  promoting SF and further AGN feedback, in accordance with the theoretical predictions of the  chaotic cold accretion model (\citealt{gaspari18}). However, these "cooling-flow" induced SF-episodes, in combination with dry mergers  lead to the build-up of less than $20\%$ of the current day stellar mass of the BCG. 
\section{Conclusion}
\label{conclusion}
Based on VLT-MUSE optical integral field spectroscopy, we investigate the BCG of the massive cool-core CLASH cluster MACS 1931.8-2635 at a redshift of z=0.35, concerning its spatially resolved star formation activity, ionisation sources, chemical abundances, gas and stellar kinematics. The optical MUSE IFS data is supplemented by sub-mm ALMA observations, allowing us  to link the properties of the  warm ionised gas to those of the cold molecular gas  component. Employing different tools on the optical data, we reliable measure  i) the fluxes of strong emission lines, which  allow us to determine quantitatively the physical conditions of the warm ionised gas, ii) the SFH and iii) the kinematics of the stellar component.\\
The principal findings of our analysis can be summarised as follows:
\begin{enumerate}
\item The  ionised and  molecular gas components are co-spatial. The normalised $\rm{H\alpha}$ to CO flux ratios are close to unity along the nebula and the peak of the CO flux intensity is located at the same position as the peak of the ionised gas flux intensity.
\item The ionised and molecular gas components are co-moving. We measure a gradient from $\sim -300 $ to $\sim 300 \: \rm{km~s}^{-1}$ with respect to the BCG rest-frame and consistently low velocity dispersions in  the order of $\sim150 - 250\:\rm{km~s}^{-1}$ for the ionised gas.
The diffuse gas confined into the tail is likely falling inward, providing additional fuel for SF and AGN feedback, in accordance  with models of chaotic cold accretion.
\item The main source of ionisation in the M1931 BCG  is a mix between star formation and other energetic processes which  give rise to LINER-like emission, the main candidates being  heating of the gas by the surrounding medium and not  ionisation by pAGB stars.  AGN ionisation dominates only in the BCG core.
\item After applying the spectral decomposition method,  we recover an SFR of $\sim 97 M_{\odot}/yr$, with the most elevated levels in the BCG core. However, star formation accounts only to $\sim$50-60$\%$ of the energetics that is required to ionise the warm gas. AGN emission accounts for 
$\sim 10\%$ of the ionised gas emission while the rest of $\sim 30 - 40\%$ of the energetics needed to excite the gas  come from mechanisms which give rise to LINER-like emission. Star formation is, thus, the main contribution to the $\rm{H\alpha}$ flux. 
\item The median values recovered from the spatially resolved maps for the electron density, electron temperature, extinction and ionisation parameter are  typical for star forming systems and in good agreement with other studies of cool-core BCGs.
\item We measure a median value for the gas phase metallicity of $12+log(O/H)\sim 8 \pm0.35$, with the lowest values observed in the BCG core. \cite{eh11} measure an ICM metallicity of $12 + log(O/H) = 8.25$, a value consistent to the gas-phase metallicity we measure in the ISM of the BCG. This can hint to the fact that the warm gas we observe in the ISM of the galaxy has condensed from the ICM. 
\item About $ 80 \%$ of the systems stellar mass has formed  more than  6 Gyr ago (i.e. at z$\gtrsim1.5$), followed by subsequent but weaker mass build-up episodes. The SFH of the M1931 BCG is in accordance with theoretical models which suggest a two phase hierarchical formation for BCGs: in-situ SF at high z followed by subsequent mass growth through dry mergers and/or "cooling-flow" induced SF episodes. Both dry-mergers and/or in-situ SF generated by ICM cooling account  to less than 20\% of the stellar mass of the M1931 system within the last 1 Gyr (i.e. at z$\lesssim0.5$).
\item The stellar kinematics reveal a dispersion dominated system, which is typical for massive elliptical galaxies.
\item The gas motions are decoupled from the stellar kinematics and are  more closely related to the bulk motions of the ICM. The  velocity dispersion of the gaseous component is approximately  two times lower than the velocity dispersion of the stellar component,  in accordance with the predictions of the precipitation model.
\end{enumerate}
Most of the Python codes / Jupyter Notebooks that we have  developed for this analysis are publicly available in an online GitHub repository (Ciocan2021-MACS1931-BCG-codes) \footnote{This code repository is archived at DOI: $\rm{10.5281/zenodo.4458242}$, and also available at \url{https://github.com/ciocanbianca/Ciocan2021_MACS1931_BCG_codes}}.

\begin{acknowledgements}
We would like to express our deep gratitude to the members of the observational extragalactic astrophysics group from the department of astrophysics, University of Vienna. Special thanks to Christian Maier and Asmus B{\"o}hm for all the valuable discussions and pieces of advice! We would also like to express our gratitude to Maria Luísa Gomes Buzzo for all her help related to analysis of the MUSE data.
We would especially like to thank the anonymous referee for providing constructive comments and help in improving the manuscript.
This work was supported through FCT grants UID/FIS/04434/2019, UIDB/04434/2020, UIDP/04434/2020 and the project "Identifying the Earliest Supermassive Black Holes with ALMA (IdEaS with ALMA)" (PTDC/FIS-AST/29245/2017). This research made use of the following \texttt{PYTHON} packages: \texttt{Astropy} \citep{astropy}, \texttt{numpy} \citep{numpy}, \texttt{matplotlib} \citep{plot}, \texttt{MPDAF} \citep{mpdaf}, \texttt{CMasher} \citep{cmasher}.
\end{acknowledgements}


\begin{appendix}
\label{appendix1}

\section{Emission line flux maps}
\begin{figure*}
  \centering
\begin{subfigure}
    \centering
    \includegraphics[width=0.3\textwidth,angle=0,clip=true]{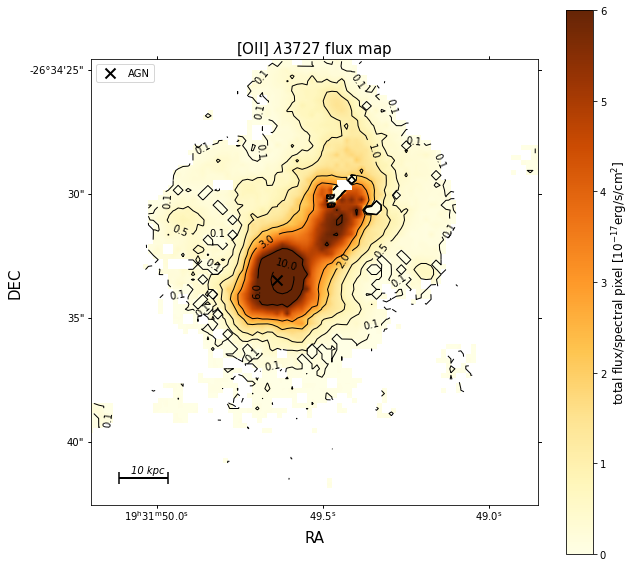}
    \centering
\end{subfigure} 
\begin{subfigure}
    \centering
    \includegraphics[width=0.3\textwidth,angle=0,clip=true]{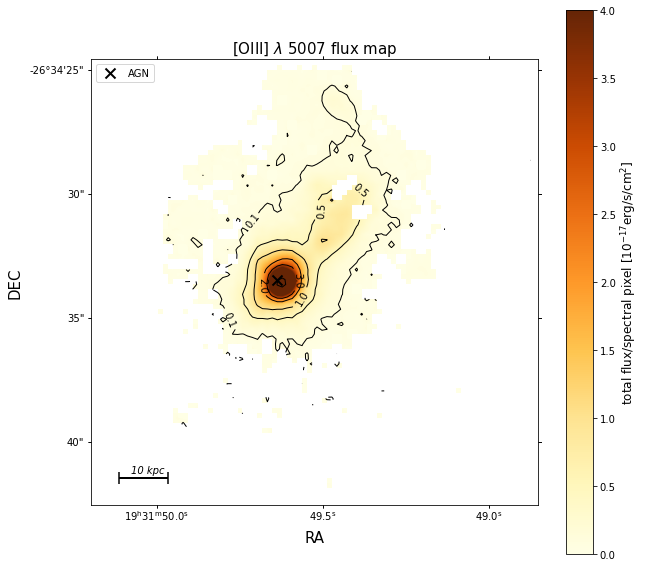}
    \centering
\end{subfigure} 
\begin{subfigure}
    \centering
    \includegraphics[width=0.3\textwidth,angle=0,clip=true]{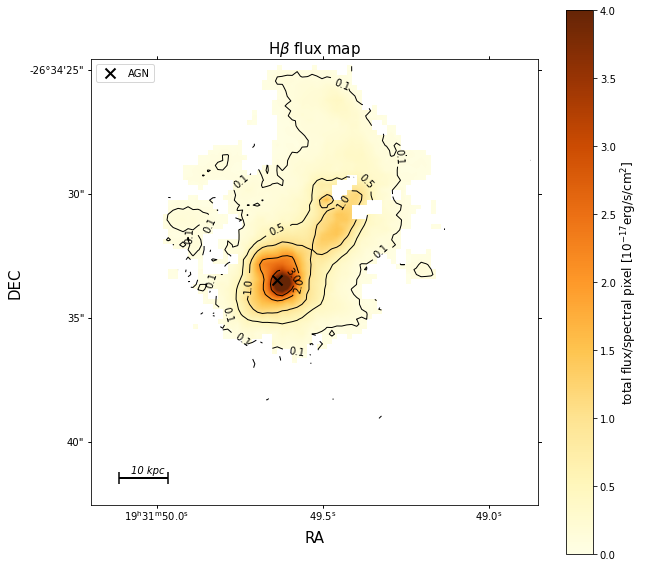}
    \centering
\end{subfigure} 
\begin{subfigure}
    \centering
    \includegraphics[width=0.3\textwidth,angle=0,clip=true]{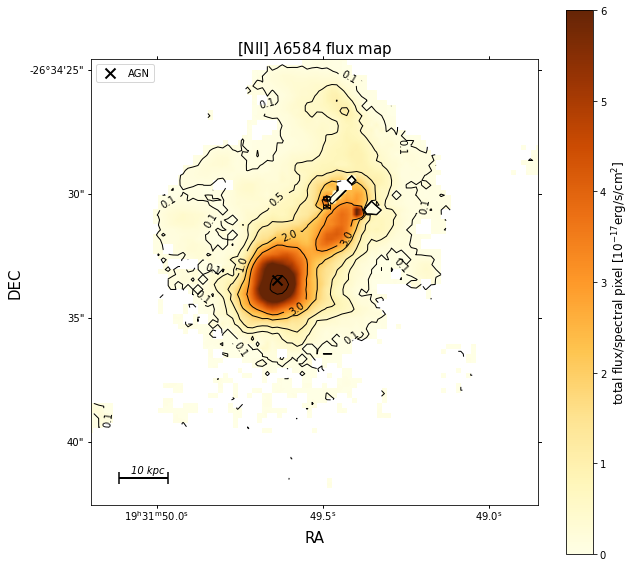}
    \centering
\end{subfigure} 
\begin{subfigure}
    \centering
    \includegraphics[width=0.3\textwidth,angle=0,clip=true]{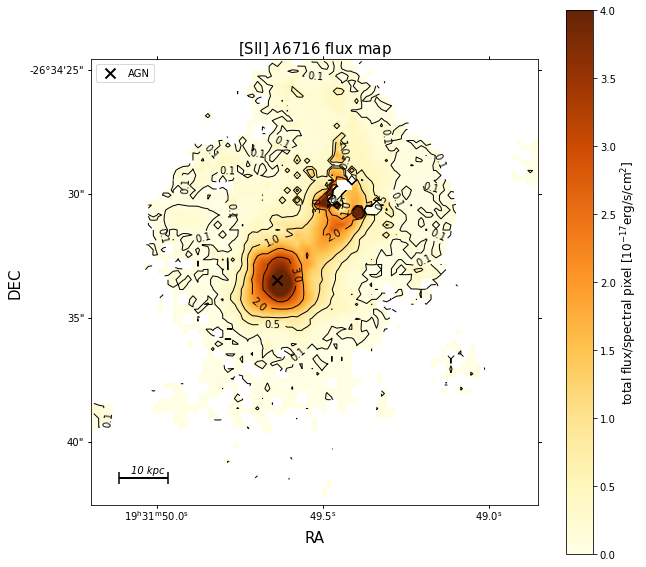}
    \centering
\end{subfigure} 
\begin{subfigure}
    \centering
    \includegraphics[width=0.3\textwidth,angle=0,clip=true]{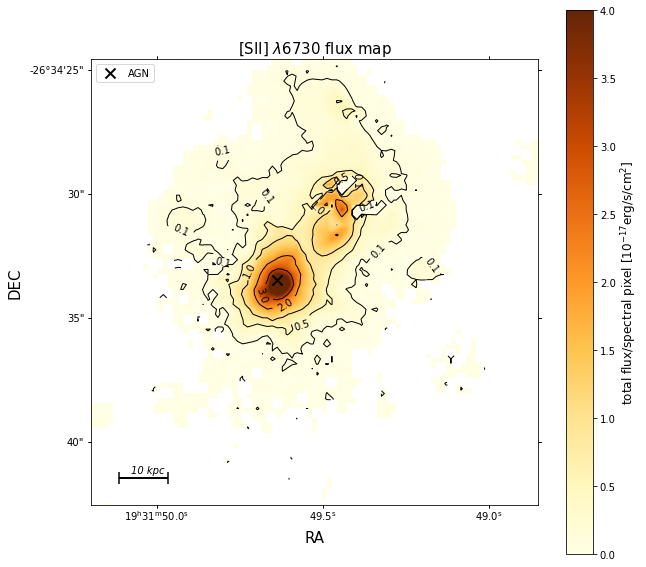}
    \centering
\end{subfigure} 
\begin{subfigure}
    \centering
    \includegraphics[width=0.3\textwidth,angle=0,clip=true]{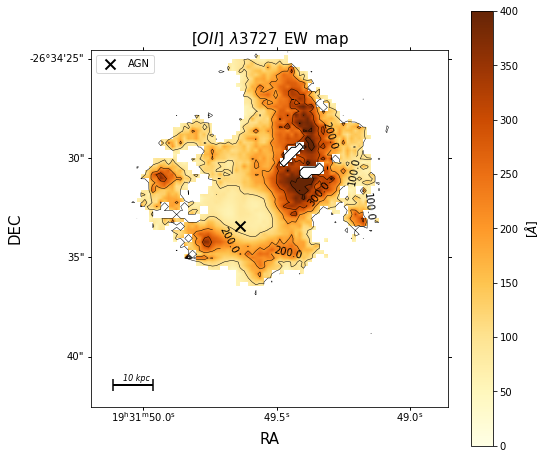}
    \centering
\end{subfigure} 
\begin{subfigure}
    \centering
    \includegraphics[width=0.3\textwidth,angle=0,clip=true]{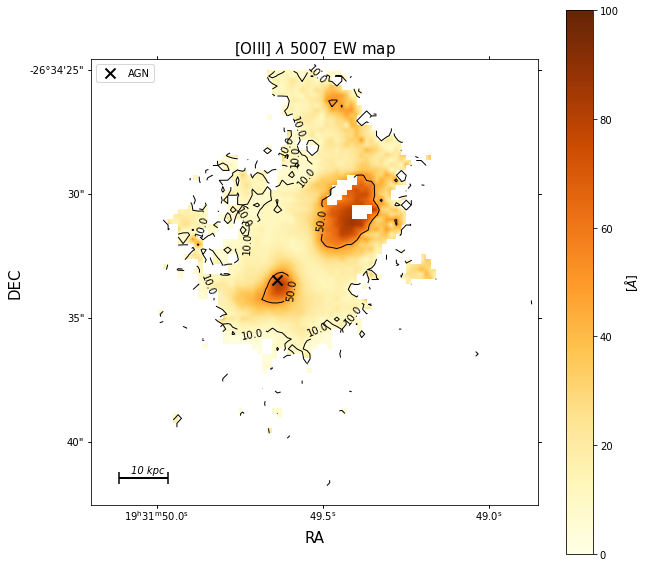}
    \centering
\end{subfigure} 
\begin{subfigure}
    \centering
    \includegraphics[width=0.3\textwidth,angle=0,clip=true]{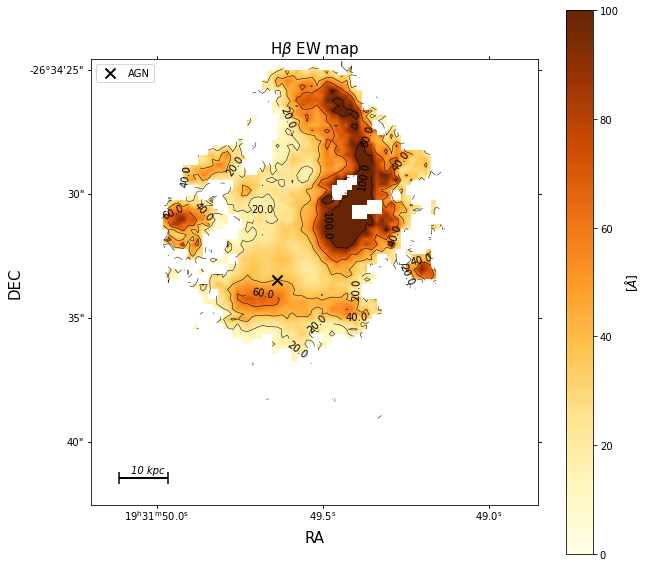}
    \centering
\end{subfigure} 
\begin{subfigure}
    \centering
    \includegraphics[width=0.3\textwidth,angle=0,clip=true]{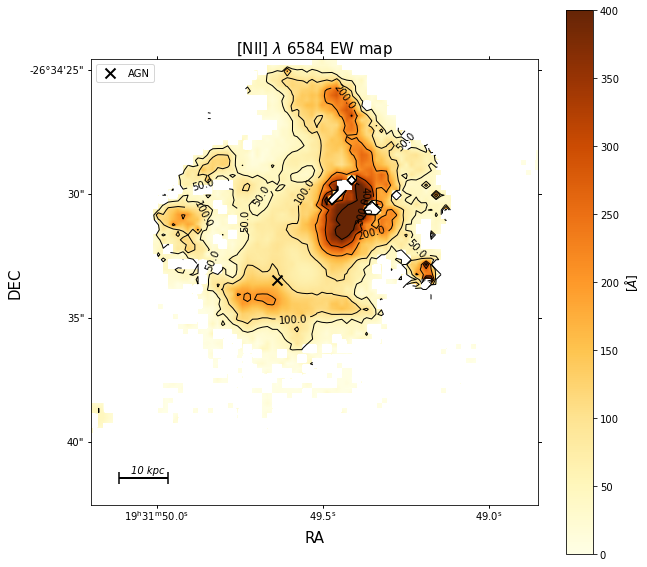}
    \centering
\end{subfigure} 
\begin{subfigure}
    \centering
    \includegraphics[width=0.3\textwidth,angle=0,clip=true]{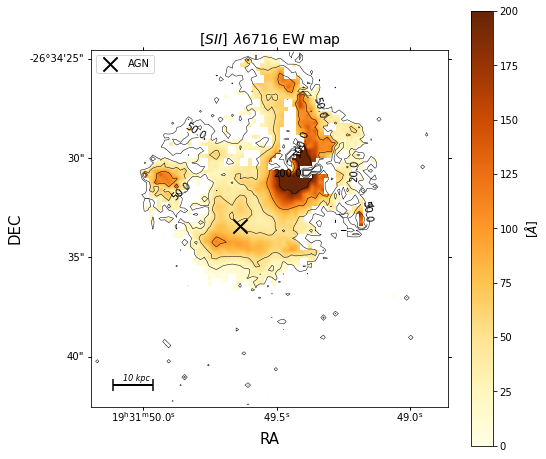}
    \centering
\end{subfigure} 
\begin{subfigure}
    \centering
    \includegraphics[width=0.3\textwidth,angle=0,clip=true]{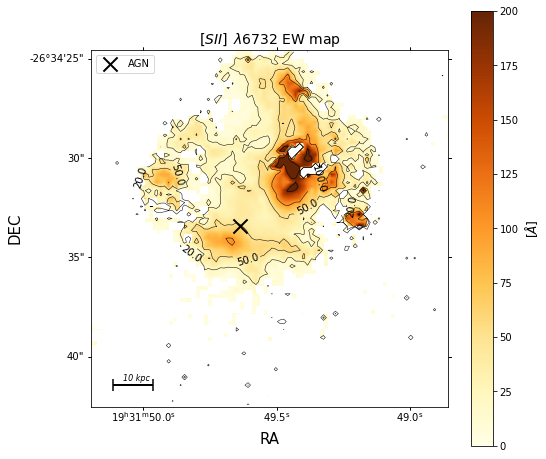}
    \centering
\end{subfigure} 
\centering
\caption{ The upper two rows display the flux maps for the  $\rm{[O\textsc{ii}]}\:\lambda 3727$,  $\rm{[O\textsc{iii}]}\:\lambda 5007$, $\rm{H\beta}$, $\rm{[N\textsc{ii}]}\:\lambda 6584 $, $\rm{[S\textsc{ii}]}\:\lambda 6718, 6732 $  gas in the M1931 BCG, while the lower two rows show the equivalent widths in units of $[\AA]$ for the same set of emission lines. The  white background  in all plots correspond to  the spaxels with a SNR<10. The cross shows the location of the AGN. The contours show different levels of flux intensity and EW.}
\label{fluxtot}
\end{figure*}

Figure \ref{fluxtot}  displays the flux maps in the upper two rows for the  $\rm{[O\textsc{ii}]}\:\lambda 3727$,  $\rm{[O\textsc{iii}]}\:\lambda 5007$, $\rm{H\beta}$, $\rm{[N\textsc{ii}]}\:\lambda 6584 $, $\rm{[S\textsc{ii}]}\:\lambda 6718, 6732 $ emission lines in units of $10^{-17} \rm{ergs \cdot s^{-1} \cdot cm^{-2}}$ . The lower two rows show the EW distribution for the same emission lines, in units of $[\AA]$. We observe similar distribution of fluxes and EWs for all the strong emission lines in the optical spectrum of the M1931 BCG.\\

\section{Emission line velocity maps}
\begin{figure*}
  \centering
\begin{subfigure}
    \centering
    \includegraphics[width=0.3\textwidth,angle=0,clip=true]{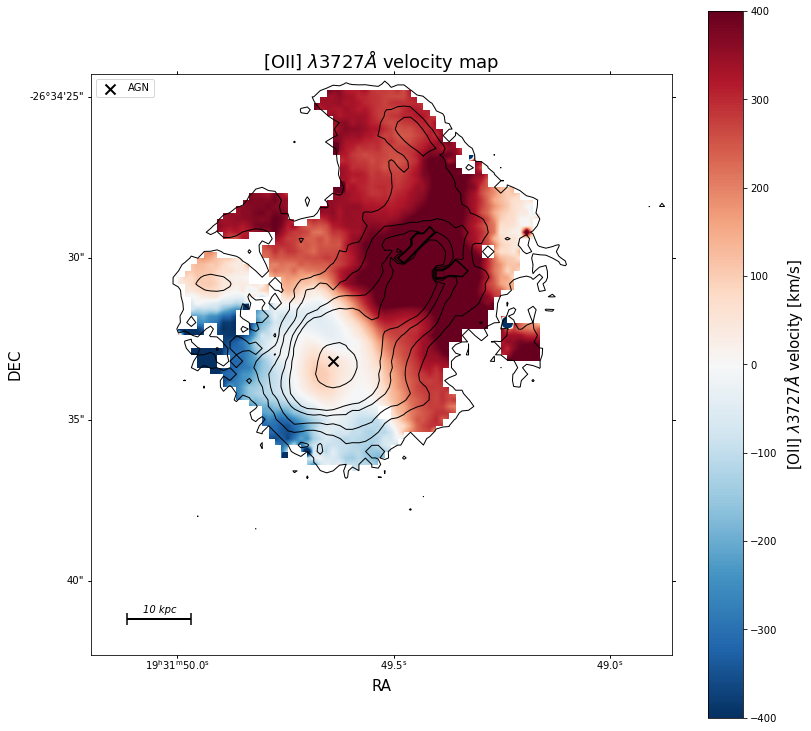}
    \centering
\end{subfigure} 
\begin{subfigure}
    \centering
    \includegraphics[width=0.3\textwidth,angle=0,clip=true]{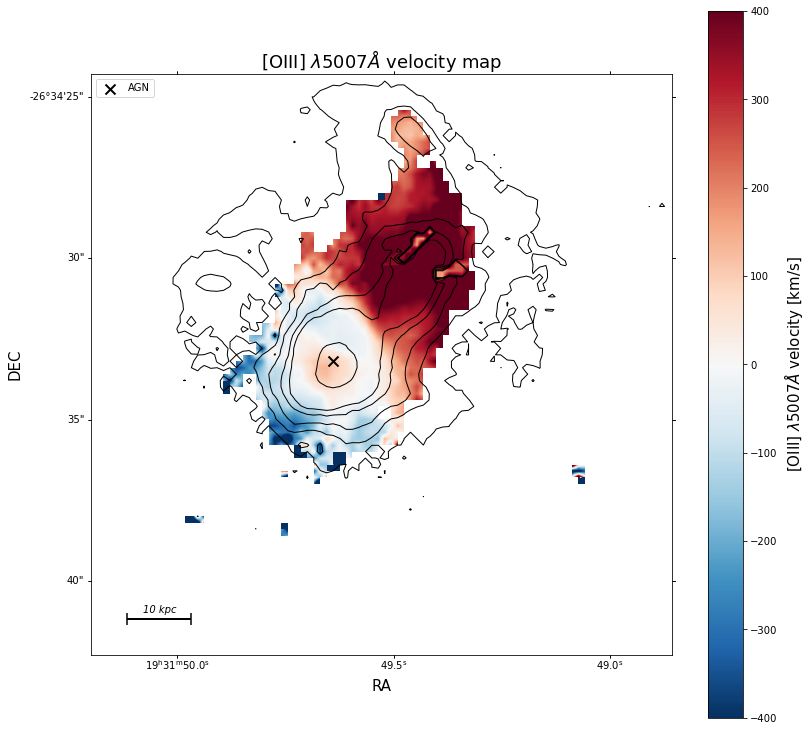}
    \centering
\end{subfigure} 
\begin{subfigure}
    \centering
    \includegraphics[width=0.3\textwidth,angle=0,clip=true]{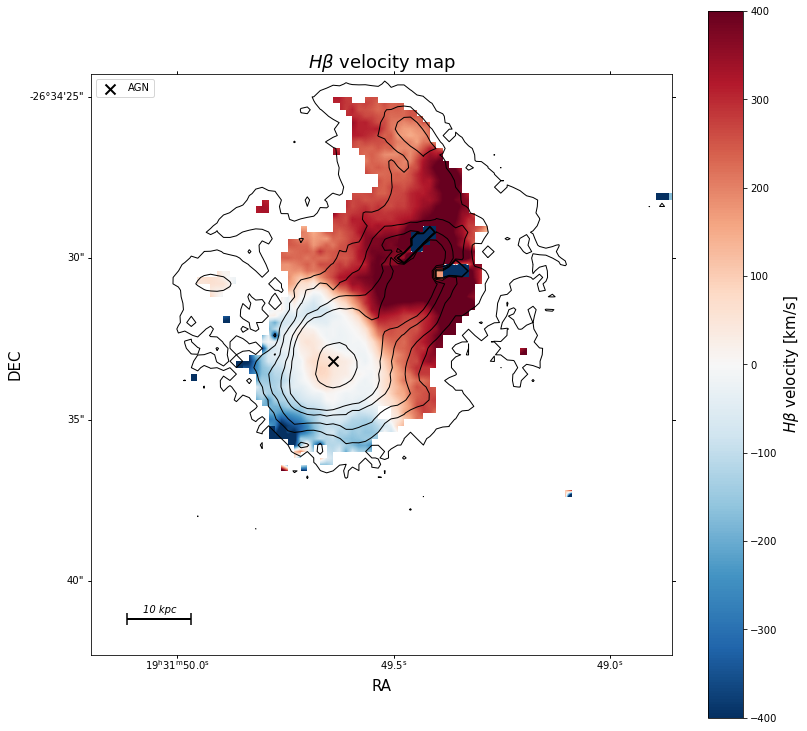}
    \centering
\end{subfigure} 
\begin{subfigure}
    \centering
    \includegraphics[width=0.3\textwidth,angle=0,clip=true]{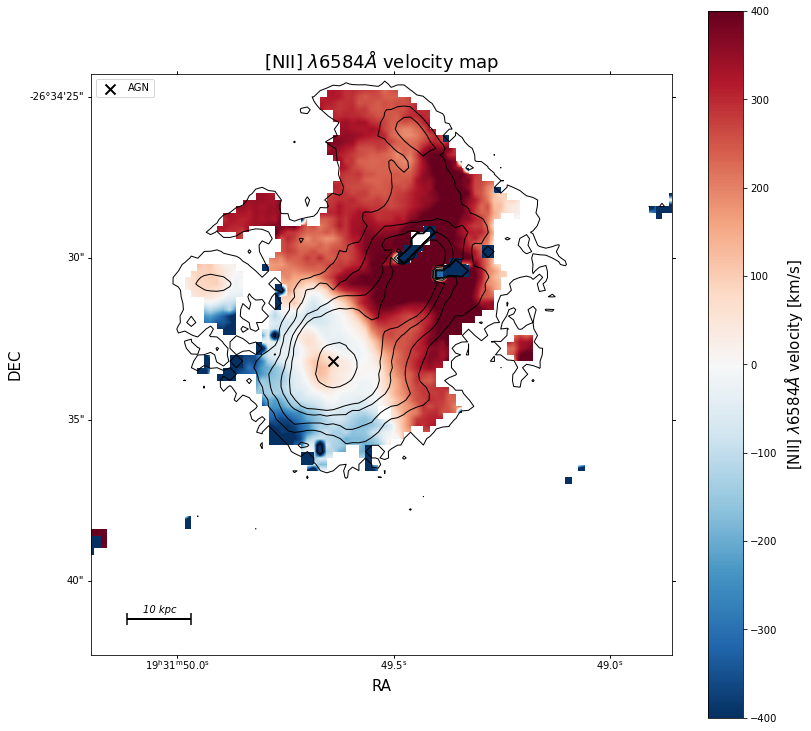}
    \centering
\end{subfigure} 
\begin{subfigure}
    \centering
    \includegraphics[width=0.3\textwidth,angle=0,clip=true]{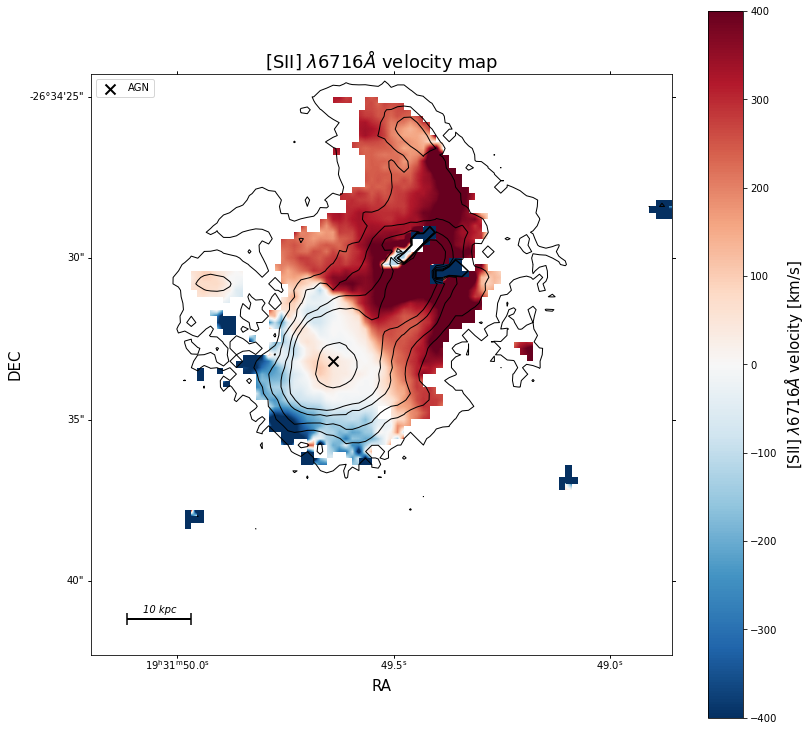}
    \centering
\end{subfigure} 
\begin{subfigure}
    \centering
    \includegraphics[width=0.3\textwidth,angle=0,clip=true]{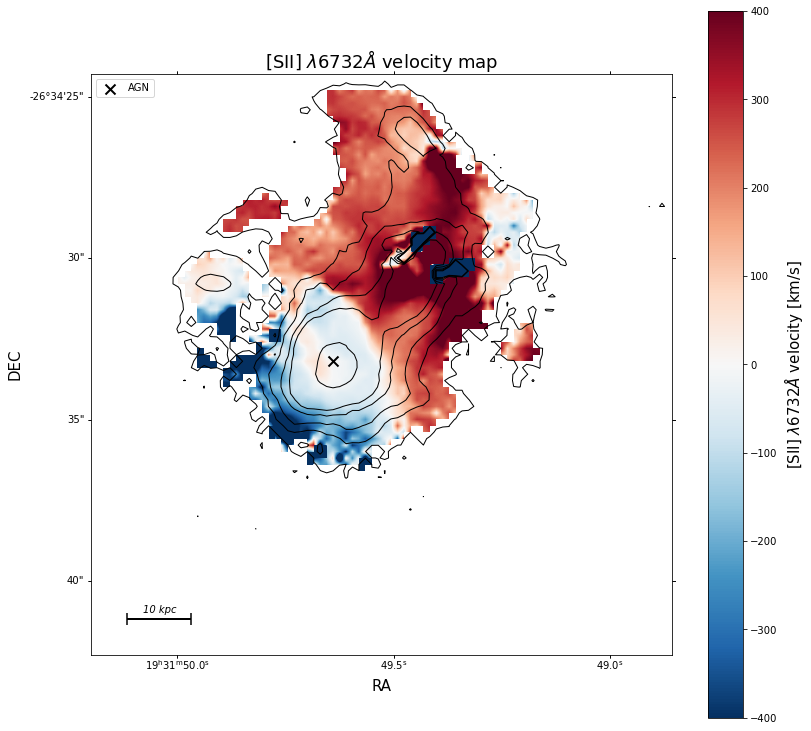}
    \centering
\end{subfigure} 
\begin{subfigure}
    \centering
    \includegraphics[width=0.3\textwidth,angle=0,clip=true]{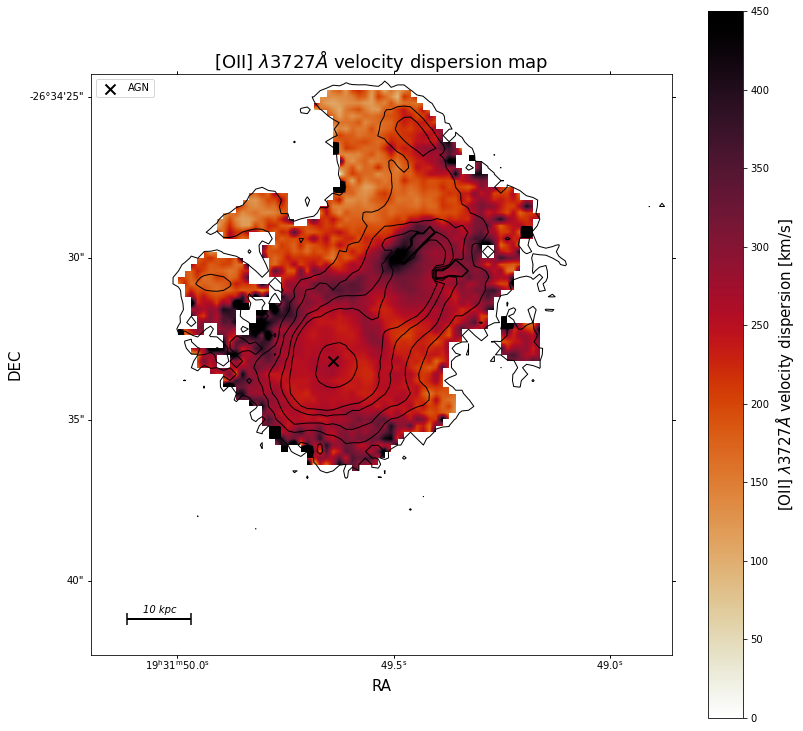}
    \centering
\end{subfigure} 
\begin{subfigure}
    \centering
    \includegraphics[width=0.3\textwidth,angle=0,clip=true]{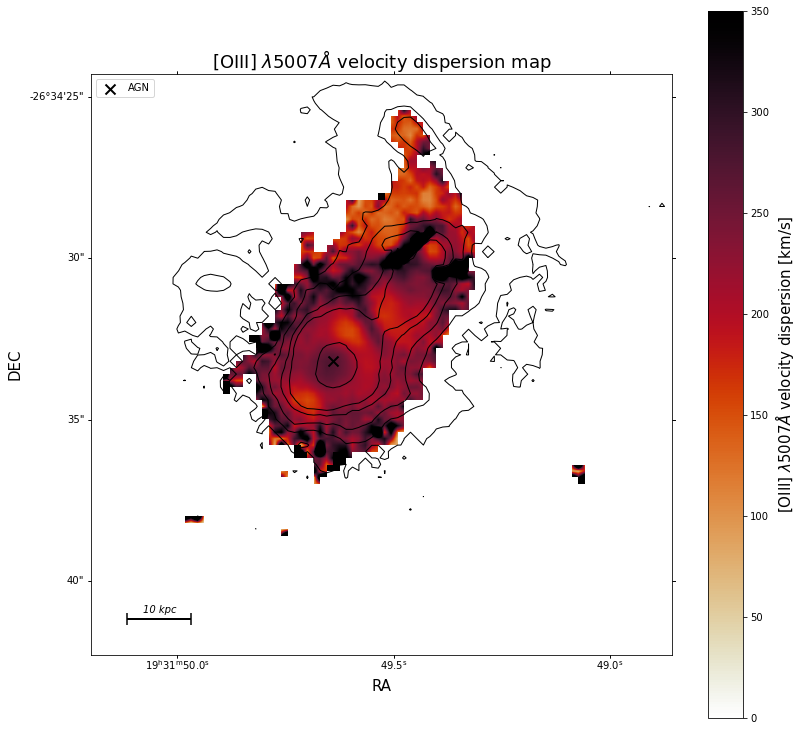}
    \centering
\end{subfigure} 
\begin{subfigure}
    \centering
    \includegraphics[width=0.3\textwidth,angle=0,clip=true]{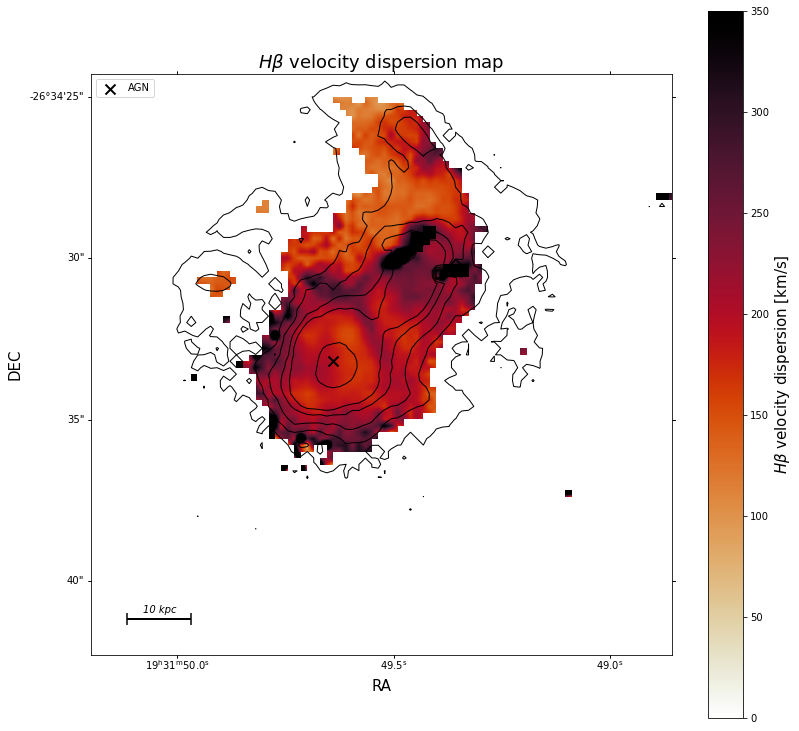}
    \centering
\end{subfigure} 
\begin{subfigure}
    \centering
    \includegraphics[width=0.3\textwidth,angle=0,clip=true]{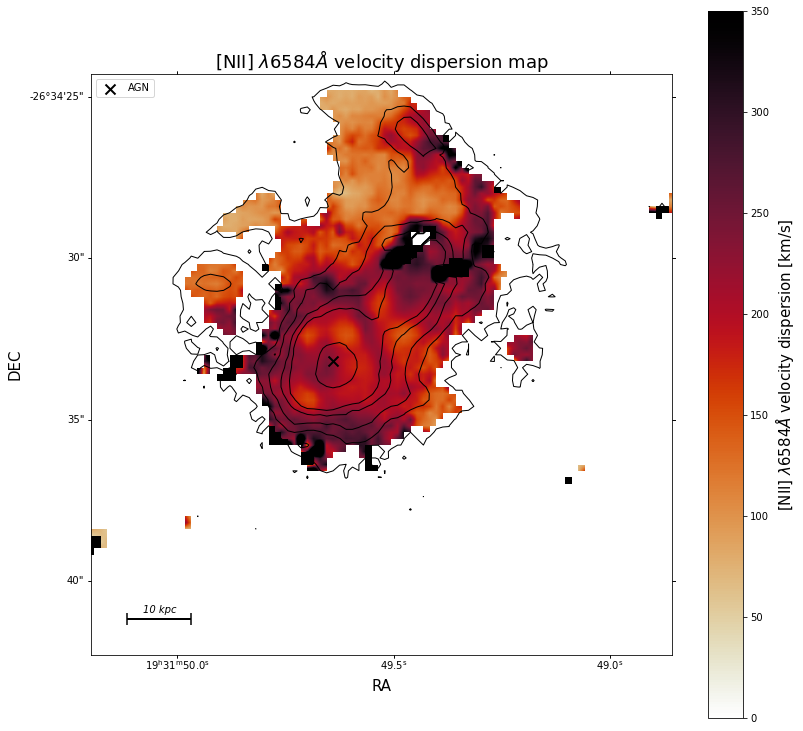}
    \centering
\end{subfigure} 
\begin{subfigure}
    \centering
    \includegraphics[width=0.3\textwidth,angle=0,clip=true]{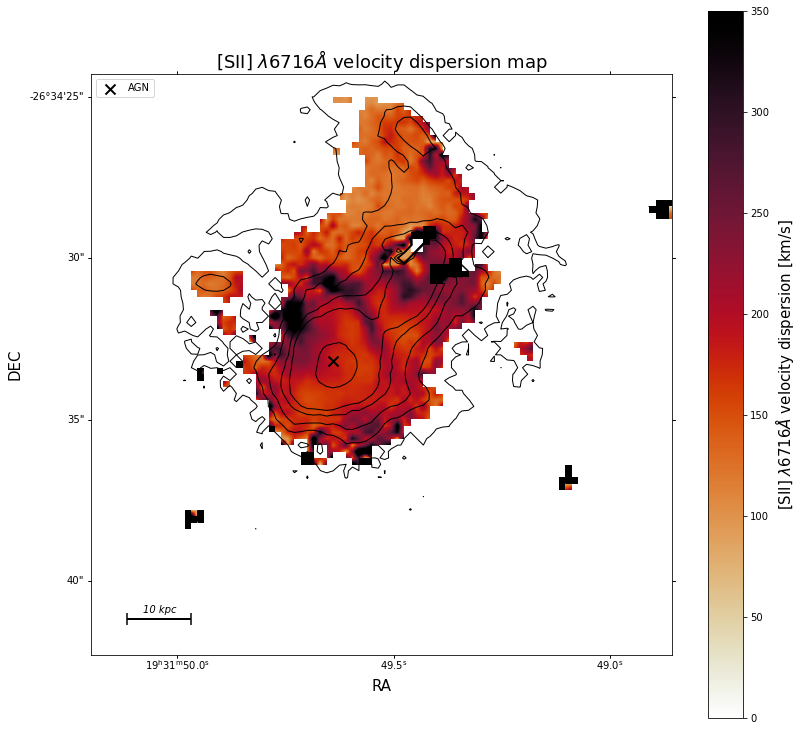}
    \centering
\end{subfigure} 
\begin{subfigure}
    \centering
    \includegraphics[width=0.3\textwidth,angle=0,clip=true]{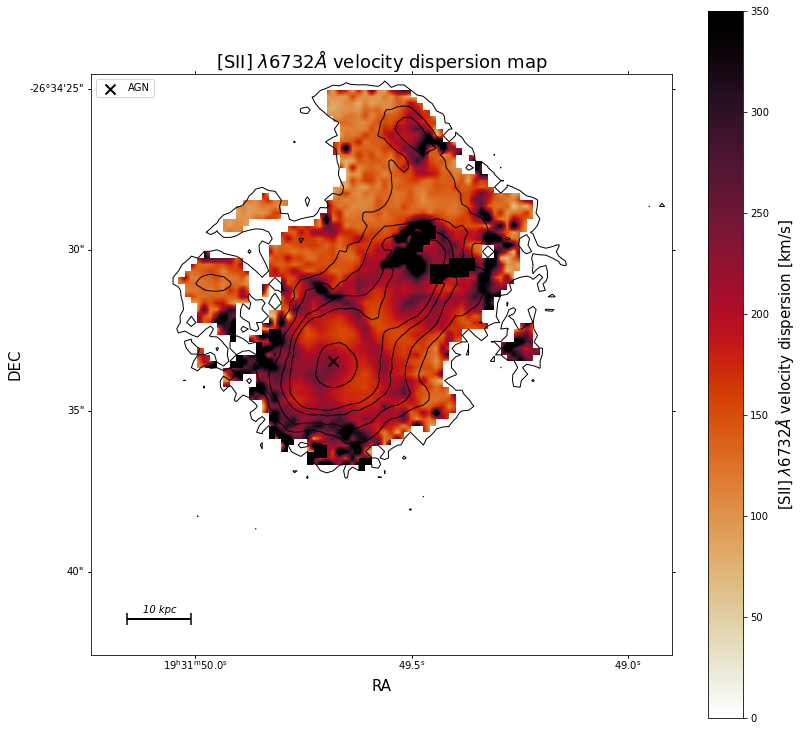}
    \centering
\end{subfigure} 
\centering
\caption{ Kinematics of the warm ionised gas. The panels from the upper two rows display the spatially resolved  radial velocity maps for the$\rm{[O\textsc{ii}]}\:\lambda 3727$,  $\rm{[O\textsc{iii}]}\:\lambda 5007$, $\rm{H\beta}$, $\rm{[N\textsc{ii}]}\:\lambda 6584 $, $\rm{[S\textsc{ii}]}\:\lambda 6718, 6732 $ gas in the BCG of the M1931 galaxy cluster. The colour-bar displays the radial velocity  of the gas with respect to the BCG rest frame, measured in [km/s].  The panels from the lower 2 rows show the spatially resolved  velocity dispersion  maps for the same emission lines, measured in [km/s]. The  white background in all plots correspond to  the spaxels with a SNR<10. The cross shows the location of the AGN. The contours show the $\rm{H\alpha}$ flux intensity.}
\label{veltot}
\end{figure*}

Figure \ref{veltot} shows the radial velocity and velocity dispersion of the $\rm{[O\textsc{ii}]}\:\lambda 3727$,  $\rm{[O\textsc{iii}]}\:\lambda 5007$, $\rm{H\beta}$, $\rm{[N\textsc{ii}]}\:\lambda 6584 $, $\rm{[S\textsc{ii}]}\:\lambda 6718, 6732 $  gas in the BCG of the M1931. We observe very similar kinematics for all the strong emission lines in the galaxy spectrum.\\

\section{Physical properties of integrated regions}
\label{appendix3}
\begin{figure*}
    \centering
    \includegraphics[width=0.5\textwidth,angle=0,clip=true]{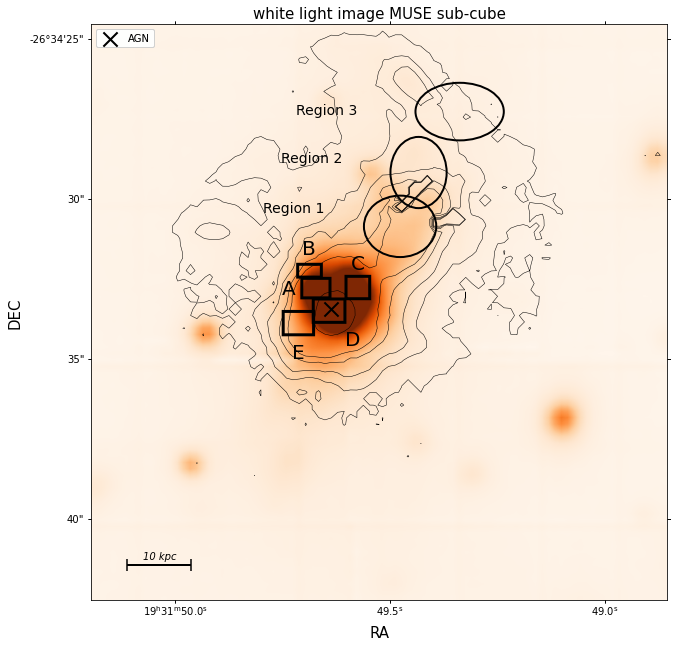}
    \centering   
    \caption{White light image of the MUSE sub-cube centred on the BCG of the M1931 galaxy cluster. The contours show the $\rm{H\alpha}$ flux intensity. The squares encompass the different  CO(1-0) source regions,  while the ellipses show the cold dust regions, as defined by \cite{fogarty19}. We have extracted these 8 regions as single spectra and computed the most important physical properties of the ionised gas, which are listed in Table \ref{tab}. }  
\label{whitelight}
\end{figure*}

Figure \ref{whitelight} shows the white light image of the MUSE sub-cube centred on the BCG of the M1931 galaxy cluster. The contours show the $\rm{H\alpha}$ flux intensity. The squares encompass the different  CO(1-0) source regions,  while the ellipses show the cold dust regions, as defined by \cite{fogarty19} - see their Figs. 4 and 6. We have extracted these 8 regions as single  integrated spectra and computed the physical properties of the ionised gas, which are listed in Table \ref{tab}. The last row of the table contains the physical properties of the ionised gas, as derived from the integrated spectrum of the 90x90 spaxels MUSE sub-cube centred on the BCG. This table lists the flux of the $\rm{H\alpha}$ line in units of [$10^{-15} \rm{ergs \cdot s^{-1} \cdot cm^{-2}}$], the systemic velocity and velocity dispersion of the $\rm{H\alpha}$ gas in [km/s], the electron density  $\rm{n_e}$ in $\rm{[cm^{-3}]}$, the electron temperature $\rm{T_e}$ in [K],  the colour excess E(B-V), the ionisation parameter log(U), the SFR in units of  $\rm{[M_{\odot}/yr]}$ without correcting for the contribution of LINER-like and AGN emission to the luminosity of the $\rm{H\alpha}$ line, as the \cite{davies17} decomposition method works only  for a spaxel by spaxel analysis and not for integrated regions,  and the oxygen abundance  12+log(O/H). The methods and calibrations used to determine the properties of the ionised gas are the same ones as used for the spaxel by spaxel analysis. It is 
worth mentioning that we could not recover the electron density and temperature using the \texttt{PyNeb} tool for regions B, 2 and 3 and for the whole sub-cube, due to the weakness of the temperature-sensitive emission lines, and also because in these regions, we probably observe the luminosity-weighted average of emission lines arising from different volumes with different physical conditions.

\begin{table*}[b]
\centering
\caption{This table contains the most important physical properties of the ionised gas, as computed for specific regions of the system.  Regions A-E correspond to the CO(1-0) source regions,  while the regions 1-3 correspond to the cold dust regions, as defined by \cite{fogarty19}.  We have extracted from the MUSE cube integrated spectra corresponding to these 8 regions, and we have computed the systemic velocity and velocity dispersion of the $\rm{H\alpha}$ gas, the electron density  $\rm{n_e}$, the electron temperature $\rm{T_e}$,  the colour excess E(B-V), the ionisation parameter log(U), the SFR $\rm{[M_{\odot}/yr]}$  and the oxygen abundance 12+log(O/H). The las row of the table refers to the integrated spectrum of the whole (90x90 spaxels) sub-cube, displayed in Fig. \ref{whitelight}}
\begin{adjustbox}{width=1\textwidth}
\centering
\small
\begin{tabular}{lcccccccccl}
\hline
Region  &   $\rm{H\alpha}$ flux $[10^{-15} \rm{ergs\cdot s^{-1} \cdot cm^{-2}}]$  &   $\rm{vel_{H\alpha}}$ [km/s]     &       $\rm{\sigma_{H\alpha}}$  [km/s]       &     $\rm{n_e\: [cm^{-3}]}$             &        $\rm{T_e}$ [K] & E(B-V) &log(U) & SFR $\rm{[M_{\odot}/yr]}$  & $\rm{12+log(O/H)}$      \\                                                                                              

\hline
region A             &     $ 1.07\pm0.0006$  &     $-109.3\pm21.9$  &        $173.1\pm3.2 $       &   $ 29.16 $   &        $ 9154.4$  &          $0.261\pm0.00009$          &        $-2.946\pm0.25$ &$ 6.4\pm0.0015$ &$7.87\pm0.26$ \\
region B            &     $0.26\pm0.0006$ &     $-160.6\pm36.9 $    &  $ 183.4\pm3.0$          &     $-$  &         $-$ &      $0.147\pm0.00032$       &        $-2.91\pm0.27$  &$1.23\pm0.0011$ &$7.84\pm0.2$6 \\
region C            &      $ 0.74\pm0.00004 $  & $    -142.3\pm59.7 $    &       $158.1\pm3.5$          &     $354.8$  &       $11704.02$ &      $0.248\pm0.00006$         &       $-2.87\pm0.2$  & $4.28\pm0.0008$&$7.79\pm0.23 $\\
region D             &  $2.91\pm0.00007$ &  $-89.26\pm24.66$  &     $ 187.6\pm2.9$   &         $785.96$  &          $8350.6$ &       $0.50\pm0.000035$          &        $ -2.94\pm0.19$ &$29.72\pm0.0028$&$7.835 \pm0.139$ \\
region E            &      $0.94\pm0.00008$  &   $-145.47\pm88.0$   & $207.7\pm2.6 $  &    $561.18$   &       $9114.7$ &        $ 0.46\pm0.00014 1$         &       $-2.96\pm0.23$  &  $8.72\pm0.0032$ & $7.88\pm0.203 $\\
region 1            &     $3.02\pm0.00010$  &    $370.43\pm65.7$   &  $187.4\pm2.9 $     &       30.5$$  &          $10125.2$ &          $ 0.152\pm0.00018$          &        $-3.08\pm0.3$  &$14.08\pm0.006$&$7.89\pm0.17$\\
region 2             &     $ 1.32\pm0.00009$  &      $333\pm65.7$   &   $250.9\pm2.2 $      &         $-$  &          $-$ &          $0.087\pm0.00035$        &    $-3.07\pm0.31 $  & $5.34\pm0.004$&$8.37\pm0.0$ \\
region 3             &   $0.24\pm0.00010$  &      $270.2\pm34.42$    &    $251.4\pm2.2 $     &       $-$   &        $-$  &         $0.085 \pm0.0009$         &    $-2.90\pm0.2$  & $0.67\pm0.0015$&$7.71\pm0.13$ \\
whole sub-cube             &     $ 26.32\pm0.00046$  &     $-78.7\pm4.6$  &        $261.5\pm2.1 $       &   $- $   &        $-$  &          $0.29\pm0.0009$          &        $-3.24\pm0.12$ &$167.3\pm0.0032$ &$8.4\pm0.13$ \\
\end{tabular}
\end{adjustbox}
\label{tab}

\end{table*}

\end{appendix}


\clearpage

\begin{thebibliography}{}

\bibitem[Alarie et al.(2019)]{3mdb} Alarie, A.;  Morisset, C.,  2019, RMxAA, 55, 377A


\bibitem[Allen et al.(2004)]{allen04}Allen S. W., Schmidt R. W., et al.,2004, MNRAS, 353, 457

\bibitem[Allen et al.(2008)]{allen08}Allen, M., G.,  Groves, B., A.,  Dopita, M., 2008, ApJS, 178, 20A

\bibitem[Akima (1970)]{Akima70}Akima, H. 1970, J. ACM (JACM), 17, 589

\bibitem[Bacon et al.(2014)]{bacon14}Bacon, R., Vernet, J., et al., 2014, MSNGR, 157,13

\bibitem[Bacon et al.(2016)]{mpdaf}Bacon, R., Piqueras, L., et al., 2016,ascl:1611.003

\bibitem[Baldwin et al.(1981)]{bald81} Baldwin, J.~A., Phillips, M.~M., \& Terlevich, R.,  PASP, 93, 5

\bibitem[Begelman et al.(1991)]{b91}Begelman, M.C.,  Fabian, A. C., 1990, MNRAS, 244P, 26B 

\bibitem[Bellstedt et al.(2016)]{bell16}Bellstedt, S., Lidman, C., Muzzin, A., et al.,  2016, MNRAS, 460, 2862B
\bibitem[Binette et al.(1994)]{binette94}Binette et al. 1994, A\&A 292, 13,

\bibitem[Bittner et al.(2019)]{gist}Bittner, A.; Falcón-Barroso, J. et al., 2019, A\&A,628A,117B


\bibitem[Breda \& Papaderos(2018)]{breda}Breda, I., Papaderos, P.,  2018; A\&A 614, A48

\bibitem[Brocklehurst et al.(1971)]{brock71} Brocklehurst, M.,1971, MNRAS.153, 471

\bibitem[Bruzual \& Charlot(2003)]{bc03}Bruzual, G.;  Charlot, S., 2003, MNRAS, 344, 1000B

\bibitem[Burke et al.(2015)]{burke13}Burke, C., Collins, C.,  2013, MNRAS, 434, 2856

\bibitem[Burke et al.(2015)]{burke15}Burke, C.,  Hilton, M., Collins, C.,  2015, MNRAS, 449, 2353B

\bibitem[Byler et al.(2019)]{byler19}Byler, N. Dalcanton, J. et al., 2019, AJ, 158, 2B

\bibitem[Calzetti et al.(2001)]{calz} Calzetti, D.,  2001, PASP, 113, 1449C
\bibitem[Cappellari \&  Copin (2003)]{vor} Cappellari, M., Copin, Y.  2003, MNRAS, 342, 345C

\bibitem[Cappellari \& Emsellem(2004)]{ppxf}Cappellari, M.,  Emsellem, E., 2004,PASP, 116, 138C 

\bibitem[Cardelli et al.(1989)]{c89} Cardelli, J., A., Clayton, G., C.,  Mathis, J., S., 1989, ApJ, 345, 245C

\bibitem[Casey (2012)]{casey12}Casey, C. M. 2012, MNRAS, 425, 3094

\bibitem[Cerulo et al.(2019)]{cerulo19}Cerulo, P.,  Orellana, G. A.,  Covone, G.,  2019, MNRAS, 487, 3759

\bibitem[Chabrier(2003)]{chabrier03} Chabrier G., 2003, PASP, 115, 763


\bibitem[Cid Fernandes et al.(2005)]{starlight} Cid Fernandes, R.;  Mateus, A.  Sodré, L., et al.,  2005, MNRAS, 358, 363C 

\bibitem[Collins et al.(2009)]{collins09}Collins C. A., Stott J. P., Hilton M., Kay S. T., et al.,  2009, Nature, 458, 603

\bibitem[Cresci et al.(2017)]{cresci17}Cresci, G., Vanzi, L.,  Telles, E., et al.  2017, A\&A, 604A, 101C 

\bibitem[Davies et al.(2017)]{davies17} Davies, R. L.;  Groves, B.;  Kewley, L. et al., 2017, MNRAS, 470, 4974D


\bibitem[Delgado-Inglada et al.(2014)]{di14}Delgado-Inglada, G.,  Morisset, C.,  Stasińska, G.,  2014, MNRAS,440, 536D

\bibitem[De Lucia et al.(2007)]{d07}De Lucia G., Blaizot J., 2007, MNRAS, 375, 2

\bibitem[D{\'i}az-Garc{\'i}a \& Knapen(2020) ]{dk20}D{\'i}az-Garc{\'i}a, S.;  Knapen, J. H., 2020, A\&A, 635A, 197D

\bibitem[Donahue et al.(1991)]{dh91} Donahue, M.,  Voit, G. M., 1991, ApJ, 381, 361D 

\bibitem[Voit and Donahue(1997)]{dh97}Voit G. M., Donahue M., 1997, ApJ, 486, 242

\bibitem[Donahue et al.(2000)]{dh00}Donahue, M.,  Mack, J., Voit, G. M., et al., 2000, ApJ, 545, 670D 

\bibitem[Donahue et al.(2011)]{dh11}Donahue, M., de Messières, G., E.,  O'Connell, R., W.,  Voit, G. M., et al.,  2011, ApJ, 732, 40D

\bibitem[Donahue et al.(2015)]{dh15}  Donahue, M.,  Connor, T., et al., 2015, ApJ, 805, 177

\bibitem[Dopita et al.(2005)]{dopita05}Dopita, M. A., Groves, B. A., Fischera, J., et al. 2005, ApJ, 619, 755

\bibitem[Dopita et al.(2010)]{dh10}Donahue, M.,  Bruch, S., et al. 2010, ApJ,715,881D

\bibitem[Edwards et al.(2020)]{edwards20}Edwards, L. O. V.,  Salinas, M., Stanley, S., Holguin W., et al.  2020, MNRAS, 491, 2617E
\bibitem[Ehlert et al.(2011)]{eh11} Ehlert, S.;  Allen, S. W., et al.,  2011, MNRAS, 411, 1641E 

\bibitem[Fabian \& Nulsen (1977)]{f77} Fabian A. C., Nulsen P. E. J., 1977, MNRAS, 180, 479


\bibitem[Ferland(2013)]{cloudy} Ferland  G. J. et al., 2013, Rev. Mex. Astron. Astrofis., 49, 137

\bibitem[Fogarty et al.(2015)]{fogarty15}Fogarty, K., Postman, M., et al., 2015, ApJ, 813, 117F 

\bibitem[Fogarty et al.(2017)]{fogarty17}Fogarty, K., Postman, M., et al., 2017, ApJ, 846, 103

\bibitem[Fogarty et al.(2019)]{fogarty19}Fogarty, K., Postman, M., et al. 2019, ApJ, 879, 103F

\bibitem[Gaspari et al.(2013)]{gaspari13}Gaspari M., Ruszkowski M., Oh S. P., 2013, MNRAS, 432, 3401

\bibitem[Gaspari et al.(2018)]{gaspari18}Gaspari, M., McDonald, M., Hamer, S. L., et al. 2018, ApJ, 854, 167

\bibitem[Gomes et al.(2016)]{gomes16}Gomes, J. M.;, Papaderos, P.,   Kehrig, C.,  Vílchez, J. M.,  Lehnert, M. D., 2016, A\&A, 588, A68 

\bibitem[Gomes \& Papaderos(2017)]{fado}Gomes, J. M.;  Papaderos, P., 2017, A\&A,603A, 63G 

\bibitem[Girardi et al.(1996)]{96}Girardi, L., Bressan, A., Chiosi, C., Bertelli, G., Nasi, E. 1996, A\&AS, 117,
113

\bibitem[Gutkin et al. (2016)]{gutkin16} Gutkin, J.,  Charlot, S.,  Bruzual, G.,2016, MNRAS,462,1757G

\bibitem[Hamer et al.(2016)]{hamer} Hamer S. L., Edge A. C. et al. 2016, MNRAS 460, 1758, 1789 
\bibitem[Hunter (2007)]{plot}Hunter, J. D. 2007, Computing In Science \& Engineering, 9, 90

\bibitem[Iani et al.(2019)]{iani19} Iani, E., Rodighiero, G., Fritz, J., Cresci, G., et al. 2019, MNRAS, 487,5593I

\bibitem[Kauffmann et al.(2003)]{kaufm03} Kauffmann, G., Heckman, T. M., Tremonti. C. et al., 2003, MNRAS, 346, 1055

\bibitem[Kennicutt(1998b)]{ken98} Kennicutt, R., C., Jr. 1998, ARA\&A, 36, 189

\bibitem[Kennicutt(1998a)]{ken98a}Kennicutt, R., C., Jr.,1998,ApJ,498,541K

\bibitem[Kennicutt et al.(2012)]{ken12} Kennicutt, R. C., Evans, N. J. 2012, ARA\&A, 50, 531

\bibitem[Kewley et al.(2001a)]{kewley01} Kewley, L.J. et al.,  2001, ApJS, 132, 37

\bibitem[Kewley et al.(2001b)]{kew01}Kewley, L. J.;  Dopita, M. A et al., 2001, ApJ, 556, 121K 

\bibitem[Kewley et al.(2013)]{kewley13a} Kewley, L. J., Dopita, M. A., Leitherer, C., et al. 2013, ApJ, 774, 100

\bibitem[Kewley et al.(2019)]{kewley19}Kewley, L., J., Nicholls, D.,C., Sutherland, R., S,. et al. 2019,ARA\&A, 57, 511K

\bibitem[Kirkpatrick \&McNamara(2015)]{km15}Kirkpatrick, C. C.;  McNamara, B. R.,  2015, MNRAS, 452, 4361K 

\bibitem[Lavoie et al.(2016)]{lavoie16}Lavoie, S.,  Willis, J. P., Democles, J.,  et al.,  2016, MNRAS, 462, 4141

\bibitem[Li et al.(2014b)]{li14}Li, Y., \& Bryan, G. L. 2014b, ApJ, 789, 153

\bibitem[Li et al.(2015)]{li15} Li, Y., Bryan, G. L., Ruszkowski, M., et al. 2015, ApJ, 811, 73

\bibitem[Li et al.(2017)]{li17}Li, Y., Ruszkowski, M., \& Bryan, G. L. 2017, ApJ, 847, 106

\bibitem[James et al.(2009)]{james09}James, B. L.;  Tsamis, Y. G.;  Barlow, M. J., et al., 2009, MNRAS, 398, 2J

\bibitem[Lauer et al.(2014)]{lauer14} Lauer, T. R.,  Postman, M.  Strauss, M. A., ET AL.  2014, ApJ, 797, 82

\bibitem[Loubser et al.(2009) ]{l09}Loubser S. I., Sánchez-Blázquez P., Sansom A. E., Soechting I. K.,
2009, MNRAS, 398, 133

\bibitem[Loubser et al.(2013) ]{l13} Loubser, S. I.; Soechting, I. K., 2013, MNRAS, 431.2933L

\bibitem[Luridiana et al.(2015)]{pyneb}Luridiana, V.,  Morisset, C.,  Shaw, R. A.,  2015, A\&A, 573A, 42L 

\bibitem[Maiolino \& Mannucci(2019)]{m19}Maiolino, R.;  Mannucci, F., 2019, A\&A, Rv, 27, 3M 

\bibitem[Maraston \& Str{\"o}mb{\"a}ck(2011)]{maratson}Maraston C., Str{\"o}mb{\"a}ck G., 2011, MNRAS, 418, 2785

\bibitem[McDonald et al.(2012a)]{mc12}McDonald, M., Veilleux, S., \& Rupke, D. S. N. 2012, ApJ, 746,153 

\bibitem[McDonald et al.(2012b)]{mc12b}McDonald, M.,  Bayliss, M.,  Benson, B. A., et al.,  2012, Natur, 488, 349M

\bibitem[McNamara \& Nulsen (2007)]{mcn07} McNamara B. R., Nulsen P. E. J., 2007, ARA\&A, 45, 117

\bibitem[McNamara et al.(2014)]{mcn14}McNamara, B. R.;  Russell, H. R.;  Nulsen, P. E. J. et al.,  2014, ApJ, 785, 44M 


\bibitem[Merten et al.(2015)]{merten15}Merten, J.;  Meneghetti, M.;  Postman, M., et al. 2015, ApJ, 806, 4M 

\bibitem[Moll{\'a} et al.(2009)] Moll{\'a} M., Garc{\'i}ıa-Vargas M. L., Bressan A., 2009, MNRAS, 398, 451

\bibitem[O’Dea et al.(2008)]{dea08}O’Dea, C. P., Baum, S. A., Privon, G., et al. 2008, ApJ, 681, 1035

\bibitem[Olivares et al.(2019)]{olivares19}Olivares, V.,  Salome, P.,  Combes, F.,  Hamer, S., et al.,  2019, A\&A, 631A, 22O

\bibitem[Olsson et al.(2010)]{olsson10}Olsson, E., Aalto, S., Thomasson, M., \& Beswick, R. 2010, A\&A, 513, A11
\bibitem[Ostriker \& Hausman(1977)]{o77}Ostriker J. P., Hausman M. A., 1977, ApJ, 217, L125


\bibitem[Osterbrock and Ferland(2006)]{of06} Osterbrock D. E., Ferland G. J., 2006, Astrophysics of gaseous nebulae and active galactic nuclei,  2006, agna.book,O

\bibitem[Papaderos et al.(2013)]{polis13}Papaderos, P.,  Gomes, J. M.,  Vílchez, J. M.,  Kehrig, C., et al.  2013, A\&A, 555, L1 

\bibitem[Persad \& Sharma(2015)]{ps15}Prasad D., Sharma P., Babul A., 2015, ApJ, 811, 108

\bibitem[P{\'e}rez-Montero(2014)]{pm14} P{\'e}rez-Montero et al., 2014, MNRAS, 441, 2663

\bibitem[P{\'e}rez-Montero(2017)]{pm17} P{\'e}rez-Montero, E., 2017, PASP, 129d3001P

\bibitem[P{\'e}rez-Montero(2017)]{pm19} P{\'e}rez-Montero et al., 2019, MNRAS, 483, 3322

\bibitem[Pettini \& Pagel(2004)]{pp4}Pettini, M.,Pagel, B. E. J.,  2004, MNRAS, 348L, 59P

\bibitem[Postman et al.(2012)]{postman12} Postman, M., Coe, M., Benitez, N., et al. 2012, ApJS, 199, 25

\bibitem[Proxauf et al.(2014)]{p14}Proxauf, B.;  {\"O}ttl, S.;  Kimeswenger, S.,  2014, A\&A, 561A, 10P 

\bibitem[Prugniel et al.(2007)]{prugniel07}Prugniel P., Soubiran C., Koleva M., Le Borgne D., 2007, ArXiv Astrophysics e-prints,

\bibitem[Rines et al.(2007)]{rines07}Rines, K., Finn, R. et al. 2007, ApJ,665L,9R


\bibitem[Rodr{\'i}guez-Merino et al.(2005)]{meriono05}Rodr{\'i}guez-Merino L. H., Chavez M., Bertone E., Buzzoni A., 2005, ApJ, 626, 411

\bibitem[Rosati et al.(2014)]{rosati14} Rosati, P., Balestra, I., Grillo, C., et al.,  2014, The Messenger, 158, 48


\bibitem[Salpeter(1955)]{salpeter}Salpeter E. E., 1955, ApJ, 121, 161

\bibitem[Santos et al.(2016)]{santos16}Santos, J. S.,  Balestra, I. et al., 2016, MNRAS, 456L, 99

\bibitem[Sarzi et al. (2006)]{gandlaf} Sarzi, M.,  Falcón-Barroso, J.,  Davies, R. L.  2006, MNRAS, 366,1151S 


\bibitem [Schawinski et al.(2007) ]{sw17} Schawinski, K., Thomas, D., 2007, MNRAS, 382, 1415

\bibitem[Sharp \& Bland-Hawthorn(2010)]{sharp10}Sharp \& Bland-Hawthorn 2010, ApJ 711, 818

\bibitem[Shields(1992)]{shields92} Shields, J. C. 1992, ApJL, 399, L27

\bibitem[Soto et al.(2016)]{soto}Soto K. T., Lilly S. J., et al., 2016, MNRAS, 458, 3210

\bibitem[Sparks et al.(2012)]{sparks12}Sparks, W. B., Pringle, J. E.,   Carswell, R. F.,  Donahue, M., et al.,  2012, ApJ, 750L, 5S 

\bibitem[Storn \& Price (1997)]{storn97} Storn \& Price, 1997, Journal of Global Optimization 11: 341–359

\bibitem[Stasinska et al.(2008)]{stasinska08}Stasinska et al. 2008, MNRAS 391, L29

\bibitem[Tremblay et al.(2012)]{tremb12}Tremblay, G. R.; O'Dea, C. P., et al. 2012, MNRAS,424,1042T

\bibitem[Tremblay et al.(2015)]{tremb15}Tremblay, G. R.;  O'Dea, C. P.;  Baum, S. A.,  et al. 2015, MNRAS, 451, 3768T 

\bibitem[Tremblay et al.(2018)]{tr18}Tremblay, G. R.,  Combes, F.; Oonk, J. B. R., 2018, ApJ, 865, 13T

\bibitem[ Robitaille \& Tollerud(2013)]{astropy} Robitaille, T. P., Tollerud, E. J., et al. 2013, A\&A, 558, A33

\bibitem[Thomas \& Dopita(2018)]{neb}Thomas, A. D.,  Dopita, M., et al., 2018, ApJ, 856, 89T 

\bibitem[Umetsu et al.(2016)]{umetsu16}Umetsu, K., Zitrin, A.,   Gruen, D., et al.,  2016, ApJ, 821, 116 

\bibitem[Vantyghem et al.(2016)]{Vantyghem16}Vantyghem, A. N., McNamara, B. R., Russell, H. R., et al. 2016, ApJ, 832, 148

\bibitem[Vale Asari et al.(2016)]{bond} Vale Asari, N.; Stasinska, G.; Morisset, C.,  2016, MNRAS, 460, 1739

\bibitem[Veilleux \& Osterbrock(1987)]{vel87} Veilleux \& Osterbrock 1987, ApJ Suppl. 63, 295

\bibitem[Vikhlinin et al.(2005)]{v05} Vikhlinin A., Markevitch M., Murray S. S., Jones C., Forman W., Van Speybroeck L., 2005, ApJ, 628, 655

\bibitem[van der Velden (2020)]{cmasher} van der Velden, E. 2020, Journal of Open Source 1 Software, 5(46), 2004 

\bibitem[van der Walt et al.(2011)]{numpy}van der Walt, S., Colbert, S. C., \& Varoquaux, G. 2011, Computing in Science Engineering, 13, 22

\bibitem[Voit \& Donahue (2011)]{voit11}Voit, G. M.,  Donahue, M.,  2011, ApJ, 738L., 24V 

\bibitem[Voit \& Donahue (2015)]{voit15b}Voit, M., Donahue, M., et al., 2015, ApJ, 799L,1V

\bibitem[Voit et al.(2017)]{voit17}Voit, G. M., Meece, G., Li, Y., et al. 2017, ApJ, 845, 80
\bibitem[Webb et al.(2015)]{webb15} Webb, T. M. A., Muzzin, A.,  Noble, A., et al.,  2015, ApJ, 814, 96W 

\bibitem[Weilbacher et al.(2020)]{pipeline}Weilbacher P. M., Palsa, R., Streicher O., et al., 2020, arXiv:200608638W 

\bibitem[Zhang et al.(2016)]{zhang16}Zhang et al. 2016, MNRAS, 466, 3217



\end{thebibliography}
\end{document}